\documentclass{jfm}

\usepackage{graphicx}
\usepackage{caption}
\usepackage{subcaption}
\usepackage{hyperref} 
\usepackage{natbib,amsmath}
\usepackage{mathtools}
\usepackage[dvipsnames]{xcolor}

\ifCUPmtlplainloaded \else
  \checkfont{eurm10}
  \iffontfound
    \IfFileExists{upmath.sty}
      {\typeout{^^JFound AMS Euler Roman fonts on the system,
                   using the 'upmath' package.^^J}%
       \usepackage{upmath}}
      {\typeout{^^JFound AMS Euler Roman fonts on the system, but you
                   dont seem to have the}%
       \typeout{'upmath' package installed. JFM.cls can take advantage
                 of these fonts,^^Jif you use 'upmath' package.^^J}%
      }
  \else
  \fi
\fi

\ifCUPmtlplainloaded \else
  \checkfont{msam10}
       \iffontfound
    \IfFileExists{amssymb.sty}
      {\typeout{^^JFound AMS Symbol fonts on the system, using the
                'amssymb' package.^^J}%
       \usepackage{amssymb}%
         \let\leq=\leqslant
         
      }{}
  \fi
\fi

\ifCUPmtlplainloaded \else
  \IfFileExists{amsbsy.sty}
    {\typeout{^^JFound the 'amsbsy' package on the system, using it.^^J}%
     \usepackage{amsbsy}}
    {\providecommand\boldsymbol[1]{\mbox{\boldmath $##1$}}}
\fi

\newsavebox{\astrutbox}
\sbox{\astrutbox}{\rule[-5pt]{0pt}{20pt}}

  \newcommand{\rmd}{{\rm d}}
\newcommand{\bx}{{ \boldsymbol{x} }}

\newcommand{\tbA}{{ \widetilde{\boldsymbol{A}} }}

\newcommand{\bE}{{\mathbb{E}}}
\newcommand{\bW}{{\wt{\mathbf{W}}}}
\newcommand{\bu}{{ \boldsymbol{u}}}

\newcommand{\bom}{{\mbox{\boldmath $\omega$}}}
\newcommand{\hd}{\hat{\rmd}}

\newcommand{\wt}{\widetilde}

\newcommand{\tsigma}{{\tilde{\sigma}}}

\newcommand{\bm}{{\boldsymbol{{\rm p}}}}
\newcommand{\bn}{\hat{{\boldsymbol{n}}}}

\newcommand{\grad}{{\mbox{\boldmath $\nabla$}}}
\newcommand{\btimes}{{\mbox{\boldmath $\times$}}}

\newcommand{\bSigma}{{\mbox{\boldmath $\Sigma$}}}
\newcommand{\btau}{{\mbox{\boldmath $\tau$}}}

\newcommand{\be}{\begin{equation}}
\newcommand{\ee}{\end{equation}} 
\newcommand{\lb}{\label}

\def\XXint#1#2#3{{\setbox0=\hbox{$#1{#2#3}{\int}$}
     \vcenter{\hbox{$#2#3$}}\kern-.5\wd0}}

\title[Stochastic Lagrangian Dynamics]{Stochastic Lagrangian Dynamics of Vorticity.\\
II. Channel-Flow Turbulence}

\author[G. L. Eyink et al.]%
{G. L. Eyink$^{1,2,3}$, A. Gupta$^{1}$, and T. Zaki$^{3}$}

\affiliation{$^1$Department of Applied Mathematics \& Statistics, The Johns Hopkins University, Baltimore, MD 21218, USA\\[\affilskip]
$^2$Department of Physics \& Astronomy, The Johns Hopkins University, Baltimore, MD 21218, USA\\[\affilskip]
$^3$Department of Mechanical Engineering, The Johns Hopkins University, Baltimore, MD 21218, USA}

\pubyear{2020}
\volume{???}
\pagerange{???}
\date{?; revised ?; accepted ?. - To be entered by editorial office}
\begin{document}

\maketitle

\begin{abstract}
We here exploit a rigorous mathematical theory of vorticity dynamics for Navier-Stokes 
solutions in terms of stochastic Lagrangian flows and their stochastic Cauchy invariants,  
that are conserved on average backward in time. This theory yields exact expressions 
for the vorticity inside the flow domain in terms of the vorticity at the wall, as it is 
transported by viscous diffusion and by nonlinear advection, stretching and rotation. 
 As a concrete application, we exploit an online database of a 
turbulent channel-flow simulation at $Re_\tau=1000$ \citep{graham2016web} to determine
the origin of the vorticity in the near-wall buffer layer. Following an experimental
study of \cite{sheng2009buffer}, we  identify typical ``ejection'' and ``sweep'' events in the buffer layer 
by local minima/maxima of the wall-stress. In contrast to their conjecture, however, we find that 
vortex-lifting from the wall is not a discrete event requiring only $\sim 1$ viscous time and 
$\sim 10$ wall units, but is instead a distributed process taking place over a space-time region at least 
$1\sim 2$ orders of magnitude larger in extent. We show that Lagrangian chaos observed in the 
buffer layer can be reconciled with saturated vorticity magnitude by a process of  ``virtual reconnection'':
although the Eulerian vorticity field in the viscous sublayer has only a single sign of spanwise component, 
opposite signs of Lagrangian vorticity evolve by rotation and cancel by viscous destruction. Our 
analysis reveals many unifying features of classical fluids and quantum superfluids. We argue that 
``bundles'' of quantized vortices in superfluid turbulence will also exhibit stochastic Lagrangian dynamics
and will satisfy stochastic conservation laws resulting from particle relabelling symmetry. 
\end{abstract}




%

\section{Introduction}

Vorticity is well-recognized to play a fundamental role in turbulent flows and the ultimate origin 
of the vorticity observed in most terrestrial fluid turbulence lies at solid walls or flow boundaries. 
It therefore becomes a basic question to understand how vorticity in the interior of the flow 
evolved from vorticity generated at the wall. Recent mathematical work of  
\cite{ConstantinIyer08,ConstantinIyer11} has provided new exact tools to answer this 
question in terms of stochastic Lagrangian particle trajectories evolved backward in time.
In the previous work  of \cite{eyink2020stochasticI}---hereafter referred to as paper I---it was 
shown that the mathematical representations of \cite{ConstantinIyer08,ConstantinIyer11} have 
a simple fluid-dynamical interpretation in terms of the ``vortex-momentum density'' associated 
to a continuous distribution of infinitesimal vortex-rings, which is the basis of the   
\cite{kuzmin1983ideal}-\cite{oseledets1989new} formulation of the incompressible Navier-Stokes 
equation. For smooth ideal Euler solutions, the vortex-momentum density is Lie-transported by the fluid flow 
as a 1-form and its curl, the vorticity, is transported as a 2-form \citep{oseledets1989new,tur1993invariants}.
This Lie-transport leads naturally to the vorticity invariants of \cite{cauchy1815theorie} for incompressible
Euler solutions and to generalized Cauchy invariants for the vortex-momentum density 
\citep{tur1993invariants,besse2017geometric}. The mathematical
theory of \cite{ConstantinIyer08,ConstantinIyer11} provides corresponding ``stochastic 
Cauchy invariants'' for Navier-Stokes solutions that are conserved on average by 
the stochastic Lagrangian flow backward in time and these invariants provide an explicit  
representation of the vortex-momentum density and the vorticity in terms of boundary data.
We furthermore discussed in paper I some relations of the \cite{ConstantinIyer08,ConstantinIyer11} 
stochastic Lagrangian representation with the Eulerian theory of 
\cite{lighthill1963boundary}-\cite{morton1984generation} for vorticity-generation 
at solid walls and with an exact statistical result of 
\cite{taylor1932transport}-\cite{huggins1994vortex} for the ``vorticity flux tensor''.
The latter is an anti-symmetric tensor $\Sigma_{ij},$ which represents the flux 
of the $j$th-component of the vorticity in the $i$th coordinate direction and which was 
observed by \cite{huggins1994vortex} (following \cite{taylor1932transport} for 2D pipe flow) 
to have a mean value directly proportional to the mean pressure-gradient 
in the $k$th direction, with $i,$ $j,$ $k$ all distinct. This relation applies to drag
generally for any flows driven by imposed pressure-gradients and/or freestream 
velocity, such as turbulent shear layers and wakes \citep{brown2012turbulent}.  The ``vorticity 
source'' of \cite{lighthill1963boundary} and \cite{morton1984generation} is defined at the wall
by the vector $\sigma_i=\Sigma_{ij} \hat{n}_j,$ summed over repeated index $j,$ where 
$\bn$ is the outward-pointing unit normal vector on the boundary
\citep{lyman1990vorticity,eyink2008turbulent}. 

The main result that we exploit in the present work is the relation [paper I, eq.(2.50)] 
\be  \bom(\bx,t) = \bE\left[\widetilde{\bom}_s(\bx,t)\right], \quad s<t  \lb{eqII2-1} \ee
which expresses the vorticity vector at space-time point $(\bx,t)$ as an expectation $\bE$ of the {\it stochastic Cauchy invariant} 
$\widetilde{\bom}_s(\bx,t)$ [paper I, eq.(2.51)]. This is a random vector that can be evaluated along 
stochastic Lagrangian particle trajectories satisfying 
\be \hd \tbA^s_t(\bx) = \bu(\tbA^s_t(\bx),s) \, \rmd s + \sqrt{2\nu}\, \hd\bW(s), \quad  s<t; \qquad \tbA^t_t(\bx)=\bx,  
\lb{eq2-32} \ee
which are released at $(\bx,t)$ and move backward in time $s.$ Here $\bW(s)$ is a random Wiener process that 
represents diffusion by molecular viscosity. For every realization of the process there is a largest time $s=\widetilde{\sigma}_t(\bx)$
at which the stochastic particle first hits the wall, backward in time. Each particle is stopped at the wall where 
it first hits and in that particular realization $\widetilde{\bom}_s(\bx,t)=\widetilde{\bom}_{\widetilde{\sigma}_t(\bx)}(\bx,t)$ for 
$s<\widetilde{\sigma}_t(\bx),$ thus remaining fixed at earlier times. If one considers $s\ll t,$ then the inequality
$s<\widetilde{\sigma}_t(\bx)$ will hold with near certainty and, in that case, the formula \eqref{eqII2-1} represents
the interior vorticity in terms of wall vorticity which is transported to $(\bx,t)$ by advection, diffusion and stretching.    
A numerical Monte Carlo Lagrangian algorithm was also developed in paper I to calculate
realizations of the stochastic Cauchy invariants and their statistics, given an Eulerian 
space-time solution of the incompressible Navier-Stokes equation. 

Here we shall exploit 
that approach to make a first-of-its-kind numerical study of stochastic Lagrangian dynamics 
of vorticity in a turbulent channel-flow at high Reynolds number. If we denote Cartesian 
coordinate directions as streamwise $x,$ wall-normal $y,$ and spanwise $z,$ then statistical 
homogeneity in $x$ and $z$ and reflection-symmetry in $y$ provide simplifications in long-time
averages. In particular,  the mean flux of spanwise vorticity vertically from the wall becomes independent of 
wall-normal location as a consequence of conservation, $\partial_y\overline{\Sigma}_{yz}=0,$ and is given by 
\be   \overline{\Sigma}_{yz}=
\overline{v'\omega_z'-w'\omega_y'} -\nu \frac{\partial \overline{\omega}_z}{\partial y}
= \frac{\partial\overline{p}}{\partial x} = -\frac{u_*^2}{h}<0.\lb{eq4-1} \ee
where $f':=f-\overline{f}$ defines fluctuation away from the long-time mean value, where velocity vector 
$\bu=(u,v,w),$ where $u_*$ is the friction velocity, and where $h$ is the channel half-height
\citep{huggins1994vortex,eyink2008turbulent}. 
According to eq.\eqref{eq4-1} the constant flux of spanwise vorticity away from the wall is numerically 
equal to the rate of mean downstream pressure drop, which characterizes turbulent drag and dissipation.   
In this paper we shall study the Lagrangian dynamics of vorticity in the buffer layer of 
turbulent channel-flow, which is conventionally taken to be the layer of fluid at heights $y^+$ over the 
range $5<y^+<30$ \citep{tennekes1972first}, with $y^+$ in dimensionless wall units 
(see section \ref{sec:numerics}).  We have both pragmatic and fundamental reasons 
to focus on the buffer layer. Since all interior flow vorticity traces its origin back to a solid wall, the 
``youngest'' vorticity must lie closest to the wall. This makes the near-wall region numerically easiest to study 
by our approach. There is also motivation to understand the turbulent physics close to the 
wall, since one can expect that control of vorticity creation and transport at the earliest stage will
be most efficient in reducing drag \citep{koumoutsakos1999vorticity,zhao2004turbulent}. On the other 
hand, nonlinear contributions to the dynamics are subleading (on average) in the viscous sublayer 
$y^+<5,$ and nonlinear contributions to mean vorticity transport first become substantial in the buffer layer.

\begin{figure}
  \centering
  \includegraphics[width=0.5\textwidth]{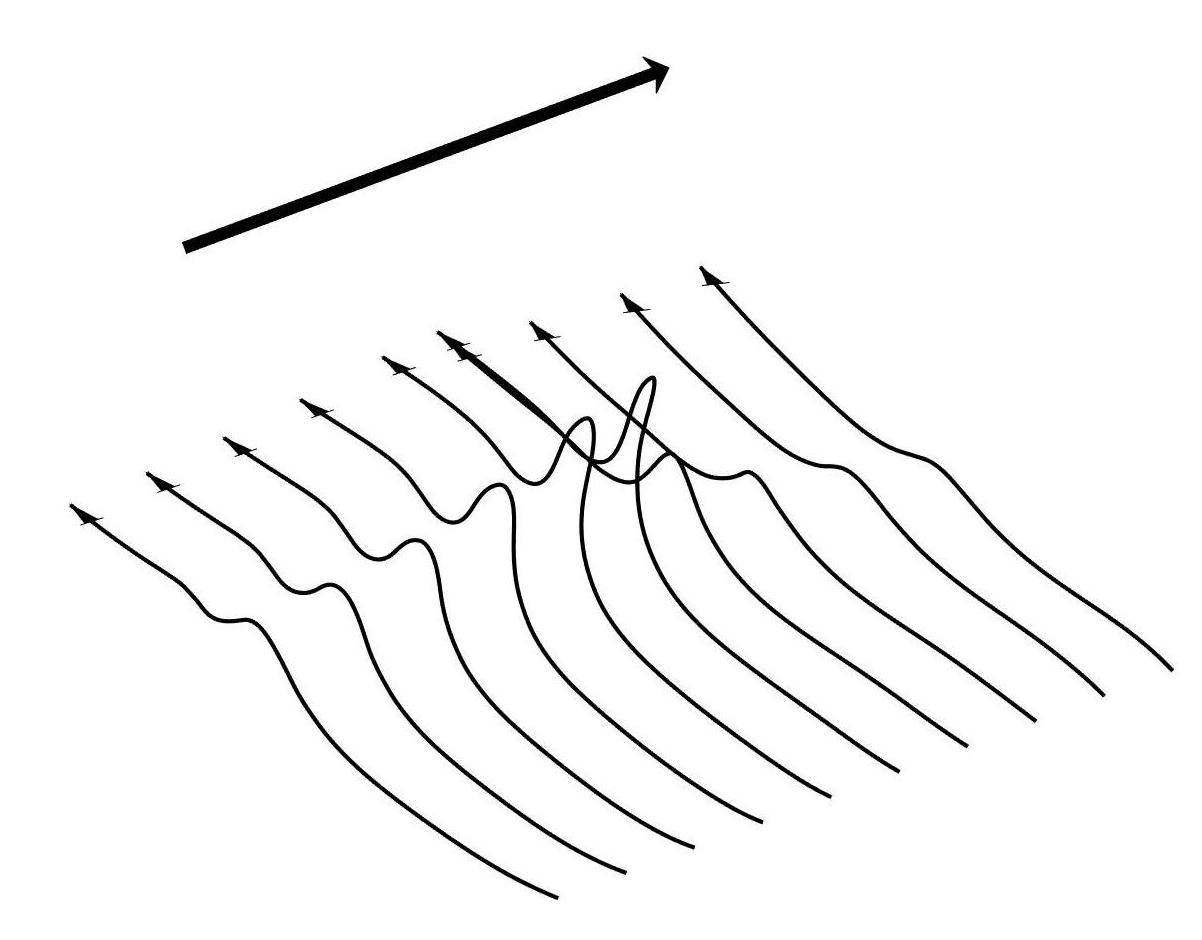}
  \caption{A typical array of vortex lines pointing spanwise and lifting in an arch over a low-speed streak 
  at a wall-stress minimum during an ``ejection'' event in the buffer-layer of a turbulent boundary layer. 
  The wide arrow indicates the direction of the mean flow.}  \label{fig:JustLines}
\end{figure}

Our study is directly motivated by laboratory experiments 
of \cite{sheng2009buffer}, who investigated buffer-layer structures in a turbulent square-duct channel flow 
at $Re_\tau=1470$ using a technique of digital holographic microscopy that yields well-resolved measurements 
of three-dimensional velocity and velocity-gradient fields.  Conditional sampling based on local 
wall shear-stress maxima and minima revealed two types of structures that generate such extreme 
stress events. In accord with a common terminology, these may be called  ``sweeps'' and ``ejections'',
respectively. The latter type of flow event generates arrays of vortex lines with a simple ``hairpin'' structure
that rise in an arch above the location of the wall-stress minimum. See Fig.~\ref{fig:JustLines}, which
represents well the typical geometry of vortex lines observed by \cite{sheng2009buffer} during an ``ejection''.   
These raised vortex lines with a non-vanishing vertical component  of vorticity are the signature closest 
to the wall of a contribution by nonlinear stretching and rotation to transport of spanwise vorticity upward from the wall. 
Such an orderly spatial array of lines as illustrated in  Fig.~\ref{fig:JustLines} invites interpretation in terms of a 
similarly simple temporal progression, with each line apparently ``moving'' forward and evolving into its 
downstream neighbor. Indeed, \cite{sheng2009buffer} on the basis of such spatial arrays of lines 
(see their Figures 7 \& 21) have proposed an ``abrupt lifting'' of vorticity in just one viscous time or, spatially,  
in a short distance of 10 wall units, above the location of a local stress-minimum.

As we shall show by 
detailed Lagrangian analysis exploiting the stochastic Cauchy invariants, the deceptively simple Eulerian 
picture of vortex-lines in Fig.~\ref{fig:JustLines} is in fact the outcome of a hidden, violent conflict between intense 
nonlinear stretching and rotation of vorticity vector elements on the one hand, and vigorous destruction  
of vorticity by viscosity on the other. Previous work of \cite{johnson2017analysis} has demonstrated 
existence of Lagrangian chaos in channel-flow turbulence, which leads to rapid exponential stretching 
of material line elements. This poses a theoretical puzzle, however, because such stretching should lead to 
unbounded growth of vorticity but the channel flow nevertheless attains a statistical steady state with a 
saturated magnitude of vorticity.   The obvious mechanism that limits vortex-stretching is viscous destruction
\citep{taylor1938production}, but advection and diffusion of same-sign vorticity 
cannot quench the growth due to stretching. We find that vorticity vector elements in the 
buffer layer are not only exponentially magnified, but also strongly rotated, so that they often point
opposite to the negative spanwise direction pictured in Fig.~\ref{fig:JustLines}. Cancellation 
of this oppositely-directed vorticity by viscous diffusion leads to the regular geometry 
in Fig.~\ref{fig:JustLines}. Furthermore, we find that the vortex lifting is not an ``abrupt'' discrete 
event, but is instead a highly distributed process spread over more than 100 viscous times 
and 1000 wall units.

These results are obtained by a numerical study using an online computational dataset of 
channel-flow turbulence at $Re_\tau=1000$ \citep{graham2016web}. The accuracy of 
this database to study buffer-layer physics has been carefully evaluated and documented in
paper I, as briefly summarized in section \ref{sec:numerics}. Our study not only yields 
new insights into Lagrangian vorticity dynamics of pressure-driven wall-bounded flows, but 
also reveals as well many common features of classical fluids and quantum superfluids, 
especially for wall-bounded turbulent flows through pipes and channels. To mention here just a 
few of these similarities, the \cite{kuzmin1983ideal}-\cite{oseledets1989new} representation of the 
classical fluid by the vortex-momentum density is closely related to intuitive pictures of 
``vortex-tangles'' in superfluids as superpositions of small vortex-rings 
\citep{campbell1972critical,schwarz1982generation,kuzmin1999vortex}.
Perhaps most importantly, \cite{huggins1994vortex} has emphasized that eq.\eqref{eq4-1} 
is the exact analogue of the ``Josephson-Anderson relation'' in quantum fluids;  
see \cite{josephson1962possible,anderson1966considerations}, and \cite{packard1998role,varoquaux2015anderson} 
for reviews. This relation provides the accepted explanation for effective drag in an otherwise 
non-dissipative superfluid by cross-stream motion of quantized vortex lines. For further 
discussions of this analogy, see also \cite{huggins1970energy} and \cite{eyink2008turbulent}. 
We argue briefly in the conclusion of the present paper that such similarity should extend to Lagrangian 
dynamics of vorticity and that motion of ``bundles'' of quantized vortex lines in turbulent superfluids 
should be also intrinsically stochastic. 


The contents of this paper are outlined as follows. Section \ref{sec:numerics} summarizes
the essentials of our numerical methods, which are described more completely in paper I.  
Section \ref{sec:chflow} explains how the two specific flow events were selected for examination (\ref{sec:identify})
and then describes both the ejection (\ref{sec:EulerEject}) and sweep (\ref{sec:EulerSweep}) events 
in detail in conventional Eulerian terms. Our novel Lagrangian analysis is presented in section 
\ref{sec:lagrange}, where we first choose specific vorticity vectors for quantitative study (\ref{sec:select}) 
and then determine their dynamical origin at the wall for both the ejection (\ref{sec:lag_eject}) 
and the sweep (\ref{sec:lag_sweep}). The final section \ref{sec:conclusion} summarizes 
our results and conclusions, especially on common features of wall-bounded turbulence 
in classical fluids and quantum superfluids. Additional material that complements the discussion 
in the main text is provided as Supplemental Materials (SM). 

\begin{table}  \label{tab:table1}
    \begin{center}
    \vspace{-10pt} 
    \caption{Simulation Parameters for Turbulent Channel-Flow Dataset}
    \vspace{10pt} 
    \begin{tabular}{c  c  c  c  c  c  c   c  c   c  c  c  c   c   c    c} 
      $N_x$ && $N_y$ && $N_z$ && $Re_\tau$ & $dp/dx$ & $\nu$ & $u_*$ & $U_{bulk}$ && $\Delta x^+$ && $\Delta z^+$ & $\Delta t$\\
      \vspace{-8pt} \\
      2048 && 512 && 1536 && 1000 & $-2.5\times 10^{-3}$ & $5\times 10^{-5}$  & $5\times 10^{-2}$ & 1.00 && 12.3 && 6.1 & $1.3\times 10^{-3}$ \\
      \vspace{1pt} 
     \end{tabular}
  \end{center}
\end{table}

\section{Numerical Methods}\lb{sec:numerics}   

We review here for completeness some necessary information about our computational 
approach from paper I,  which should be consulted for full details.   
The Johns Hopkins Turbulence Databases (JHTDB) channel-flow dataset \citep{graham2016web} is exploited for the empirical study in this paper.
This data was generated from a Navier-Stokes simulation in a channel using a pseudospectral method in the plane parallel to the 
walls and a seventh-order B-splines collocation method in the wall-normal direction \citep{lee2013petascale}. For numerical solution, the 
Navier-Stokes equations were formulated in wall-normal velocity-vorticity form \citep{kim1987turbulence}. Pressure was computed by 
solving the pressure Poisson equation only when writing to disk, which was every five time steps for 4000 snapshots, enough for about one domain 
flow-through time. The simulation domain $[0,8\pi h] \times [-h,h] \times [0,3\pi h],$ $h=1,$ was discretized using a spatial grid of 
$2048 \times 512 \times 1536$ points. Time advancement 
was made with a third-order low-storage Runge-Kutta method and dealiasing was performed with 2/3 truncation \citep{orszag1971elimination}. 
A constant pressure head was applied to drive the flow at $Re_\tau = 1000$ ($Re_{bulk} = \frac{2hU_{bulk}}{\nu} = 40\,000$) with 
bulk velocity near unity. As is common, we shall indicate with a superscript ``+'' non-dimensionalized quantities in viscous wall units,
with velocities scaled by friction velocity $u_*$ and lengths by viscous length $\delta_\nu=\nu/u_*=10^{-3}.$ Also as usual, we define 
$y^+=(h\mp y)/\delta_\nu$ near $y=\pm h.$ In these units,
the first $y$-grid point in the simulation is located at distance $\Delta y_1^+ = 1.65199\times 10^{-2}$ from the wall, while in the center 
of the channel $\Delta y_c^+ = 6.15507.$ Other numerical parameters are summarized in Table 1. 

In paper I we developed and tested a numerical Monte Carlo Lagrangian algorithm to calculate the 
stochastic Cauchy invariants and their statistics,  by discretizing the stochastic ODE \eqref{eq2-32} with 
a step-size $\Delta s$ and by averaging over $N$ independent realizations $\bW^{(n)}(s),$ $n=1,...,N$ of the 
Wiener process. We showed in that work for the two specific cases from the JHTDB channel-flow 
dataset studied in the present paper that $\Delta s=10^{-3}$ and $N=10^{7}$ sufficed to give converged results 
for the mean and variance of the stochastic Cauchy invariant over a time interval $-100<\delta s^+<0$ with $\delta s:=s-t.$    
In particular, it was shown that the mean conservation law \eqref{eqII2-1} holds for those two cases to within 
a few percent over that time interval,  which is a quite stringent test of validity of our numerics. The residual 
errors in the mean conservation can be explained by some near-wall under-resolution of the numerical 
JHTDB data, indicated by the sizable $\Delta x^+$ and $\Delta z^+$ values in Table 1, and by errors in the 
finite-difference approximation of velocity-gradients within the database. To test that hypothesis, we also spatially 
coarse-grained the relation \eqref{eqII2-1} over $n_i$ grid-spacings in each of the coordinate directions $i=x,y,z,$ since 
such coarse-grained fields from the JHTDB dataset should represent better a coarse-grained Navier-Stokes 
solution. We verified that such coarse-graining noticeably improves mean conservation, in particular 
for $(n_x,n_y,n_z)=(4,0,4).$ 

In sections \ref{sec:identify} and \ref{sec:select} below, we describe the criteria that we used to select 
the two test cases for study in this work. In particular, we discuss in section \ref{sec:identify} how a pair of events, 
an ``ejection'' and a ``sweep'', were identified in the Eulerian dataset, following the work of \cite{sheng2009buffer}. 
In section \ref{sec:select} we then describe how we selected a space-time point $(\bx,t)$ in each event 
for comparative study by stochastic Lagrangian analysis. We also show there that coarse-graining the 
JHTDB fields with $(n_x,n_y,n_z)=(4,0,4)$ does not change qualitatively the Eulerian and Lagrangian 
picture of the two events. These results, together with those presented in paper I,  validate both 
our Monte Carlo numerical method to calculate the stochastic Cauchy invariant and also the adequacy 
of the JHTDB channel-flow database to resolve the physics of the turbulent buffer layer.
 
\section{Eulerian Vorticity Dynamics in the Buffer Layer}\lb{sec:chflow} 

\subsection{Identification of Ejections and Sweeps}\lb{sec:identify} 

Following the approach of \cite{sheng2009buffer}, we selected events where the viscous shear stress 
$\tau_{xy}=\nu (\partial u/\partial y)$ at the wall has local minima and local maxima with magnitudes 
satisfying a threshold condition. For this purpose, we downloaded the stress field at the entire 
top wall of the channel-flow database at the final archived time $t_f=25.9935$ and searched 
for local minima and maxima. Note that we used the data at the top wall because the bottom-wall data was 
temporarily unavailable when we began our study; in order to present our results below we always 
rotate the figures $180^\circ$ around the streamwise axis, so that the top wall is exchanged with 
the bottom wall. In SM we provide a plot of the normalized stress field $\tau_{xy}^+=\tau_{xy}/u_*^2$
over the entire channel wall and a PDF of its values, which range from $-2.55$ to $+7.54$ and 
have mean value unity. To find local minima and maxima, we used a fast peak-finder for 
2D scalar data \citep{natan2013fast}. We found that the points identified by this code for field 
$\tau_{xy}$ were indeed local maxima and for $-\tau_{xy}$ were local minima, but that weaker
maxima/minima were often missed. We therefore do not regard the output of this algorithm to be
completely reliable to obtain statistics of the local extrema, but it suffices for our purpose of 
identifying specific local maxima/minima. Nevertheless, we do provide in the SM for the interest 
of readers the obtained PDF's of the stress values at the positions both of the local minima and also of the 
local maxima. The PDF of the local-minimum stress values shows a large peak at $\tau_p^+\doteq 0.6,$ 
while the PDF of the local-maximum stress values shows a large peak at $\tau_p^+\doteq 1.8.$ 
Interestingly, the condition that \cite{sheng2009buffer} applied to identify ``extreme stress events''
was $\tau_{xy}^+<0.6$ for local-minima and $\tau_{xy}^+>1.8$ for local-maxima, in good agreement 
with these peaks values. We therefore searched for two typical events of this type, namely, 
for a local-minimum of stress with value $\tau_{xy}^+\doteq 0.6$ and for a local-maximum of 
stress with value $\tau_{xy}^+\doteq 1.8.$ We also looked for such local-extremum points 
which were relatively isolated from others. After examining several candidates, we selected as 
representative the two local-extrema with database space-time coordinates given in Table 2. 
The reader will note that these coordinates correspond, as stated above, to points on the 
top wall of the channel.  
The reader should transform results in paper I according to 
$(\omega_x,\omega_y,\omega_z)\to (\omega_x,-\omega_y,-\omega_z)$
for consistency with visualizations in the present paper. In particular, mean spanwise
vorticity $\overline{\omega}_z$ under this transformation becomes negative.

\begin{table} 
    \begin{center}
    \vspace{-10pt} 
    \caption{Coordinates of Local-Minimum and Local-Maximum Wall-Stress Events}
    \vspace{10pt} 
    \begin{tabular}{c  c  c  c  c  c} 
                      &&  $x$  &  $y$  &  $z$  &  $t$   \\
      local-minimum &&  21.107576 & 1.000000 & 7.565593 & 25.9935 \\
      \vspace{-8pt} \\
      local-maximum && 0.711767 & 1.000000  & 0.724039  & 25.9935\\
      \vspace{1pt} 
     \end{tabular}
  \end{center}
   \label{tab:table2}
\end{table}

We shall refer to the local-minimum stress event as an ``ejection'' and to the local-maximum 
stress event as a ``sweep.'' This terminology is in agreement with the classification of 
\cite{sheng2009buffer} in their Table 2, but it differs somewhat from the most common 
characterization of such structures based on quadrant analysis in the $(u',v')$ velocity 
plane, with connected regions of Q2 fluctuations designated as ``ejections'' and regions 
of Q4 fluctuations as ``sweeps'' \citep{jimenez2013near}. As we shall see from a detailed 
Eulerian description of these two selected events in the following sections 
\ref{sec:EulerEject}-\ref{sec:EulerSweep}, our use of the 
terms ``ejection'' and ``sweep'' is not unrelated with the traditional quadrant analysis. We have purposely
avoided using the term ``burst'' to describe either of our two events, although ``ejections'' have
sometimes in the past been equated with ``bursts''. In more current understanding, however, 
buffer-layer ``bursting'' is believed to be associated with quasi-periodic breakdown of unstable 
coherent structures presumably described by traveling-wave solutions of Navier-Stokes 
equations (\cite{jimenez2013near}, sections IV.A-B; \cite{park2018bursting}). The quasi-period of this 
bursting is expected to be $\simeq 400$ viscous times $t_\nu=\nu/u_*^2$ with duration $\simeq 100$
viscous times, during which the traveling structure evolves from a low wall-stress to high wall-stress
state. E.g. see \cite{jimenez2013near}, Figure 10 or \cite{park2018bursting}, Figure 6.  Therefore 
``ejections" and ``sweeps" in our sense may both be associated with buffer-layer bursting, at 
different stages in the evolution of the burst.  Our interest here is mainly in analyzing the Lagrangian 
dynamics of vorticity within these two buffer-layer events and not in determining their relation 
with ``bursting.'' 

In subsections \ref{sec:EulerEject}-\ref{sec:EulerSweep} below we first provide a detailed description 
of these events in standard Eulerian terms. This does illuminate some connections with ``bursting'',
but our primary purpose is to describe these two events in terms of standard Eulerian theory 
of vorticity generation at walls and to compare with prior numerical and experimental results. 
Among these, we wish to compare our chosen events with those selected in \cite{sheng2009buffer}
by identical criteria and to verify that our events have the same characteristic features. 
In particular, we shall show that our ``ejection'' event is quite typical of those studied by \cite{sheng2009buffer}
and used by them to postulate the ``abrupt lifting'' of vortex-lines. Here it is appropriate to make a remark 
on the relative importance of nonlinear dynamics and viscous diffusion for vorticity transport in the buffer layer. 
As stressed in the Introduction, an ``abrupt lifting'' event requires nonlinear stretching and rotation of spanwise 
vorticity in order to create a vertical arch. However, linear viscous diffusion dominates the mean vertical flux 
of spanwise vorticity not only in the viscous sublayer and buffer layer but even into the log-layer! In fact, the 
simple relation 
\be \overline{v'\omega_z'-w'\omega_y'} =  \frac{\partial}{\partial y}\left(\,\overline{-u' v'}\,\right) \lb{eq4-2} \ee
implies that the nonlinear contribution to spanwise vorticity flux is positive for $y^+<y^+_p,$ \\ the location 
of the peak Reynolds shear stress $\overline{-u' v'}$ \citep{klewicki2007physical,eyink2008turbulent}. 
In the JHTDB channel-flow data at $Re_\tau=1000,$ this peak occurs at about $y_p^+\doteq 50$
\citep{graham2016web}. The net effect of the nonlinear terms for $y^+<y^+_p$ is thus to transport 
spanwise vorticity {\it opposite} to the conserved total flux $\overline{\Sigma}_{yz}$
in eq.\eqref{eq4-1}, and viscous transport must be even more negative to compensate. We shall see 
also in the Lagrangian description that viscous diffusion plays an essential role in buffer-layer vorticity transport, 
even for extreme-stress events where nonlinear terms are enhanced.



\subsection{Eulerian Description of Ejection Event}\lb{sec:EulerEject}

\begin{figure}
\centering 
\includegraphics[width=0.8\textwidth]{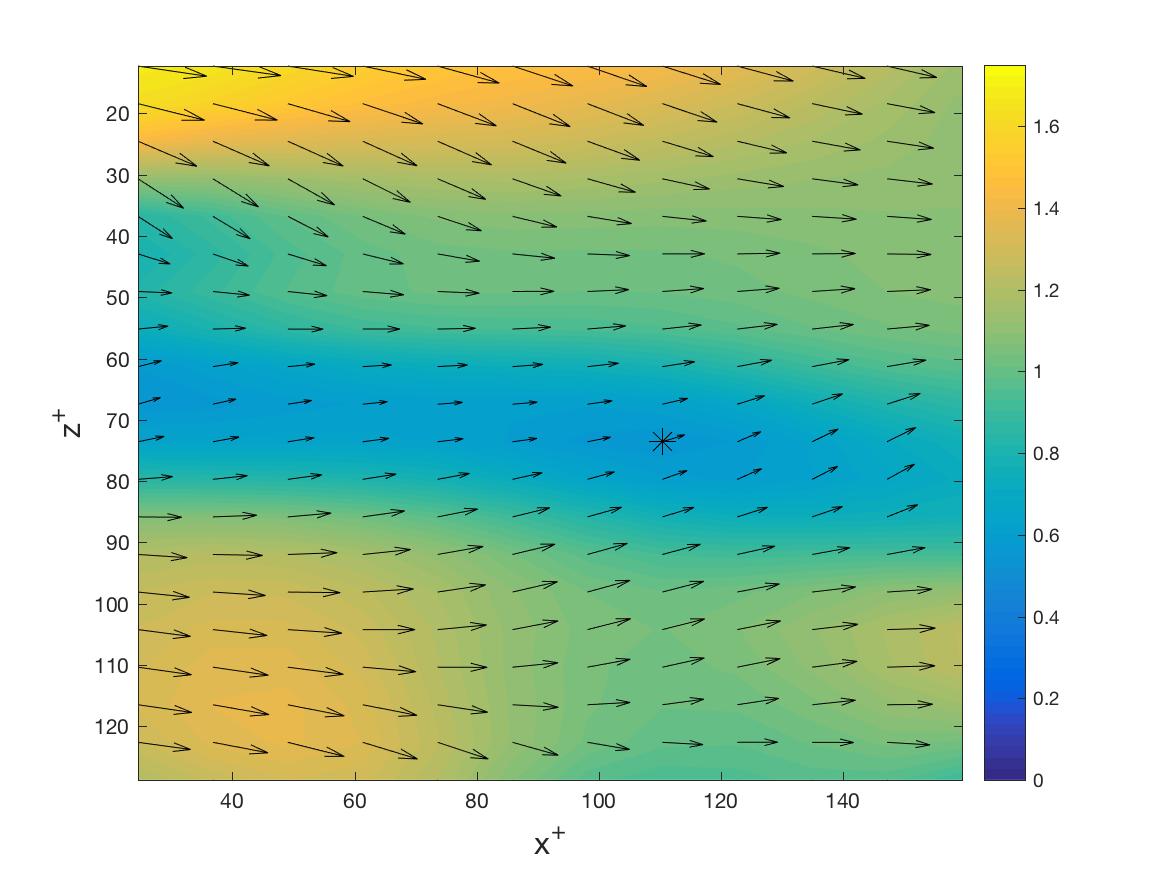}
\caption{Field of viscous shear stress $\tau_{xy}^+=(\partial u^+/\partial y^+)$ at the wall $y^+=0$, in wall units so that 
the average is unity. Black arrows represent the two-dimensional in-wall stress vector $\btau_W.$ The 
asterisk ``$*$'' marks the location of the selected stress local-minimum.} 
\label{fig4-stressL} 
\end{figure} 

\begin{figure}
\begin{subfigure}[b]{\textwidth}
  \centering
  \includegraphics[width=.8\linewidth]{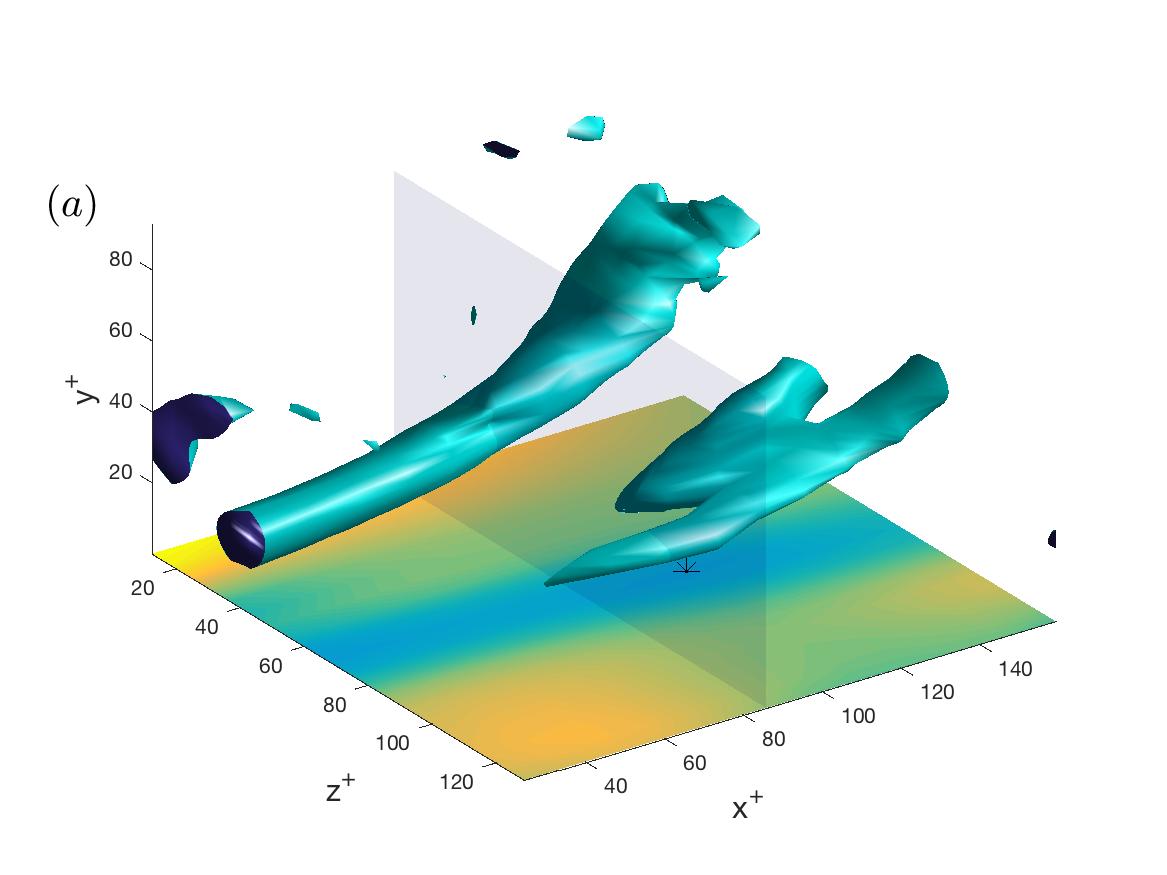}  
\end{subfigure}
\\
\begin{subfigure}[b]{\textwidth}
\centering
  \includegraphics[width=.8\linewidth]{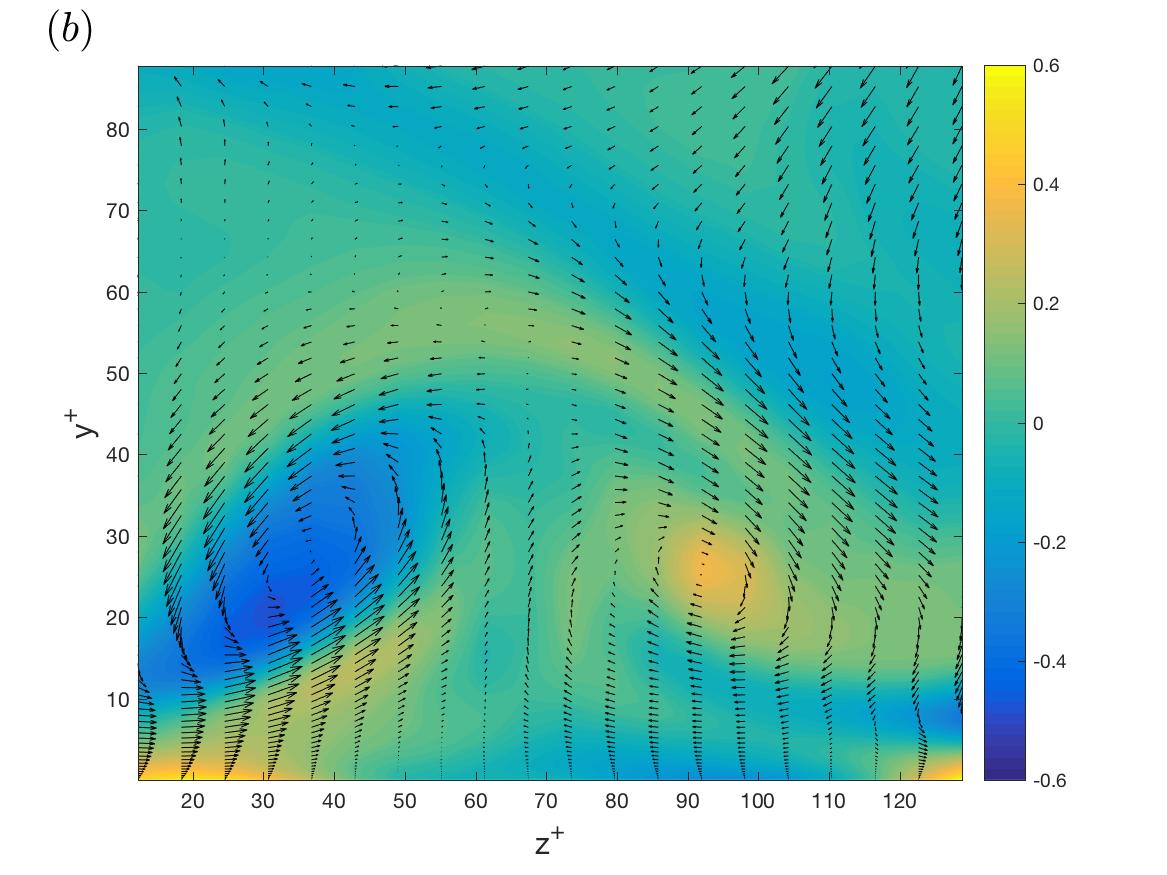}  
\end{subfigure}
\caption{(a) Isosurface $\lambda_2^+=-0.0163,$ with magnitude 4 times the local box-average value 
$\langle |\lambda_2^+|\rangle=4.09\times 10^{-3}.$
The shear-stress field from Fig.~\ref{fig4-stressL} is replotted in the bottom $x$-$z-$plane for reference.
(b) Field of streamwise vorticity $\omega_x^+$ in plane $x^+=85.9$ (transparent in panel (a)). 
The black arrows represent cross-stream velocity vectors $(w,v).$} 
\label{fig4-lam2L}
\end{figure}

The stress local-minimum that we selected for study is located within a long low-speed streak of the type commonly 
observed in near-wall turbulence, with typical spanwise separations $\delta z^+\simeq 100$ between streaks \citep{jimenez2013near}.
This environment is illustrated in Fig.~\ref{fig4-stressL}, which plots the viscous shear-stress $\tau_{xy}=\nu(\partial u/\partial y)$ 
at the wall, with magnitudes represented by the color (or shade, in greyscale), together with the location of the 
stress local-minimum as an asterisk ``$*$''. This figure also plots the two-dimensional wall stress field 
$\btau_W=\nu(\partial u/\partial y,\partial w/\partial y)$ with black arrows. The arrows indicate a near-wall, in-plane 
flow which is converging toward the streak. This convergence is consistent with a vertical flow that is upward, away from the wall,
at the streak and it agrees with results of \cite{sheng2009buffer} for the conditional average stress-field in the vicinity of such 
local-minima (see their Fig.~6(f)). More insight into the local flow conditions is provided by Fig.~\ref{fig4-lam2L}, 
which in panel (a) visualizes the coherent vortices in the vicinity of the local-minimum using isosurfaces of $\lambda_2,$ 
or the second eigenvalue of the $(\grad_x\bu)^2$ matrix \citep{jeong1995identification}. Somewhat different choices 
of $\lambda_2$-levels and different vortex visualization criteria yield similar results. Clearly observed  are two long,  
equal-strength, streamwise vortices located on each side of the low-speed streak and inclined away from the wall. Measurement of  
$\omega_x$ reveals that these vortices are counter-rotating, generating a lifting flow above the low-speed streak. This is 
illustrated in Fig.~\ref{fig4-lam2L}(b), which provides a color-plot of  $\omega_x$ in the transverse $y$-$z-$plane through 
the middle of the visualized box and which also plots as arrows the two-dimensional cross-stream flow vectors $(w,v)$ within that plane. 
This type of counter-rotating vortex-pair generating a lifting flow between them is a quite common buffer-layer configuration,
encountered in about 16\% of all of samples in the study of \cite{sheng2009buffer} and in 98\% of their realizations satisfying 
the condition $\tau_{xy}^+<0.6.$ Our selected event thus appears to be quite typical of such stress-minima 
and similar features were observed in many other local-minima stress events that we identified 
in the JHTDB channel-flow dataset satisfying the criteria $\tau_{xy}^+\simeq 0.6.$

\begin{figure}
\begin{subfigure}[b]{.5\textwidth}
  \centering
  \includegraphics[width=1.1\linewidth]{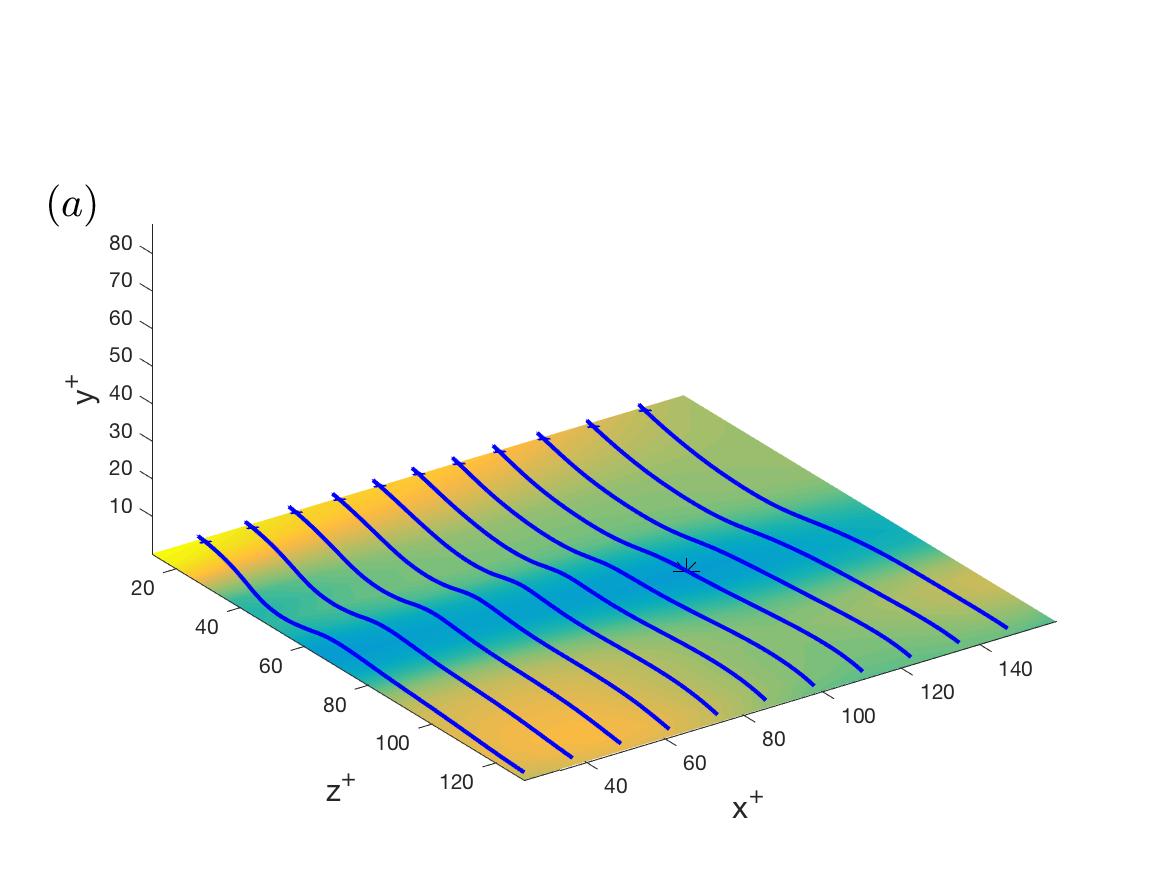}  
\end{subfigure}
\begin{subfigure}[b]{.5\textwidth}
  \centering
  \includegraphics[width=1.1\linewidth]{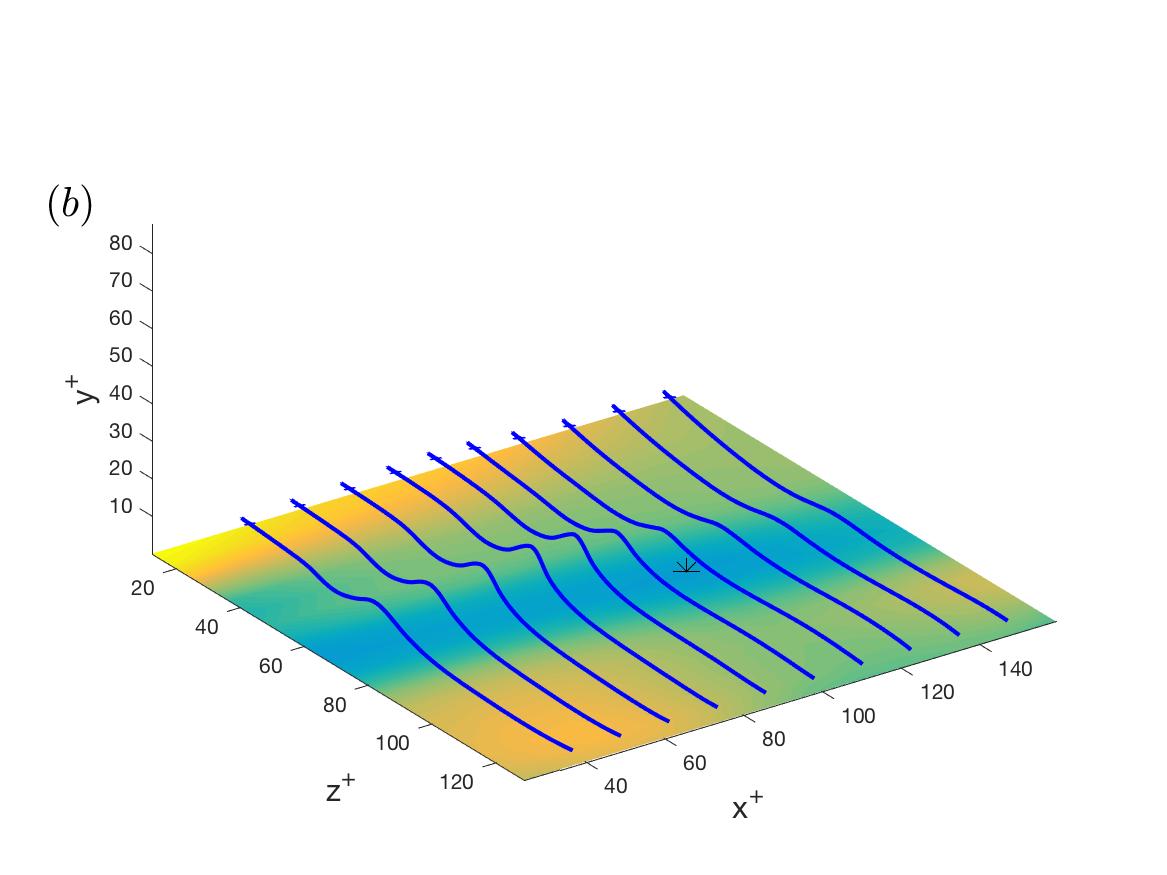}  
\end{subfigure}
\\
\begin{subfigure}[b]{.5\textwidth}
  \centering
  \includegraphics[width=1.1\linewidth]{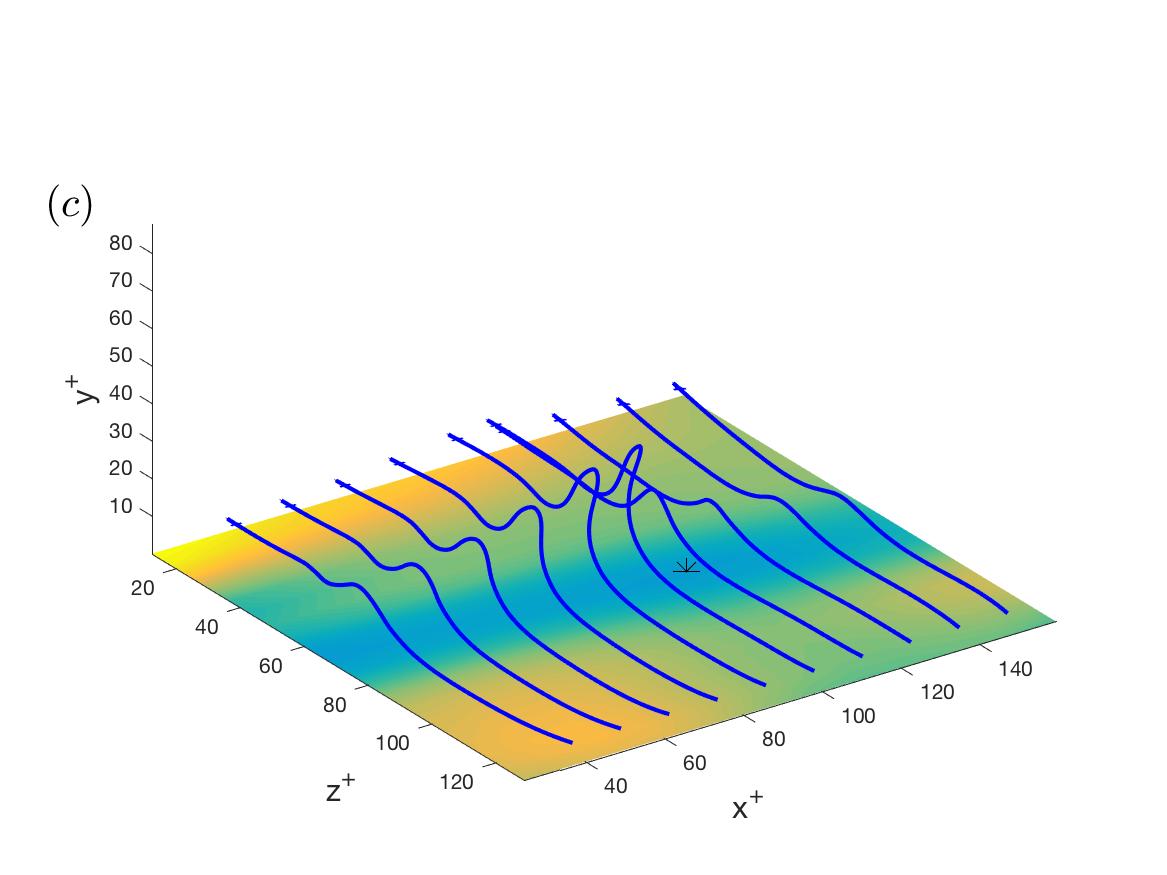}  
\end{subfigure}
\begin{subfigure}[b]{.5\textwidth}
  \centering
  \includegraphics[width=1.1\linewidth]{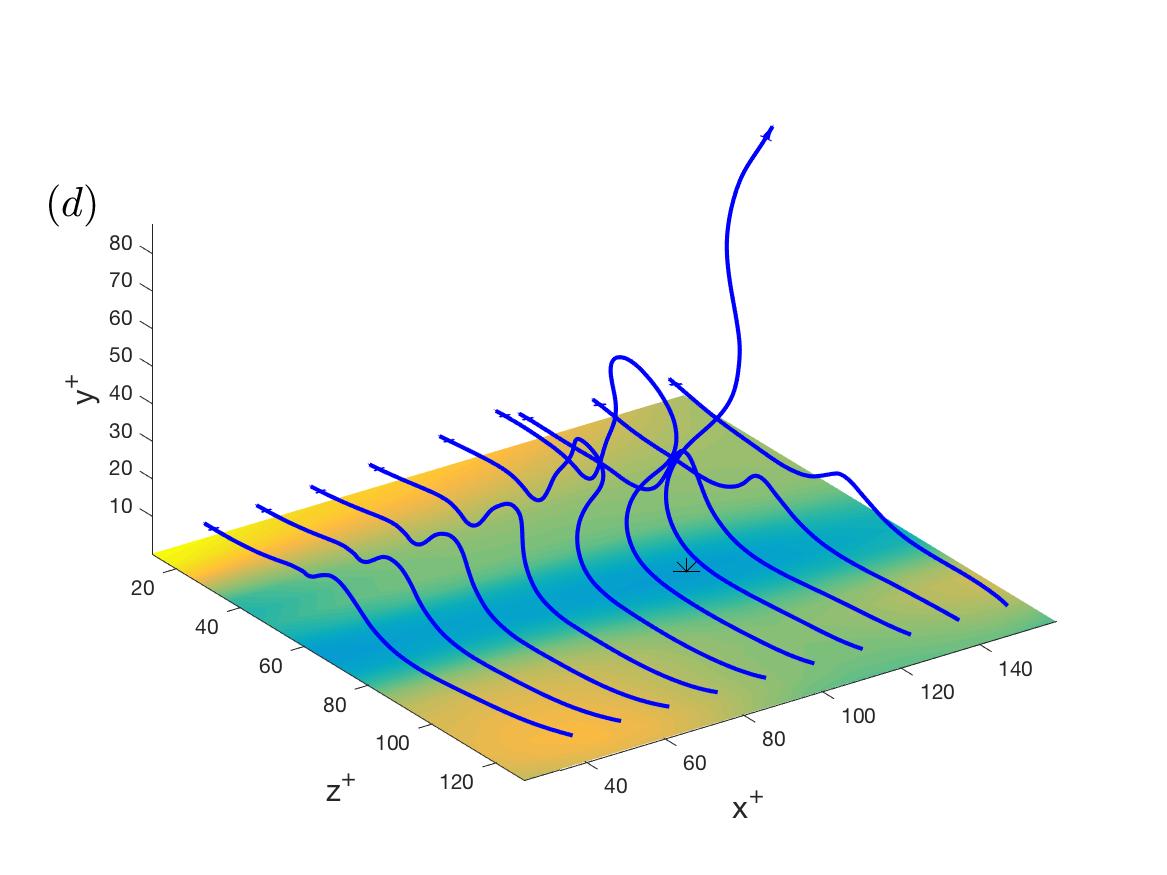}  
\end{subfigure}
\caption{Vortex lines initiated at points with streamwise separations $\Delta x^+=12.3,$ with $z^+=128.8$ and 
(a) $y^+=2,$ \ (b) $y^+=4,$ \ (c) $y^+=6,$ \ (d) $y^+=8.$ The shear-stress field from Fig.~\ref{fig4-stressL} is 
replotted in the bottom $x$-$z-$planes for reference.}
\label{fig4-linesL}
\end{figure}

This typicality is confirmed 
by Fig.~\ref{fig4-linesL} which plots vortex-lines crossing the visualized domain in the spanwise direction, initialized at evenly 
spaced streamwise locations and at initial elevations $y^+=2,4,6,8.$ These vortex lines are lifted in arches above the 
low-speed streak, with a typical ``hairpin'' geometry.  The arches are nearly vertical for 
lines initiated at $y^+=2,4$ and also for $y^+=6,8$ at points well upstream of the stress minimum. For lines 
initiated at $y^+=6$ the arches rise and bend downstream approaching the stress-minimum, while the lines 
initiated at $y^+=8$ near the stress-minimum have instead an  ``$\Omega$-vortex'' geometry and the uppermost tips are 
bent back slightly upstream. These arrays of vortex lines are typical of those observed in the vicinity of stress 
local-minima in the study of \cite{sheng2009buffer}, as illustrated in their Figures 4(a), 8(a) for individual realizations 
and in their Figure 7 for lines of the conditionally averaged vorticity field given $\tau_{xy}^+<0.6.$ Note Fig.~\ref{fig4-linesL}(c) 
plots the same vortex lines shown in Fig.~\ref{fig:JustLines} in the Introduction. One of the primary goals of this work is to elucidate 
the surprisingly complex and violent Lagrangian dynamics underlying this simple vortex-line structure. 

\begin{figure}
\begin{subfigure}[b]{\textwidth}
  \centering
  \includegraphics[width=.8\linewidth]{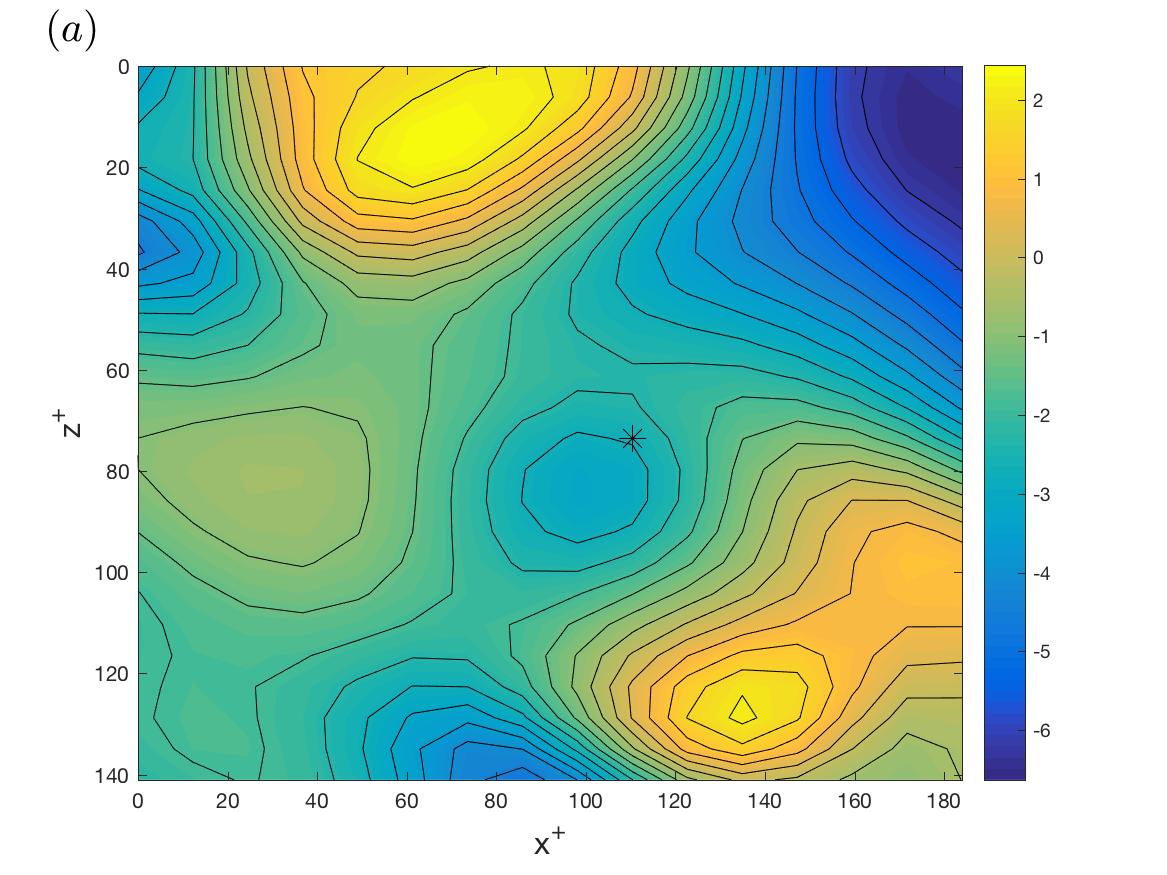}  
\end{subfigure}
\\
\begin{subfigure}[b]{\textwidth}
\centering
  \includegraphics[width=.8\linewidth]{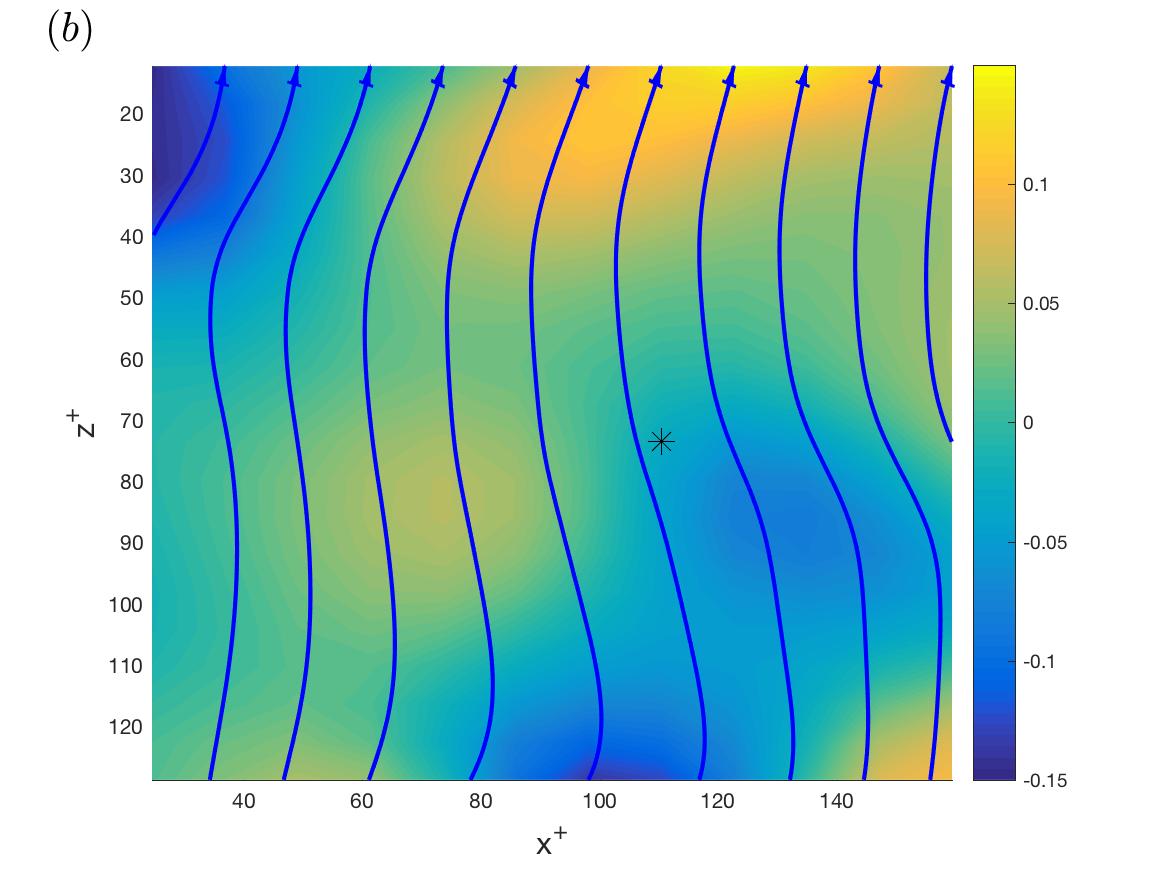}  
\end{subfigure}
\caption{(a) Pressure field $p^+$ at the wall, with selected isolines in black.  \ (b) The source field 
$-\sigma_z^+$ of the negative spanwise vorticity and selected in-wall vortex lines. The asterisk ``$*$'' 
in both panels marks the location of the selected stress local-minimum.}
\label{fig4-pressL}
\end{figure}


\begin{figure}
\begin{subfigure}[b]{\textwidth}
  \centering
  \includegraphics[width=.8\linewidth]{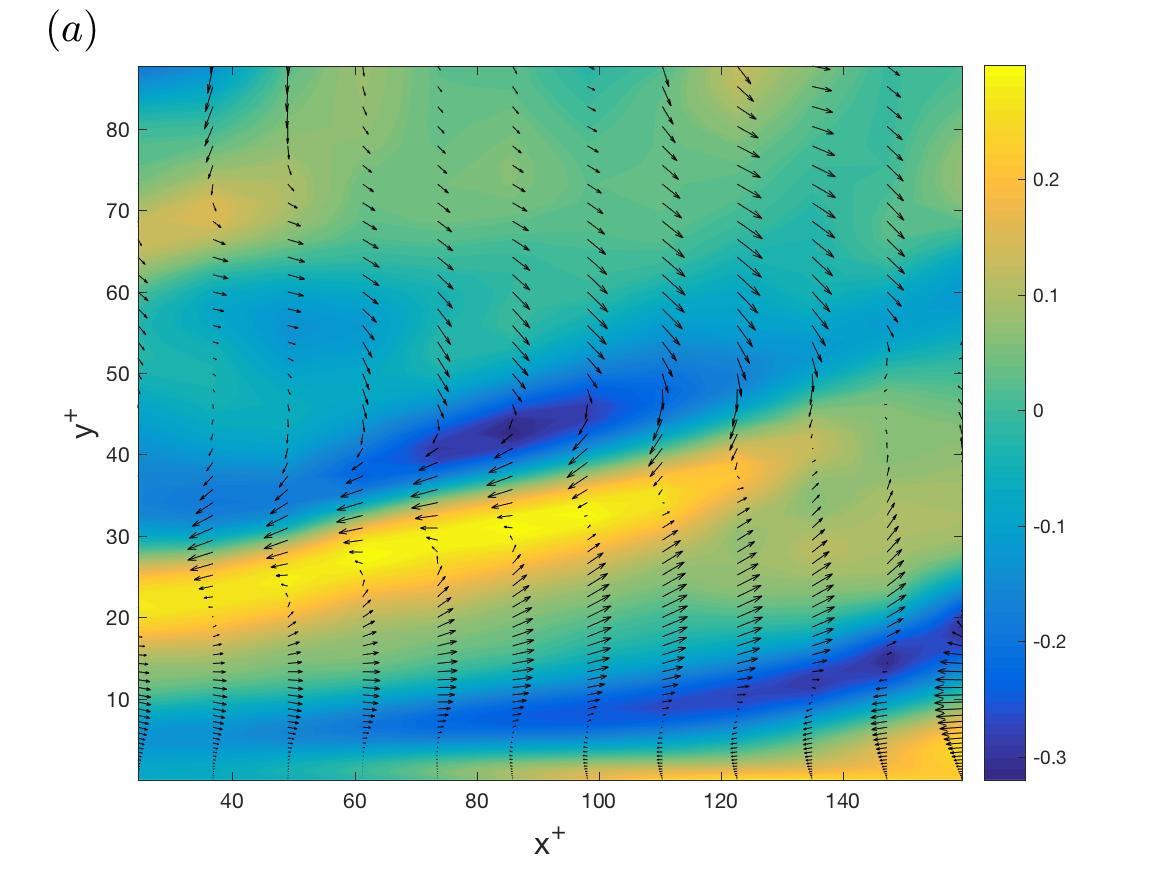}  
\end{subfigure}
\\
\begin{subfigure}[b]{\textwidth}
\centering
  \includegraphics[width=.8\linewidth]{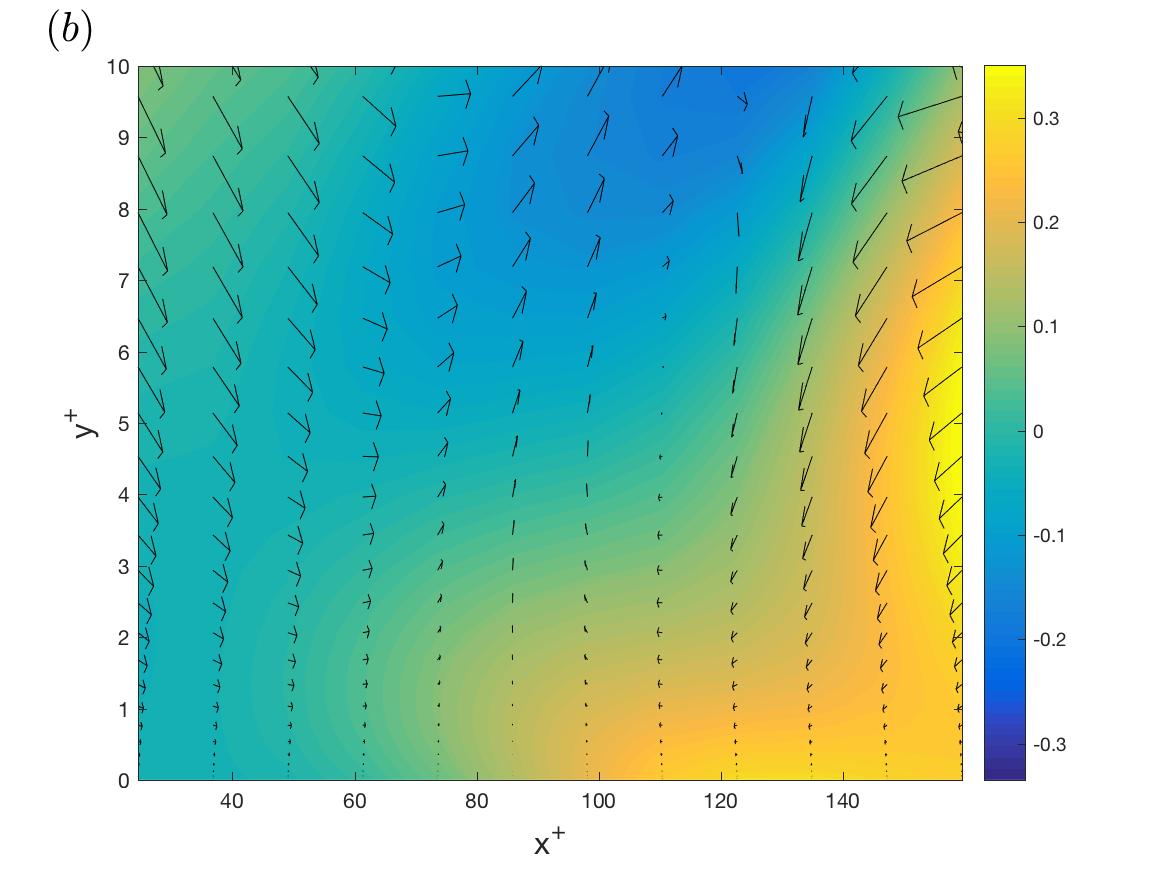}  
\end{subfigure}
\caption{(a) Field of spanwise vorticity fluctuation $\omega_z^{+\prime}$ in the $x$-$y-$plane at 
$z^+=85.9.$ The black arrows represent the vectors $\,(u',v')\,$ of the cross-section velocity fluctuation.\\
(b) Same as (a), but with fluctuations calculated relative to local planar averages at constant
$y^+$ and plotted in the near-wall region $y^+<10.$}
\label{fig4-domzL}
\end{figure}

Further insight into the local vorticity dynamics from the Eulerian perspective is provided by results 
on the in-wall pressure distribution $p(x,z)$, spanwise vorticity source $\sigma_z(x,z),$ and 
selected in-wall vortex-lines, as plotted in Fig.~\ref{fig4-pressL} for the vicinity of our stress 
local-minimum. Panel (a) of that figure reveals that our selected stress minimum is very close to a 
local pressure minimum, which in turn is flanked upstream and downstream at distances 
$\delta x^+\simeq \pm 60$ by a pair of local pressure maxima. The pressure isolines or isobars in this plot 
are the lines of instantaneous generation of tangential vorticity in the Lighthill-Morton theory,
with positive (counterclockwise) sense of rotation around pressure maxima and negative (clockwise) 
rotation around pressure minima. Of course, these isobars align only with the direction of generation 
of vorticity and the instantaneous vortex-lines within the wall are instead pointed mainly in the spanwise 
direction with small streamwise deviations, as shown in Fig.~\ref{fig4-pressL}(b). The bending of these 
lines is explained in detail by the relation $\btau_W=\nu\bn\btimes\bom_W$ between the in-wall stress 
and vorticity fields \citep{lighthill1963boundary,morton1984generation}. As a consequence, the stress 
vectors $\btau_W$ plotted in Fig.~\ref{fig4-stressL} are locally perpendicular to the in-wall vortex-lines  
in Fig.~\ref{fig4-pressL}(b) and the concavity of the lines is exactly that required to produce a 
converging flow at the low-speed streak. We plot in Fig.~\ref{fig4-pressL}(b) as well the {\it negative} 
spanwise vorticity source $-\sigma_z^+ = -\partial p^+/\partial x^+ = \partial \omega_z^+/\partial y^+$
in wall units. We included the minus sign since drag evidenced by a streamwise drop in pressure 
is associated to flux of negative spanwise vorticity away from the wall. Thus, the color/shading schemes 
in Figs.~\ref{fig4-stressL} and Fig.~\ref{fig4-pressL}(b) are consistent, with yellow/light corresponding 
to increased drag (high stress, pressure drop) and blue/dark corresponding to reduced drag 
(low stress, pressure rise).  Note however that the mean pressure drop associated to dissipative 
turbulent drag is $-\partial \overline{p}^+/\partial x^+=1/Re_\tau=10^{-3},$ whereas the instantaneous
streamwise pressure-gradients plotted in Fig.~\ref{fig4-pressL}(b) are 100 times larger in magnitude, 
spanning a range from -0.15 to +0.15. It is consistent with the results in Fig.~\ref{fig4-pressL}(b) that
instantaneous pressure-gradients at the wall scale as $u_*^2/\delta_\nu$ and thus remain $O(1)$ 
in wall-units. The average streamwise pressure drop is thus the result of 
near-cancellation between large instantaneous gradients of both signs. 

It is interesting to observe
in Fig.~\ref{fig4-pressL}(b) that a region of negative streamwise pressure-gradient occurs just upstream 
of the  stress local-minimum and a corresponding region of positive gradient occurs just downstream.
This seems to agree with experimental observations of ``bipolar'' spanwise vorticity generation by 
\cite{andreopoulos1996wall} and \cite{klewicki2008statistical}, based on conditionally-averaged 
time series of pressure and vorticity-flux, and on time-correlation functions of pressure and 
pressure-gradients.  \cite{andreopoulos1996wall} proposed a conceptual model of ejections
as rising ``mushroom vortices'' that would produce exactly such a bipolar pattern of spanwise
vorticity source at the wall. See  \cite{andreopoulos1996wall}, Figure 28(b). However, a plot in 
Fig.~\ref{fig4-domzL} of the spanwise vorticity fluctuation $\omega_z'$ and velocity vector 
fluctuation $(u',v')$ in the plane $z^+=85.9$ for our event is not consistent with such 
a picture. Panel (a) of that figure shows the entire $y^+$-range, for context, and panel (b) 
zooms into the near-wall region $y^+<10.$ To make the flow pattern more clear in Fig.~\ref{fig4-domzL}(b)
we have calculated fluctuations with respect to local planar averages at fixed distances $y^+$ from 
the wall (see SM for details). We observe that the ``bipolar'' source is produced by a fluid layer with $\omega_z'<0$ 
being lifted up and to the right from the bottom wall, while replacement fluid  with $\omega_z'>0$ is advected 
in from the right and downward toward the wall. This dynamics is very similar to that postulated 
by \cite{jimenez1988ejection}, Fig.6, as a mechanism of sublayer ejections. Those authors also 
observed thin, low-inclined layers of  $\omega_z'$ like those in our Fig.~\ref{fig4-domzL}(a) 
and interpreted them as Tollmien-Schlichting waves. 
Such shear layers are also inferred for coherent, nonlinear travelling waves (\cite{waleffe1998three}, Fig. 1).  
Finally, \cite{andreopoulos1996wall} have argued that 
``ejections which carry fluid of negative $\omega_z$ away from the wall ... are expected to be characterized 
by positive $\partial\omega_z/\partial y$. Negative $\partial\omega_z/\partial y$ is expected to be 
the distinguishing feature of sweeps...''.  The ejection event that we consider has near 
$y^+\simeq 5-10$ an upward flux of negative spanwise vorticity associated with $v'\omega_z'<0.$ 
However, there is no {\it instantaneous} balance between this advective flux and the viscous flux 
$\sigma_z=-\nu(\partial \omega_z/\partial y)$ at the wall. Thus it is not clear that the region with 
$\partial\omega_z/\partial y>0$ upstream of the stress local-minimum should be regarded 
as the ``source'' of the advective spanwise vorticity flux at $y^+\simeq 5-10$. Our Lagrangian 
analysis in section \ref{sec:lagrange} shall show indeed that there is no causal connection.  

\begin{figure}
\centering 
\includegraphics[width=0.8\textwidth]{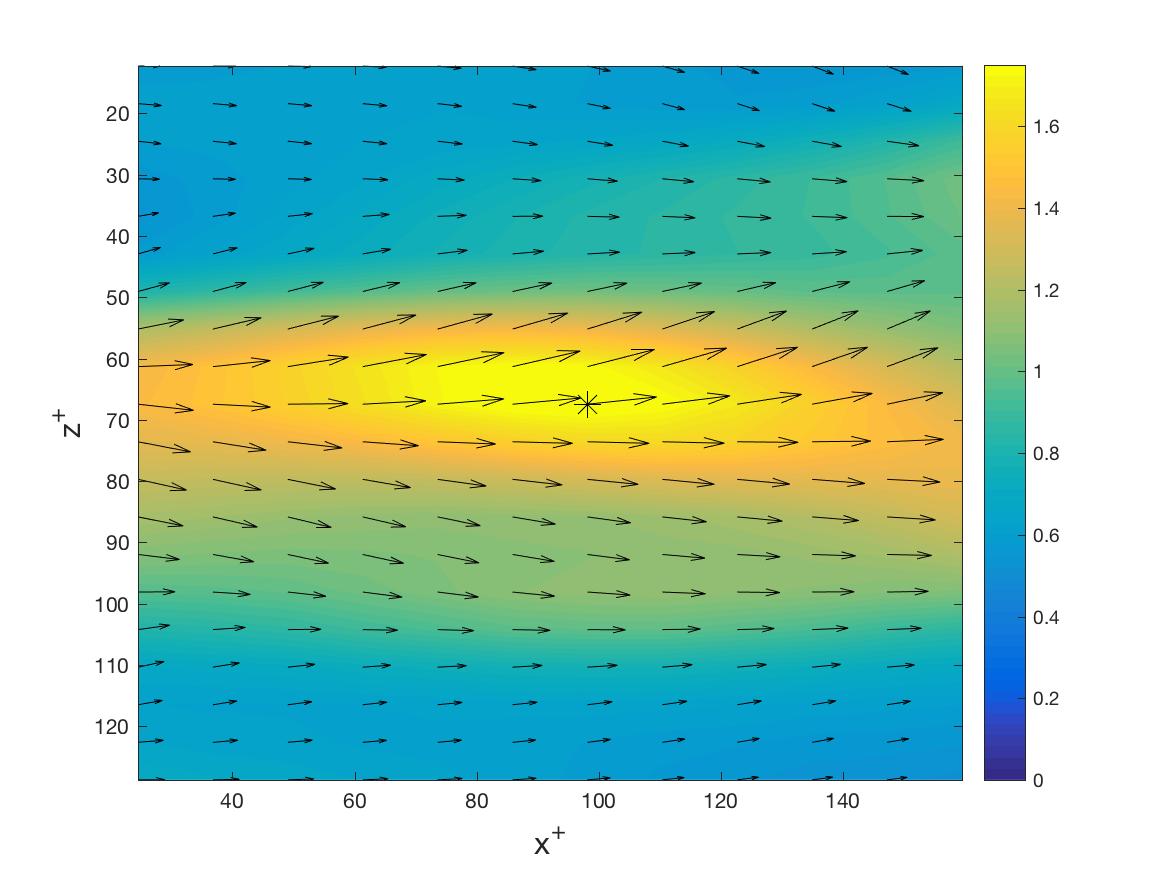}
\caption{Field of viscous shear stress $\tau_{xy}^+=(\partial u^+/\partial y^+)$ at the wall $y^+=0$, in wall units 
so that the average is unity. Black arrows represent the two-dimensional in-wall stress vector $\btau_W.$ The 
asterisk ``$*$'' marks the location of the selected stress local-maximum.} 
\label{fig4-stressH} 
\end{figure}



\subsection{Eulerian Description of Sweep Event}\lb{sec:EulerSweep} 

\begin{figure}
\begin{subfigure}[b]{\textwidth}
  \centering
  \includegraphics[width=.8\linewidth]{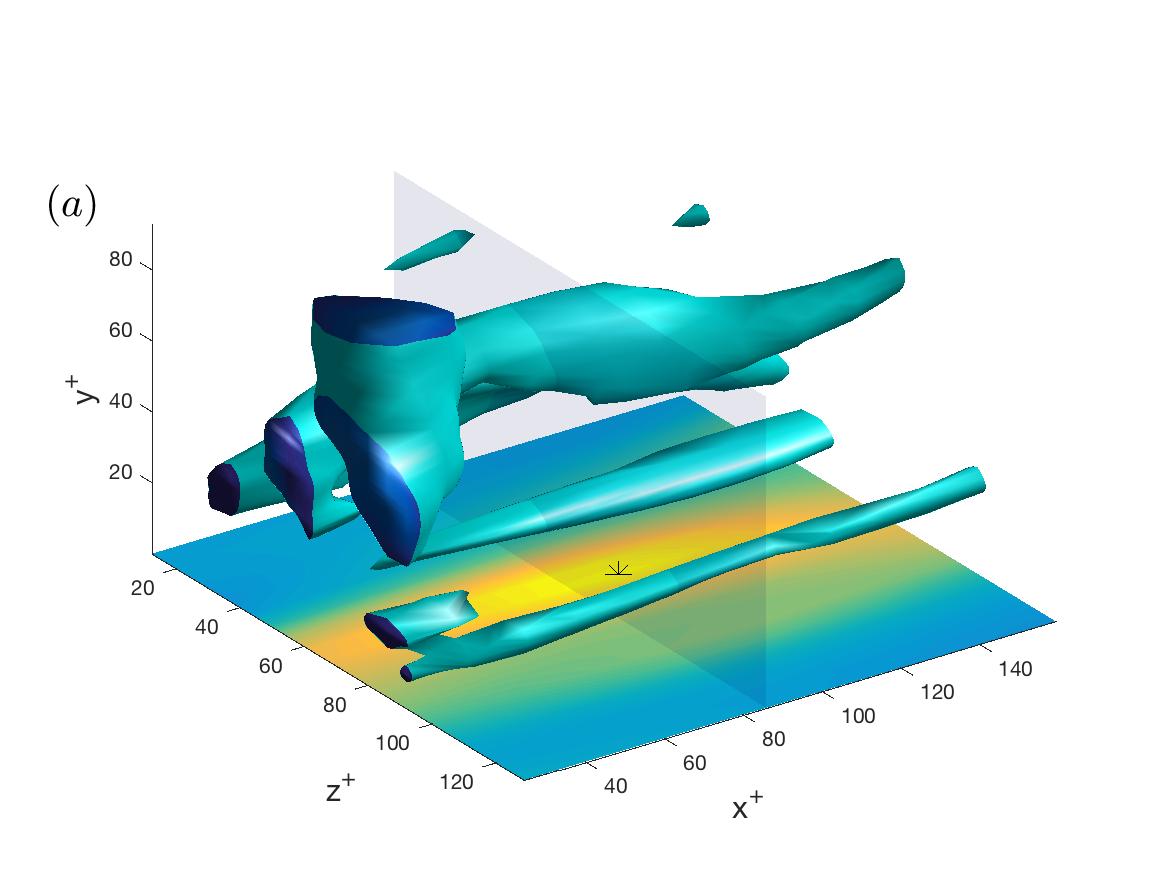}  
\end{subfigure}
\\
\begin{subfigure}[b]{\textwidth}
\centering
  \includegraphics[width=.8\linewidth]{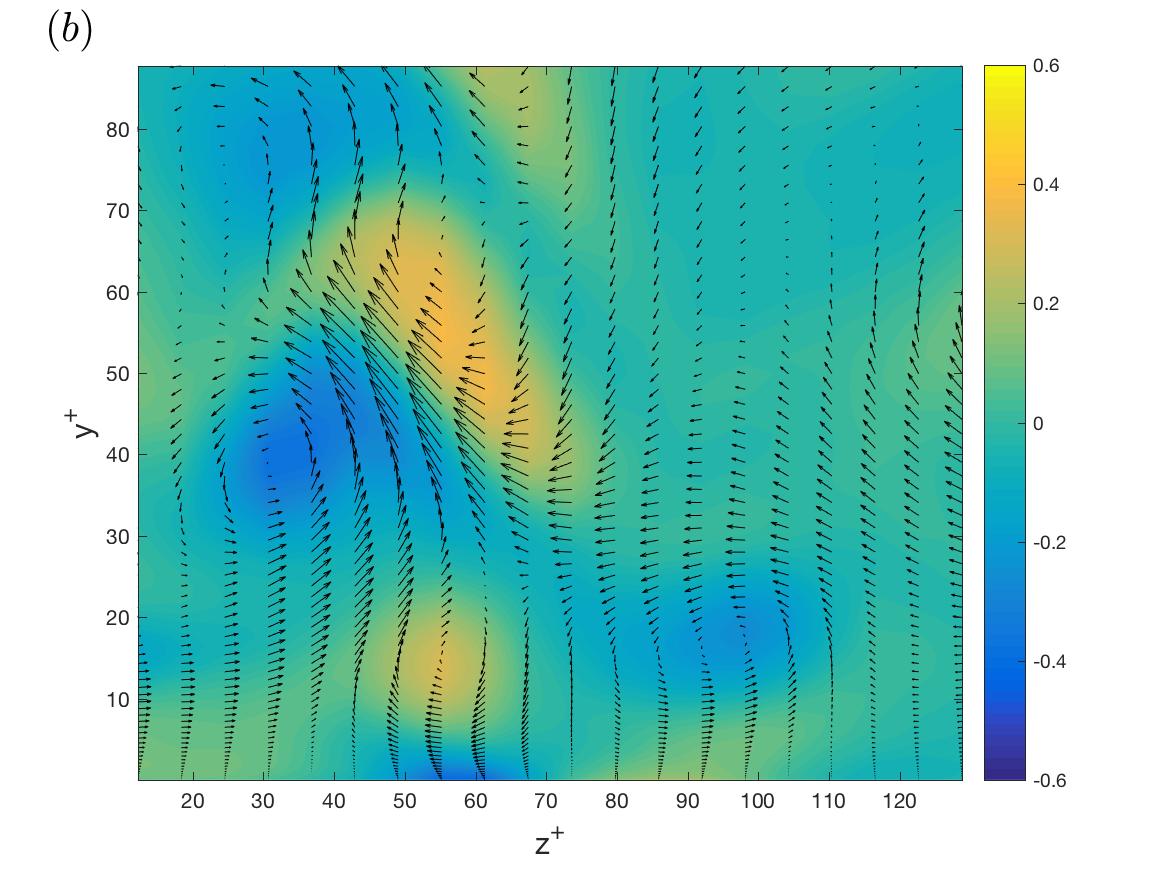}  
\end{subfigure}
\caption{(a) Isosurface $\lambda_2^+=-0.0107,$ with magnitude 3 times the local box-average value 
$\langle |\lambda_2^+|\rangle=3.57\times 10^{-3}.$ The shear-stress field from Fig.~\ref{fig4-stressH} 
is replotted in the bottom $x$-$z-$plane for reference.
(b) Field of streamwise vorticity $\omega_x^+$ in plane $x^+=85.9$ (transparent in panel (a)). 
The black arrows represent cross-stream velocity vectors $(w,v).$}
\label{fig4-lam2H}
\end{figure}

\begin{figure}
\begin{subfigure}[b]{.5\textwidth}
  \centering
  \includegraphics[width=1.1\linewidth]{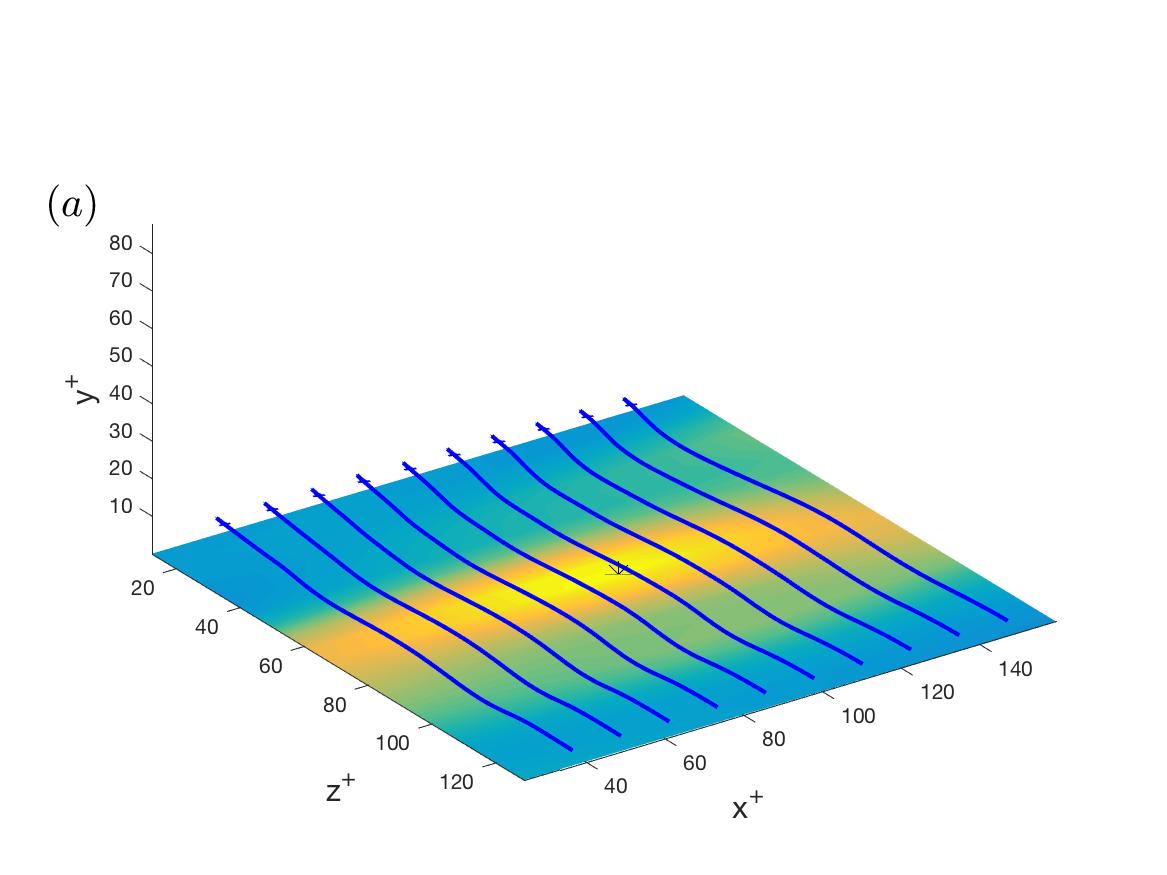}  
\end{subfigure}
\begin{subfigure}[b]{.5\textwidth}
  \centering
  \includegraphics[width=1.1\linewidth]{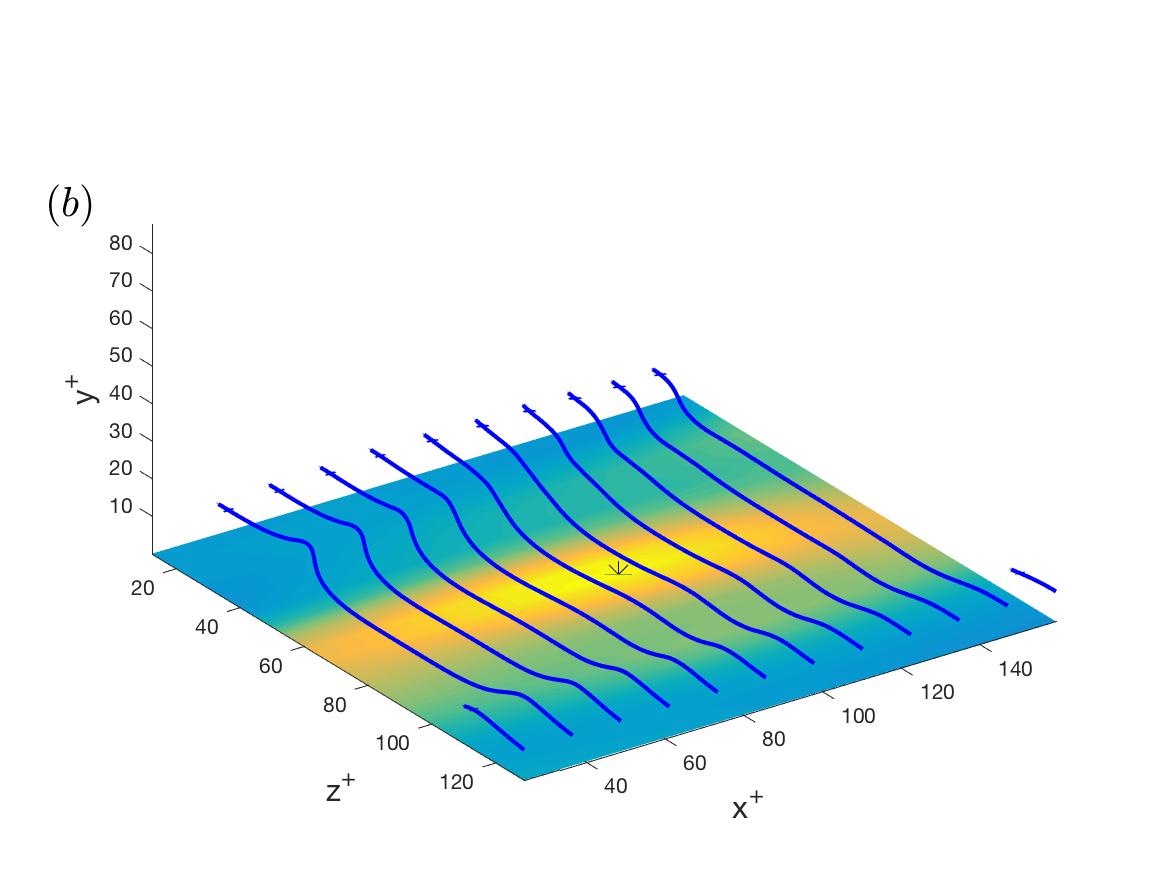}  
\end{subfigure} 
\\
\begin{subfigure}[b]{.5\textwidth}
  \centering
  \includegraphics[width=1.1\linewidth]{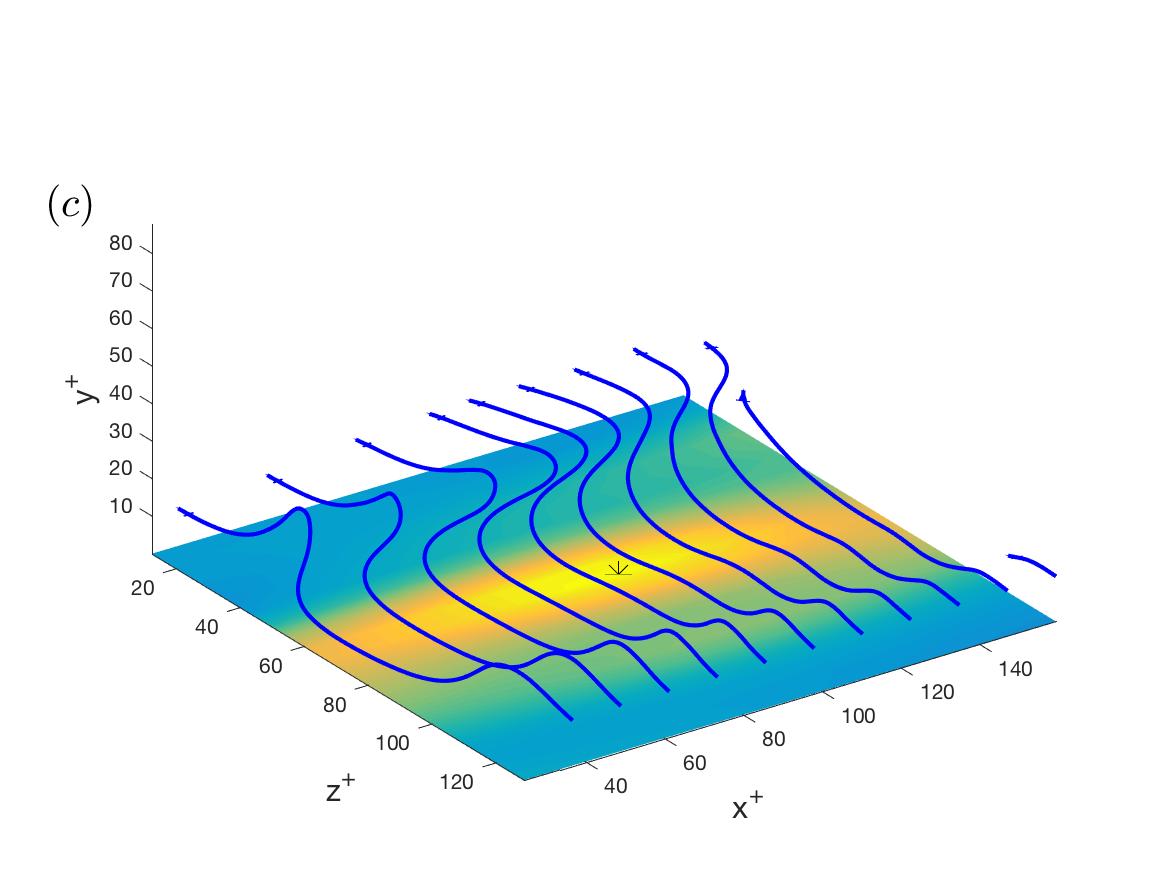}  
\end{subfigure}
\begin{subfigure}[b]{.5\textwidth}
  \centering
  \includegraphics[width=1.1\linewidth]{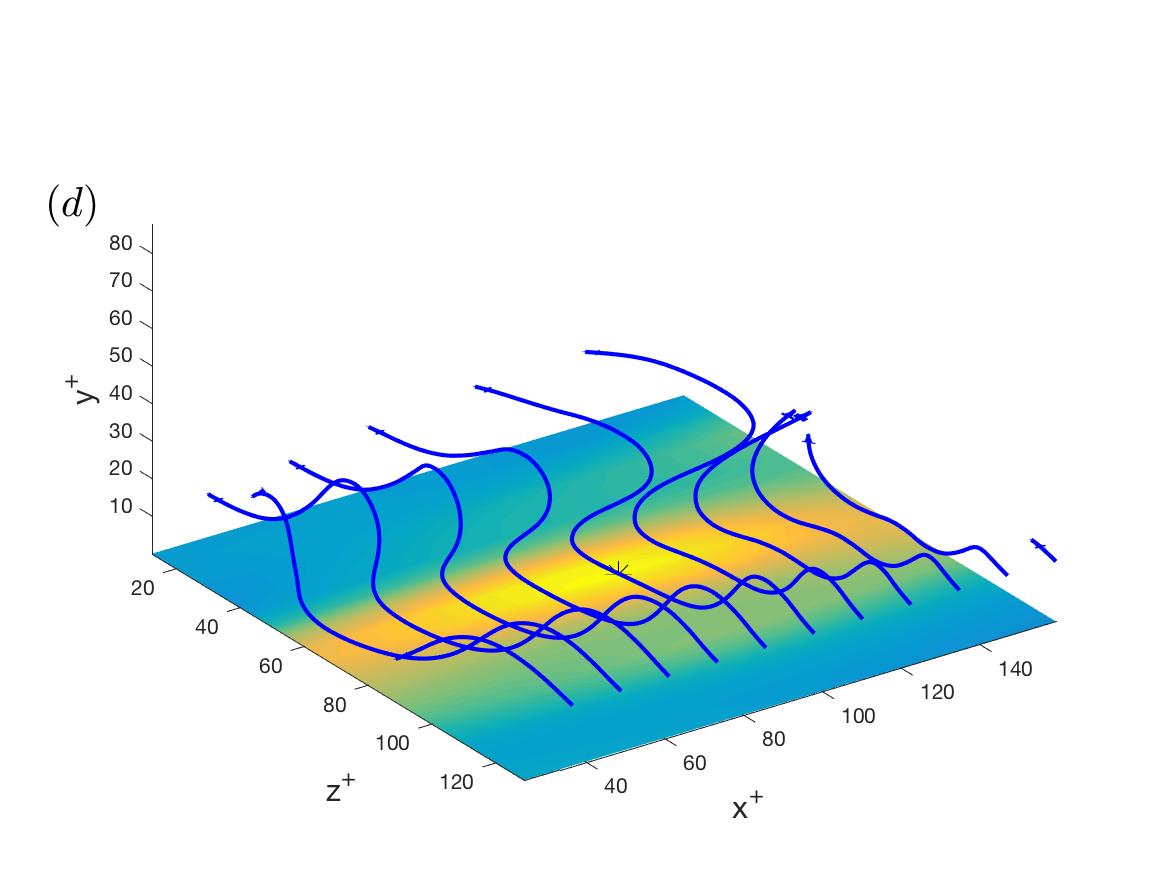}  
\end{subfigure}
\caption{Vortex lines initiated at points with streamwise separations $\Delta x^+=12.3,$ with $z^+=128.8$ and 
(a) $y^+=4,$ \ (b) $y^+=8,$ \ (c) $y^+=12,$ \ (d) $y^+=16.$ The shear-stress field from Fig.~\ref{fig4-stressH} is 
replotted in the bottom $x$-$z-$planes for reference.}
\label{fig4-linesH}
\end{figure}

\begin{figure}
\begin{subfigure}[b]{\textwidth}
  \centering
  \includegraphics[width=.8\linewidth]{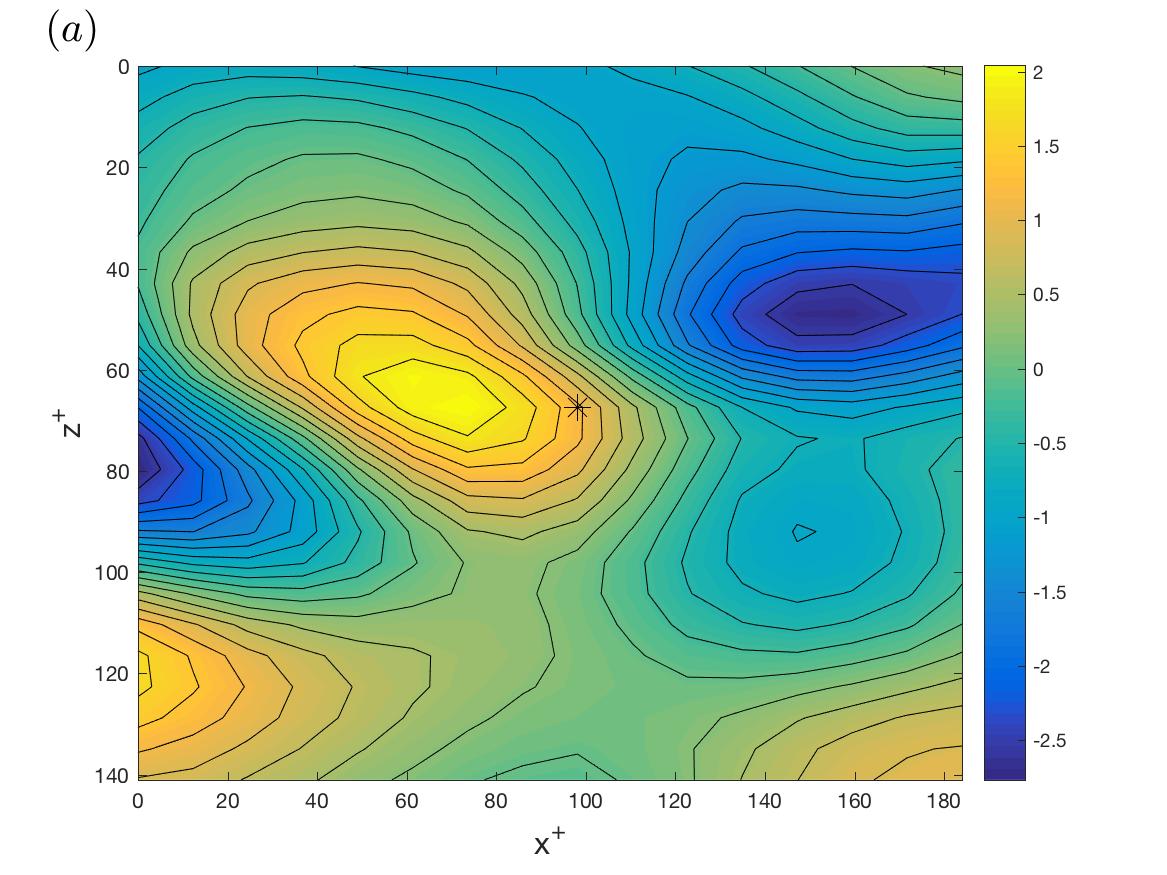}  
\end{subfigure}
\\
\begin{subfigure}[b]{\textwidth}
\centering
  \includegraphics[width=.8\linewidth]{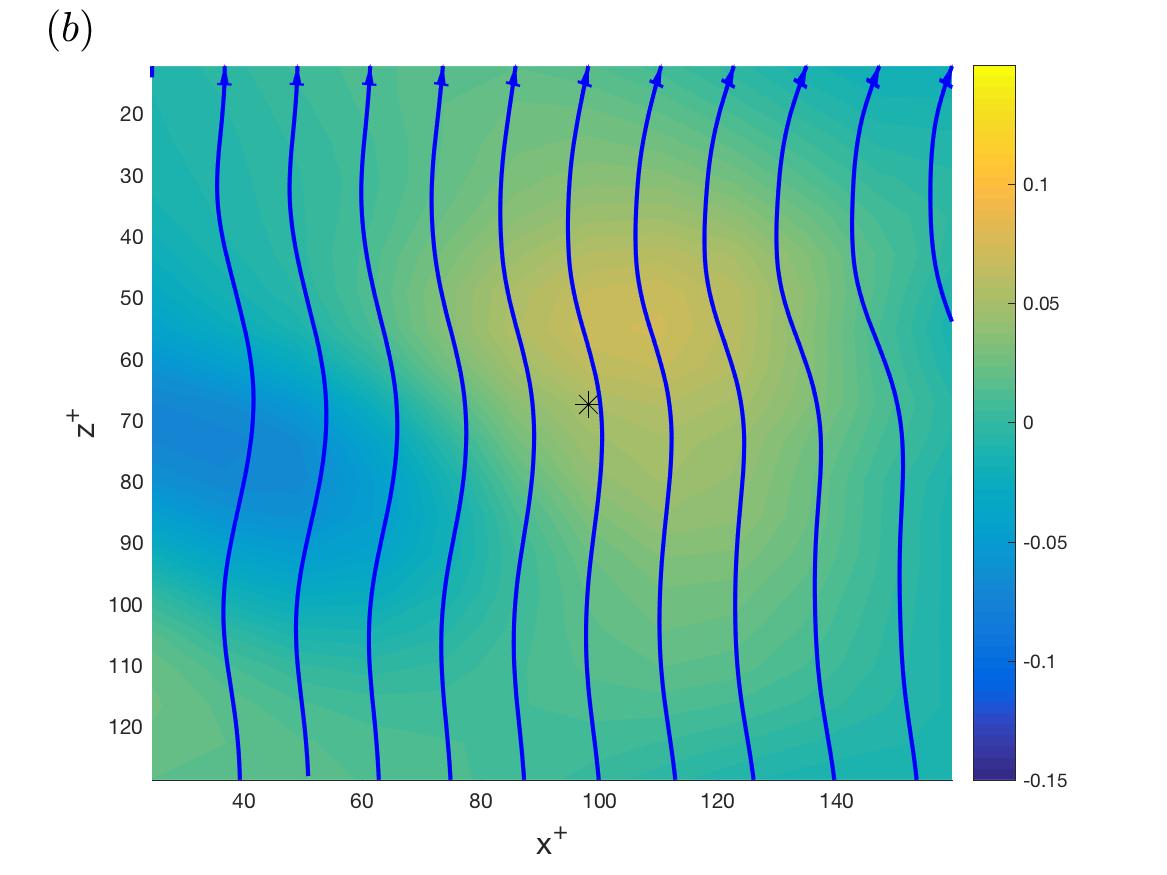}  
\end{subfigure}
\caption{(a) Pressure field $p^+$ at the wall, with selected isolines in black.  \ (b) The source field 
$-\sigma_z^+$ of the negative spanwise vorticity and selected in-wall vortex lines. The asterisk ``$*$'' 
in both panels marks the location of the selected stress local-maximum.}
\label{fig4-pressH}
\end{figure}

The stress local-maximum that we selected for study is likewise located within a high-speed streak of a type
also commonly observed in near-wall turbulence, generally shorter than the low-speed streaks in streamwise extent and 
flanking them \citep{jimenez2013near}. This neighborhood is illustrated in Fig.~\ref{fig4-stressH}, which plots the 
viscous shear-stress $\tau_{xy}=\nu(\partial u/\partial y)$ at the wall and the location of the stress local-maximum 
as an asterisk ``$*$''. The arrows representing the two-dimensional wall stress field $\btau_W$ indicate a near-wall, in-plane 
flow which is diverging from the streak. This divergence is consistent with a vertical flow that is downward toward the wall
at the streak and it agrees with results of \cite{sheng2009buffer} for the conditional average stress-field in the vicinity of such 
local-maxima (see their Fig.~6(e)). Vortex visualization in Fig.~\ref{fig4-lam2H}(a) via $\lambda_2$-isosurfaces  
shows a more complex environment than for the preceding local-minimum event. There are two or three large  
quasi-streamwise vortices at heights $y^+>20,$ but these do not seem to influence strongly the near-wall physics. 
Instead at elevations $y^+<20$ there is a pair of counter-rotating almost streamwise vortices, one on each side 
of the observed high-speed streak. The plot in Fig.~\ref{fig4-lam2H}(b) of streamwise vorticity $\omega_x$ and 
cross-stream velocity vectors $(w,v),$ in the transverse $y$-$z-$plane cutting through the middle of the visualized 
box,  shows clearly that this low-lying pair generate a downward, splatting flow between them. Unlike the pair
observed near the stress-minimum, however, this pair is asymmetrical in strength, with the rightmost member 
of the pair distinctly weaker. If we increase the magnitude of the threshold value of $\lambda_2$ 
by even 33\% to $\lambda_2^+=-0.0143$ , then no isosurface appears for this weaker vortex and 
only the single stronger vortex is observed at  $y^+<20.$ This is consistent with the findings of 
\cite{sheng2009buffer}, who did encounter counter-rotating vortex pairs generating a splatting flow between them,  
in 11\% of all of the samples in their study. However, under the condition $\tau_{xy}^+>1.8,$ only about
8\% of the realizations were of this type and all of these vortex pairs were quite asymmetrical in strength. 
Instead, 55\% of the realizations in the study of \cite{sheng2009buffer} that satisfied the condition $\tau_{xy}^+>1.8.$
had the stress maximum generated by a single low-lying vortex.  Our selected event thus exhibits typical features 
for such stress-maxima. Similar features were observed also in other local-maxima stress events that 
we identified in the JHTDB channel-flow dataset satisfying the criteria $\tau_{xy}^+\simeq 1.8.$

The vortex lines that we observe near this stress local-maximum likewise show expected features.
See Fig.~\ref{fig4-linesH}, which plots vortex-lines crossing the visualized domain in the spanwise direction, 
initialized at evenly spaced streamwise locations and at initial elevations $y^+=4,8,12,16.$ These lines 
are clearly squashed or depressed toward the wall by the downward splatting flow at the high-speed streak. 
Such arrays of vortex lines are typical of those observed in the vicinity of stress local-maxima in the study of 
\cite{sheng2009buffer}, as illustrated in their Figure 16 for individual realizations and in their Figure 17 
for lines of the conditionally averaged vorticity field given $\tau_{xy}^+>1.8.$ As they also observed,
the ``troughs" of depressed lines are wider than the corresponding ``hairpins'' above low-speed streaks.
Also, the asymmetry in strength of the streamwise vortices is clearly visible, with the ridge of lines  
located at the strong vortex obviously twisted higher than those at the weak vortex. In section \ref{sec:lagrange} 
we shall study in depth the Lagrangian dynamics of the illustrated squashed lines at the local-maximum of stress. 


\begin{figure}
\begin{subfigure}[b]{\textwidth}
  \centering
  \includegraphics[width=.8\linewidth]{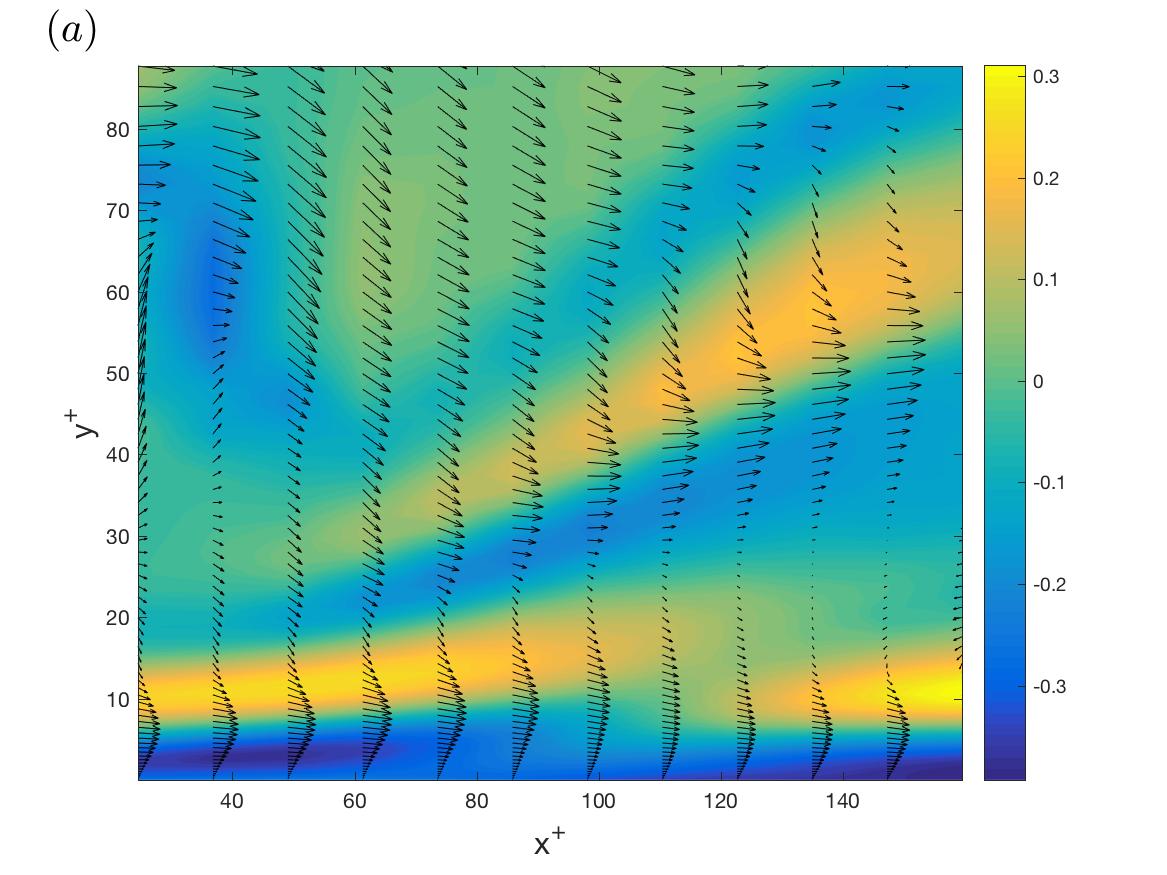}  
\end{subfigure}
\\
\begin{subfigure}[b]{\textwidth}
\centering
  \includegraphics[width=.8\linewidth]{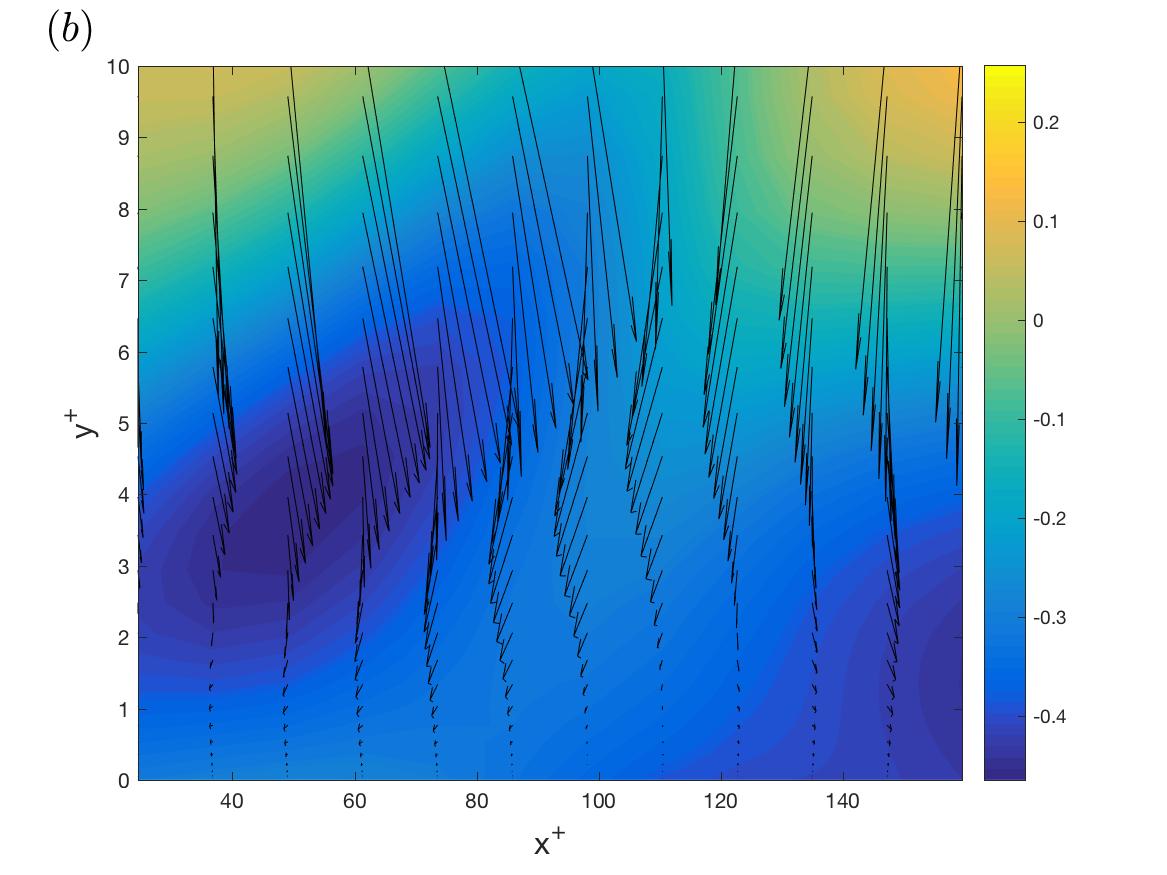}  
\end{subfigure}
\caption{Field of spanwise vorticity fluctuation $\omega_z^{+\prime}$ in the $x$-$y-$plane at 
$z^+=79.7.$ The black arrows represent the vectors $\,(u',v')\,$ of the cross-section velocity fluctuation.\\
(b) Same as (a), but with fluctuations calculated relative to local planar averages at constant
$y^+$ and plotted in the near-wall region $y^+<10.$}
\label{fig4-domzH}
\end{figure}

First, however, we consider the vorticity dynamics for this event in more detail from the Eulerian 
perspective. The plot in Fig.~\ref{fig4-pressH}(a) of the pressure field and its isolines at the wall
shows that the stress maximum is close to a local pressure maximum, with a pair of local pressure 
minima upstream and downstream at distances $\delta x^+\simeq \pm 80.$ In contrast to the 
instantaneous generation of vorticity along the isobars, the actual vortex lines at the wall plotted 
in Fig.~\ref{fig4-pressH}(b) are aligned mainly in the spanwise direction. The lines bend in the 
streamwise direction so that the locally-perpendicular stress vectors $\btau_W,$ as plotted in 
Fig.~\ref{fig4-stressH}, correspond to a near-wall flow diverging away from the high-speed streak. 
The negative spanwise vorticity source $-\sigma_z$ also plotted in Fig.~\ref{fig4-pressH}(b) 
again shows a bipolar pattern of the type inferred by \cite{andreopoulos1996wall} and 
\cite{klewicki2008statistical} from experimental data, with a region of positive streamwise pressure-gradient 
occurring just upstream of the  stress local-maximum and a region of negative gradient just downstream.
In the conceptual model of \cite{andreopoulos1996wall}, Figure 28(a), sweeps correspond to 
 inverted ``mushroom vortices'' moving toward the wall, producing just such a pattern of positive 
 spanwise vorticity source upstream and negative spanwise source downstream.  However, 
 we do not observe such a mushroom vortex here. The plot in Fig.~\ref{fig4-domzH}(a) of the spanwise 
 vorticity fluctuation $\omega_z'$ and velocity vector fluctuation $(u',v')$  in the plane $z^+=79.7$ 
 does exhibit a fluid layer near $y^+\simeq 10$ with $\omega_z'>0$ and $v'<0,$ associated with 
 a spanwise vorticity flux $v'\omega_z'<0.$ However, this occurs mainly upstream 
 from the stress local-maximum, which is exactly opposite to what is proposed in the mushroom-vortex 
 model.  Fig.~\ref{fig4-domzH}(b) shows instead underneath the primary vortex with $\omega_z'>0$ a layer 
 of strong secondary vorticity with $\omega_z'<0$ just above the wall. (Unlike for the ejection case 
 earlier, the local plane-average of streamwise velocity $u,$ used to define the fluctuation $u'$ in 
 Fig.~\ref{fig4-domzH}(b), is noticeably larger than the global average. See Fig.~\ref{fig4-domzH}(a) 
 and the SM.) This secondary layer is apparently produced  by strong interaction 
 of the primary vortex with the wall, as illustrated in Figure 26 of \cite{andreopoulos1996wall}, 
 but the primary and secondary layers are not rolled up to form the head of a mushroom vortex. 
 It is the presence of the secondary layer with $\omega_z'<0$ just above the wall that 
 makes $\partial \omega_z/\partial y<0$ upstream of the stress local-maximum,  while 
 the absence of such a secondary layer makes $\partial \omega_z/\partial y>0$ downstream.    
 The thin, inclined layers of $\omega_z'$ observed in Fig.~\ref{fig4-domzH}(a), although 
 now for a sweep rather than an ejection,  are again quite similar to those reported by 
 \cite{jimenez1988ejection}, who also emphasized the important role of opposite-sign 
vorticity induced at the wall in producing such structures. These observations suggest some 
possible common features in viscous sublayer dynamics of ejections and sweeps. 
In the following section \ref{sec:lagrange} we shall attempt to further explicate the physics 
of both types of events by stochastic Lagrangian analysis.

\section{Stochastic Lagrangian Dynamics of Vorticity in the Buffer Layer}\label{sec:lagrange}  

\subsection{Selection of Vorticity Vectors}\lb{sec:select} 

\begin{figure}
\hspace{-50pt} 
\begin{subfigure}[b]{.6\textwidth}
  \centering
  \includegraphics[width=1.1\linewidth]{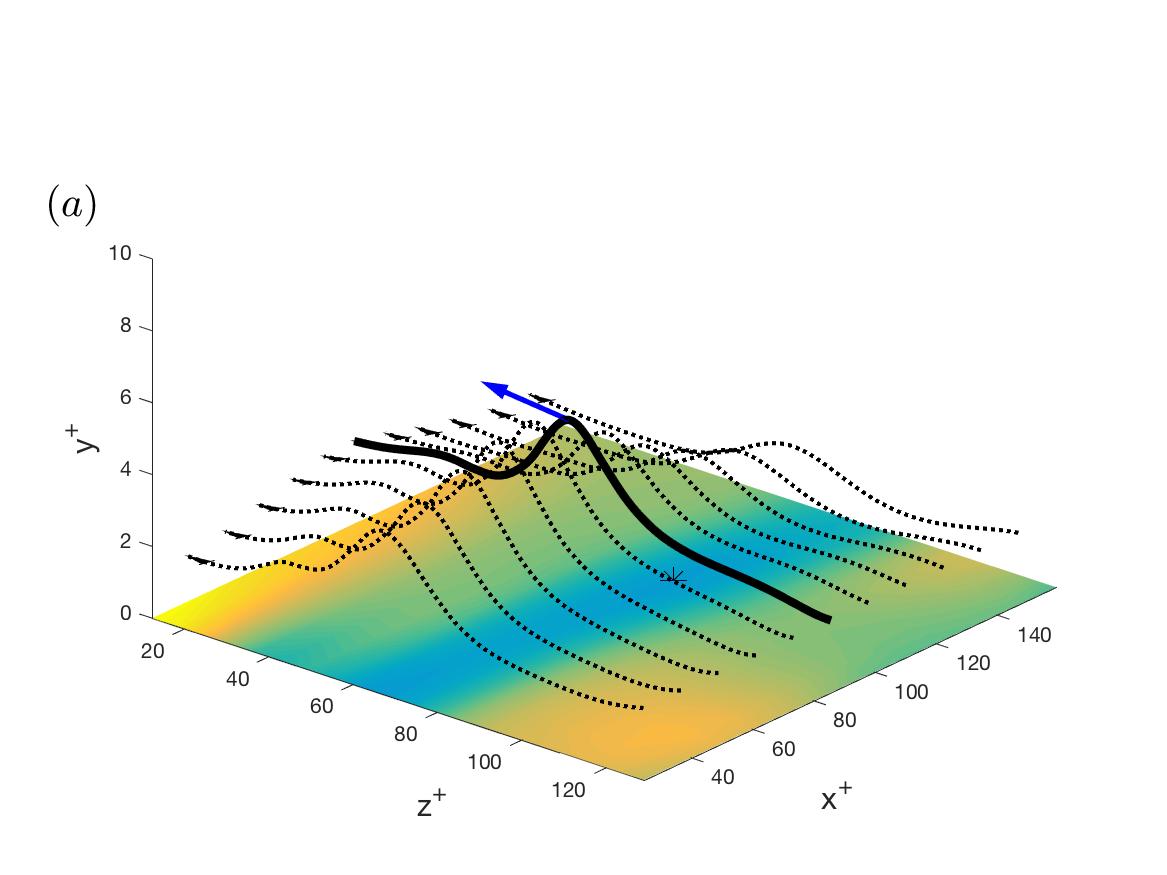}  
\end{subfigure}
\begin{subfigure}[b]{.6\textwidth}
  \centering
  \includegraphics[width=1.1\linewidth]{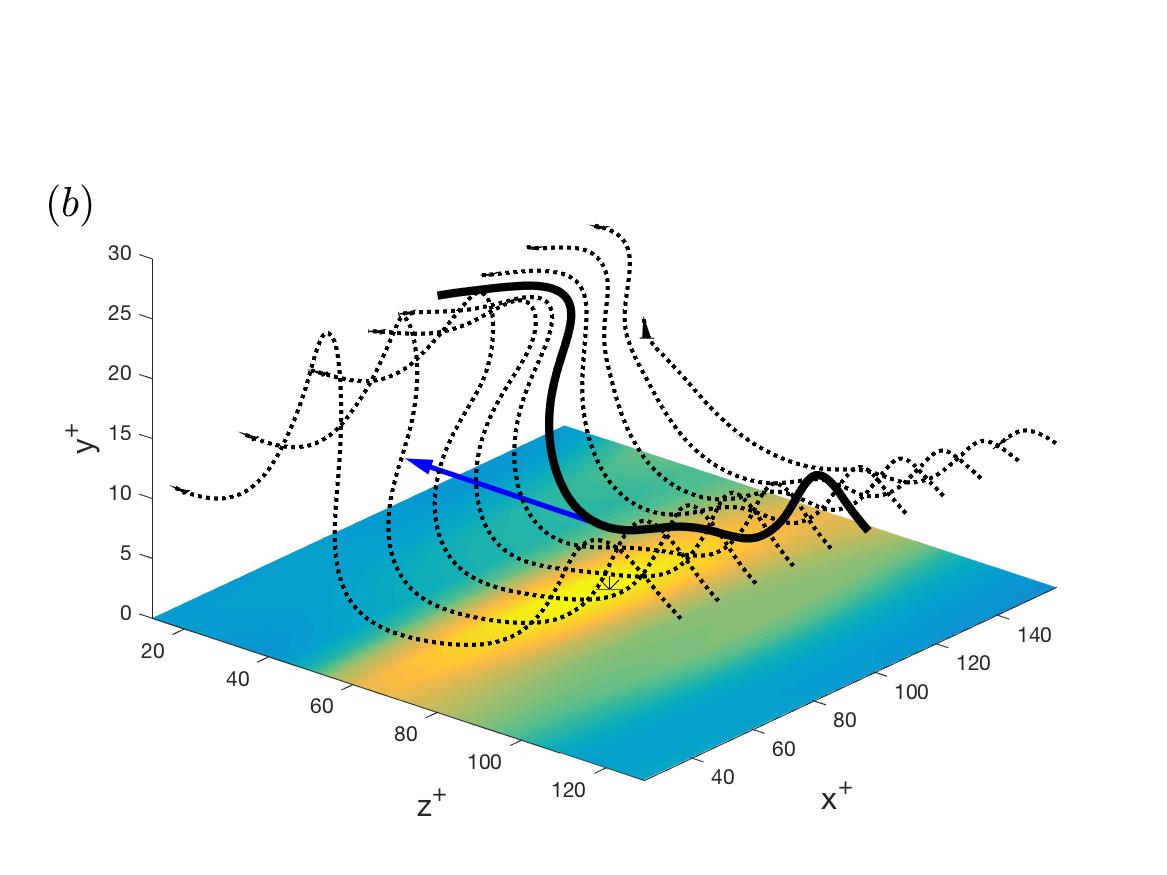}  
\end{subfigure}
\caption{The selected vortex lines, solid black, and the two selected vorticity vectors represented as arrows
for (a) the ejection and (b) the sweep event. Other vortex lines in the two flow events are plotted as dotted lines.}
\label{fig5-chosenline} 
\end{figure}

We now select for analysis specific vortex lines and specific vorticity vectors lying upon them. In this initial study,
we shall compare the Lagrangian dynamics of two vorticity vectors both pointing approximately in the negative 
spanwise direction and located at points at a similar distance from the wall, $y^+\simeq 5,$ at the bottom of the 
buffer layer.  For the ejection event in particular, we want to investigate the initial nonlinear transfer of spanwise 
vorticity from the wall, which was described by \cite{sheng2009buffer} as the process where ``spanwise vorticity 
lifts abruptly from the wall, creating initially a vertical arch''.  We have therefore considered for Lagrangian
analysis in the ejection event the lowest-lying vortex lines in Fig.~\ref{fig4-linesL}(a), which are replotted
in Fig.~\ref{fig5-chosenline}(a). These lines all start on the right at height $y^+=2$ and then pass left 
in the negative spanwise direction, rising up in a nearly vertical arch over the low-speed streak. We 
chose the middle of these lines, colored solid black, and in particular the vorticity vector at the top of 
the arch on that line at height $y^+=5.35,$ indicated by the arrow. For comparison, we considered in the 
sweep event the vortex lines which are squashed down to a comparable height above the wall by the 
splatting flow.  This is the set of lines in Fig.~\ref{fig4-linesL}(c), replotted in Fig.~\ref{fig5-chosenline}(b),  
which all start at $y^+=12$ and as they pass leftward are depressed in a trough above the high-speed streak.  
We chose the middle of these lines, colored solid black, and in particular the vorticity vector at the bottom
of the trough on that line at height $y^+=4.90,$ indicated by the arrow. The components of the two vorticity 
vectors and their position coordinates in the channel-flow database are recorded in Table 3. Note that 
the vorticity vector for the sweep has a magnitude around twice that for the ejection.
This is consistent with the argument of \cite{lighthill1963boundary} that sweeps to the wall should 
stretch and magnify spanwise vorticity, while ejections should attenuate it (see his Fig.II.22).   

Although the two vectors have different magnitudes, they are otherwise similar in orientation and 
in distance from the wall. This pair of vectors is thus well-suited to illuminate 
differences in Lagrangian dynamics that arise solely from the 
different flow conditions that exist in the two events. In particular,  our expectation is that the vorticity 
in the ejection event should have arisen recently from the wall and should be ``younger'', while the 
vorticity at the same distance from the wall in the sweep event should be ``older'' vorticity that entered 
from the wall at an earlier time, was processed by the flow, and was then returned by the splatting motion 
toward the wall. We shall see whether these expectations are borne out by our stochastic Lagrangian analysis. 

\begin{table} 
    \begin{center}
    \vspace{-10pt} 
    \caption{Coordinates of Analyzed Vorticity Vectors}
    \vspace{10pt} 
    \begin{tabular}{c  c  c  c  c  c  c  c  c} 
                      &&  $x$  &  $y$  &  $z$  &  $t$ & $\omega_x$  & $\omega_y$  & $\omega_z$ \\
      ejection  &&  21.094707  &  0.994647 & 7.563944  & 25.9935 & -2.24948978 & -0.110395804 &  22.1811829\\
      \vspace{-8pt} \\
      sweep  &&  0.715000 & 0.995100  &  0.725900  & 25.9935 &  0.05745593 &  -0.1597188  &  47.2303467\\
      \vspace{1pt} 
     \end{tabular}
  \end{center}
   \label{tab:table2}
\end{table}

\begin{figure}
\hspace{-50pt} 
\begin{subfigure}[b]{.6\textwidth}
  \centering
  \includegraphics[width=1.1\linewidth]{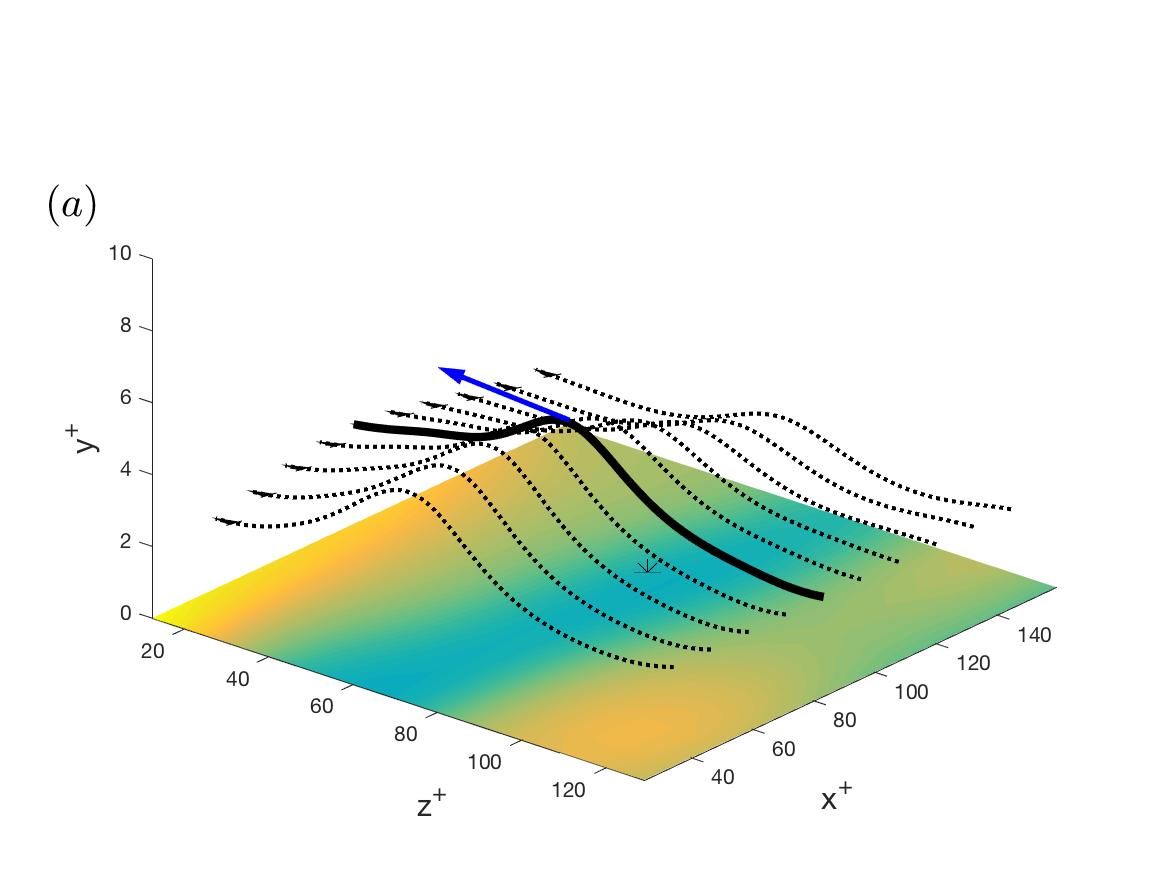}  
\end{subfigure}
\begin{subfigure}[b]{.6\textwidth}
  \centering
  \includegraphics[width=1.1\linewidth]{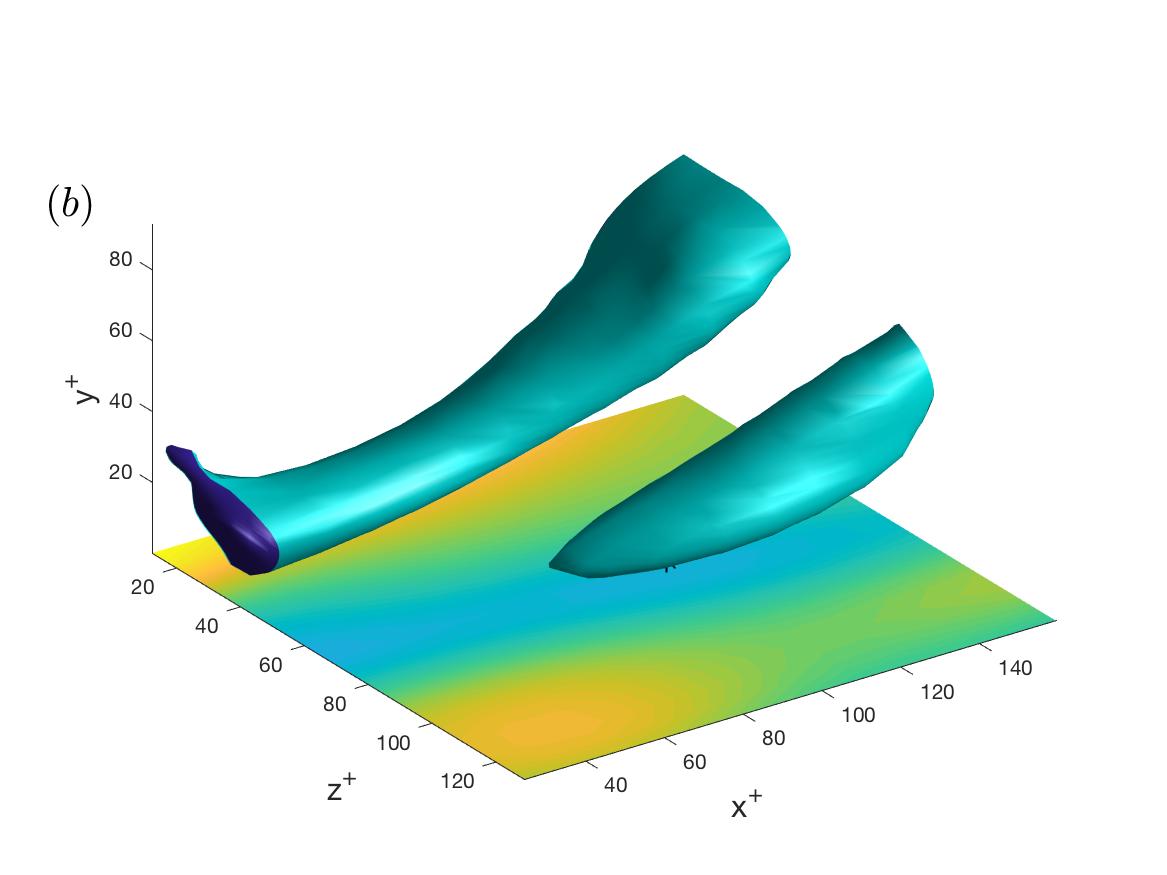}  
\end{subfigure}
\caption{(a) Vortex line (solid black line) and vorticity vector (arrow), for the coarse-grained vorticity field
at the selected point in the ejection event. Other vortex lines are plotted as dotted lines. (b) Isosurface $\lambda_2^+=-0.00815$ 
of the coarse-grained field, with magnitude 4 times the local box-average value $\langle|\lambda_2^+|\rangle = 
2.038\times 10^{-3}.$
}
\label{fig5-fchosenlineL} 
\end{figure}

We must first consider, however, the effect of filtering/coarse-graining on the structure 
and dynamics of the two events pictured in Fig.~\ref{fig5-chosenline}. As discussed in
section \ref{sec:numerics}, the mean conservation law \eqref{eqII2-1} obtained from 
the JHTDB simulation data would be expected to have improved validity when spatially 
coarse-grained over $n_i$ grid-spaces in the three coordinate directions $i=x,y,z.$ 
This expectation was verified in paper I with a box-filter; see Fig.~3 there for the ejection case 
and Fig.~5 for the sweep case. However, to justify the JHTDB channel-flow data as a 
Navier-Stokes solution accurate enough to investigate buffer-layer physics, it must 
also be shown that the filtering does not smear out the motions of interest.
We have thus attempted to choose $(n_x,n_y,n_z)$ as small 
as possible, sufficient to verify that mean conservation is improved but not so large 
that the coarse-graining obscures the essential physics. We have found that 
coarse-graining in the $y$-direction does not help to improve mean conservation, presumaby since the 
numerical resolution in the channel-flow simulation was sufficient in that direction, so that we take $n_y=0.$ 
By trial, we have found that a good choice of filtering lengths in the wall-parallel directions are 
$n_x=n_z=4,$ although somewhat smaller or larger values give similar results. Fig.~\ref{fig5-fchosenlineL}  
verifies that the coarse-grained fields for $(n_x,n_y,n_z)=(4,0,4)$ in the ejection event preserve 
the basic flow features. Plotted in Fig.~\ref{fig5-fchosenlineL}(a) is the field-line of the coarse-grained vorticity 
field which passes through the same space point as the field-line of the fine-grained vorticity that was plotted in 
Fig.~\ref{fig5-chosenline}(a) (see Table 3 for the spatial coordinates). This line starts at the right of the 
figure at vertical height $y^+ =2.7039$ and Fig.~\ref{fig5-fchosenlineL}(a) also plots an array of lines that 
start at that same height and pass in the spanwise direction.  As in Fig.~\ref{fig5-chosenline}(a), the vortex 
lines rise up nearly vertically in an arch over the low-speed streak, and the main effect of the coarse-graining 
is that the arch is broadened in the spanwise direction. The coarse-grained vorticity field at the selected point 
is $\widehat{\bom}(\bx,t)=(-1.2871, 0.3523, -29.8522)$
with magnitude increased by mixing with adjacent stronger vorticity. Fig.~\ref{fig5-fchosenlineL}(b) plots isosurfaces
of the $\lambda_2$-invariant for the coarse-grained field, which reveals a pair of quasi-streamwise, 
counter-rotating vortices flanking the low-speed streak. Compared with Fig.~\ref{fig4-lam2L}(a) for the 
fine-grained field, the vortices are somewhat smoother and weaker, but otherwise are quite similar. 
The relatively mild effects of the coarse-graining are explainable by the well-known long-range 
coherence in the streamwise direction of buffer-layer, slow-speed streaks, which extend with some 
meander for $10^3-10^4$ wall units  \citep{jimenez2013near}. 

\begin{figure}
\hspace{-50pt} 
\begin{subfigure}[b]{.6\textwidth}
  \centering
  \includegraphics[width=1.1\linewidth]{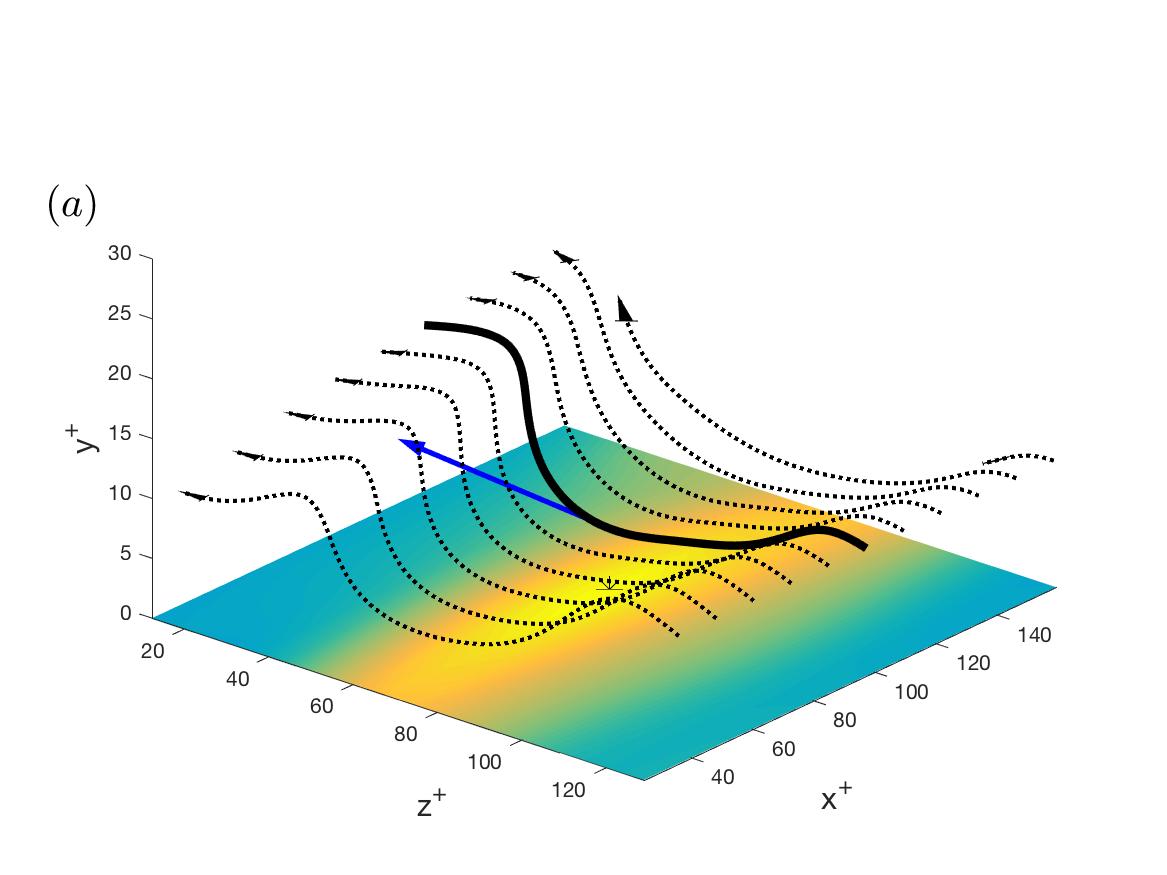}  
\end{subfigure}
\begin{subfigure}[b]{.6\textwidth}
  \centering
  \includegraphics[width=1.1\linewidth]{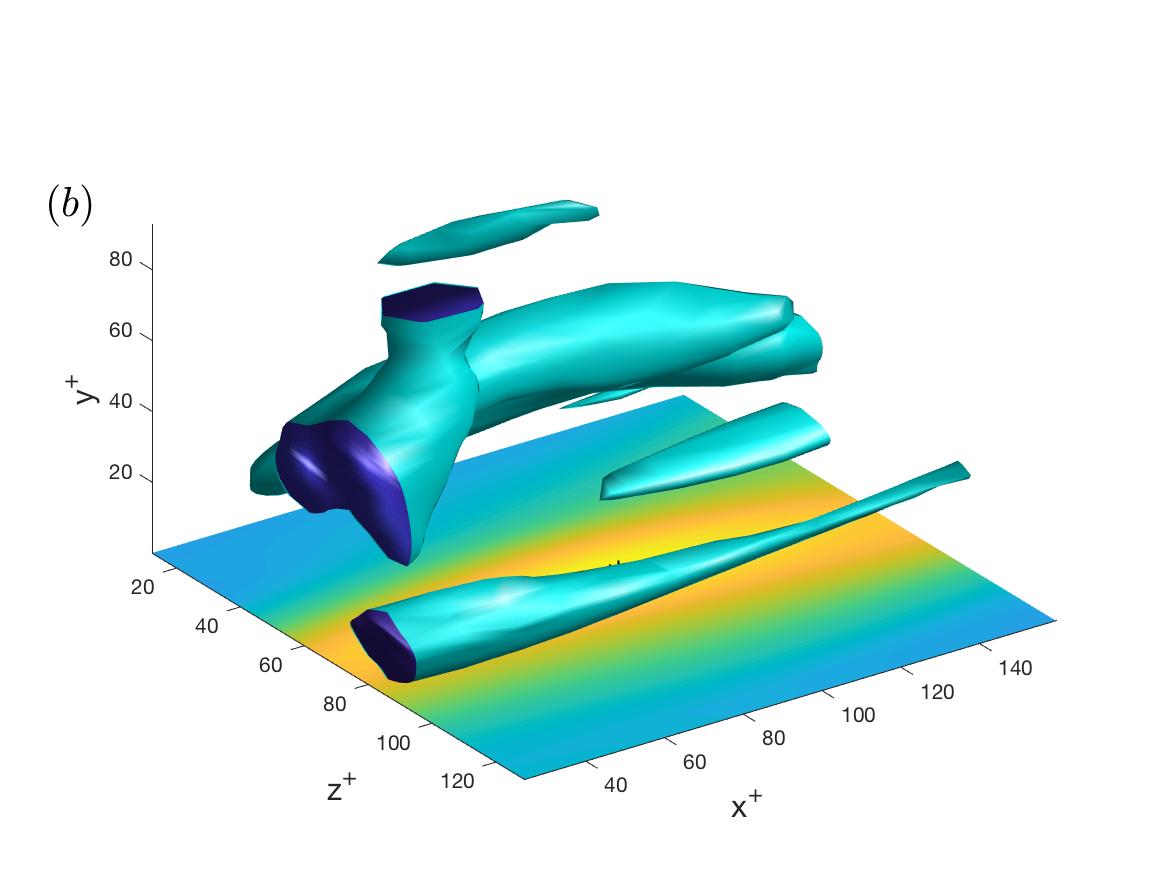}  
\end{subfigure}
\caption{(a) Vortex line (solid black line) and vorticity vector (arrow), for the coarse-grained vorticity field
at the selected point in the sweep event. Other vortex lines are plotted as dotted lines. (b) Isosurface 
$\lambda_2^+=-0.005338$
of the coarse-grained field, with magnitude 3 times the local box-average value $\langle|\lambda_2^+|\rangle = 
1.780\times 10^{-3}.$
}
\label{fig5-fchosenlineH} 
\end{figure}

Similar observations are presented in Fig.~\ref{fig5-fchosenlineH} for the sweep event, 
considering also fields coarse-grained over a box with sides of $(n_x,n_y,n_z)=(4,0,4)$
grid-lengths. At the same space point as shown in Fig.~\ref{fig5-chosenline}(b), the coarse-grained 
vorticity vector is $\widehat{\bom}(\bx,t)=(0.099327, -1.5828,  49.04330),$ which is 
not much changed from the value of the fine-grained vorticity reported in Table 3. We have plotted 
through this point the integral line of the coarse-grained vorticity, which starts at the right of the figure 
at a height $y^+=10.5622.$ This is just a bit lower than the height $y^+=12$ of the corresponding 
line for the fine-grained field. An evenly spaced array of lines of coarse-grained vorticity that start 
also at height $y^+=10.5622$ is shown in Fig.~\ref{fig5-chosenline}(b). These are depressed into 
a trough over the high-speed streak, which is shallower but just slightly wider than for the similar 
trough of lines in Fig.~\ref{fig5-chosenline}(b) for the fine-grained vorticity. Fig.~\ref{fig5-fchosenlineH}(b) 
plots isosurfaces of the $\lambda_2$-invariant for the coarse-grained field, which have a somewhat 
simpler structure but are qualitatively very similar to the isosurfaces in Fig.~\ref{fig4-lam2H}(a) for 
fine-grained vorticity. The effects of coarse-graining are even less pronounced here than for the ejection event,
presumably because the sweep is a bit broader in the spanwise direction.

We believe that these results provide sufficient {\it a priori} justification for our use of the 
JHTDB dataset to study buffer-layer vortex dynamics. We shall present below the results 
of our stochastic Lagrangian analysis of the raw/unfiltered vorticity, where mean conservation 
of the stochastic Cauchy invariant already holds to within a few percent. In the SM we present 
the corresponding Lagrangian analysis for the coarse-grained stochastic Cauchy invariant, 
with $(n_x,n_y,n_z)=(4,0,4),$ demonstrating improved conservation. All results are changed quantitatively, 
but not qualitatively, by the filtering.

\subsection{Lagrangian Description of Ejection Event}\lb{sec:lag_eject} 

\subsubsection{Origin of Vorticity at the Wall}

We now turn to the central question of the paper, the origin at the wall of the buffer-layer vorticity. 
A regular spatial array of vortex-lines such as plotted in Fig.~\ref{fig5-chosenline}(a) might suggest 
a simple dynamical process  of ``abrupt lifting'' of the vorticity away from the wall within a few viscous 
times. There are, however, {\it a priori} theoretical reasons to expect that the process is much slower
and more complex. We have already discussed in paper I the strong Lagrangian chaos in the buffer-layer 
\citep{johnson2017analysis}, which argues against a simple ideal lifting motion. Furthermore, we 
note that stochastic Lagrangian particles, moving backward in time, cannot ever reach the wall by 
fluid advection alone but only through viscous diffusion, because the wall-normal velocity drops to zero rapidly 
with decreasing distance to the wall. Diffusion is an intrinsically slow process. If wall-normal 
velocity were exactly zero, then a Brownian particle with diffusivity $\nu$ released at height $y^+$
would reach the wall at $y^+=0$ in a random time $\tsigma^+>0$ (in wall units) with probability density 
\be p(\sigma^+) = \frac{y^+}{\sqrt{4\pi\sigma^{+3}}}\exp\left(-\frac{y^{+2}}{4\sigma^+}\right), \quad \sigma^+>0 
\lb{eq5-6} \ee 
which is a special case of an ``inverse Gaussian distribution''; see \cite{chhikara1988inverse}, 
\cite{borodin2015handbook}, Part II, formula 2.02 (p.295). The mean value of the hitting time with 
distribution Eq.~\eqref{eq5-6} is infinite, which implies that it takes a very long time to reach the wall, 
with a high probability. This result does not contradict the natural expectation that vorticity created at 
the wall will diffuse across the viscous sublayer to $y^+\sim 5$ in a time of order $t_\nu=\delta_\nu^2/\nu,$
because the diffusion process is asymmetric in time. Under time-reversal, a Brownian particle that reaches the 
wall is described by a ``3-dimensional Bessel process'' forward in time (\cite{borodin2015handbook}, Part I, Chapter II.5, p.35),
which is strongly repelled from the wall at $y=0,$ because it cannot return there again. The stochastic trajectory 
released at $y^+=5$ and moving backward in time will take much longer to reach the wall, because 
it will cross and recross the level $y^+=5$ many times before ultimately reaching the wall located at $y^+=0.$

\begin{figure}
\begin{subfigure}[b]{\textwidth}
  \centering
  \includegraphics[width=.65\linewidth]{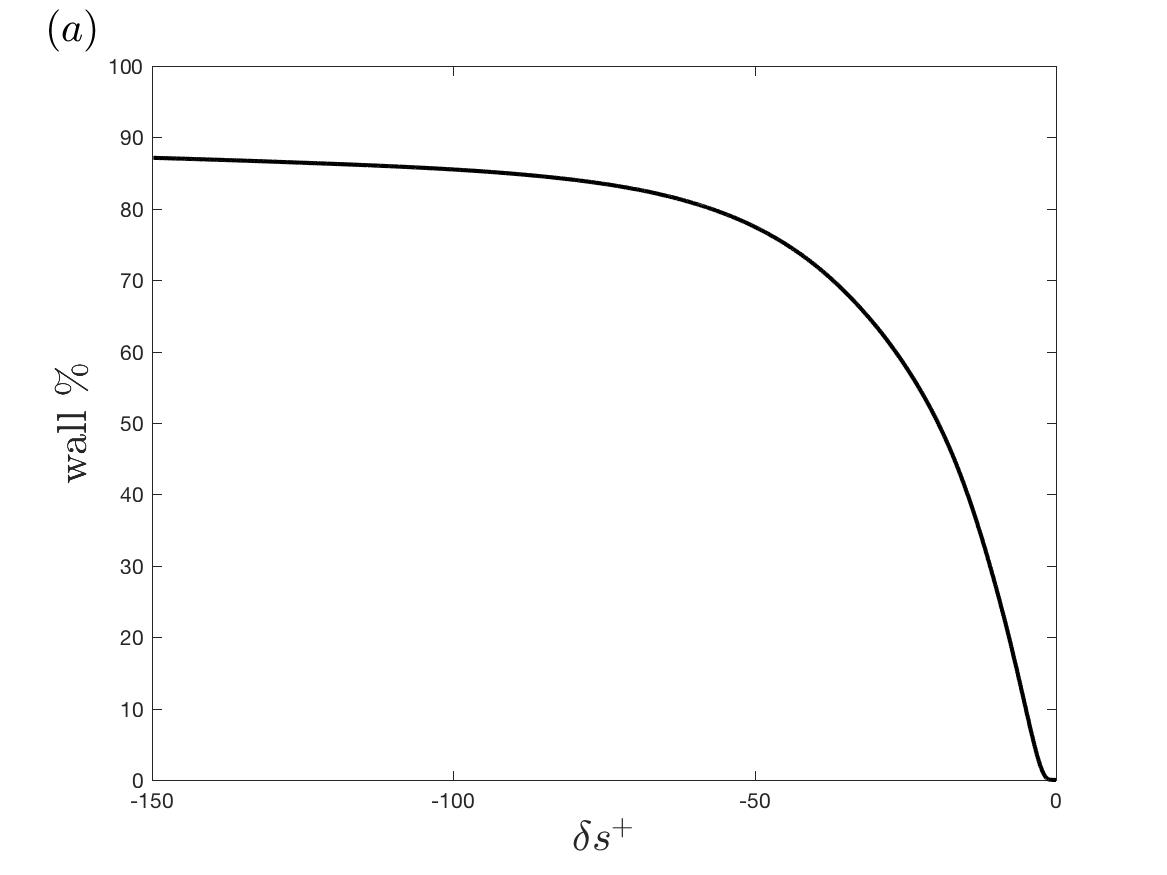}  
\end{subfigure}
\\
\begin{subfigure}[b]{\textwidth}
\centering
  \includegraphics[width=.65\linewidth]{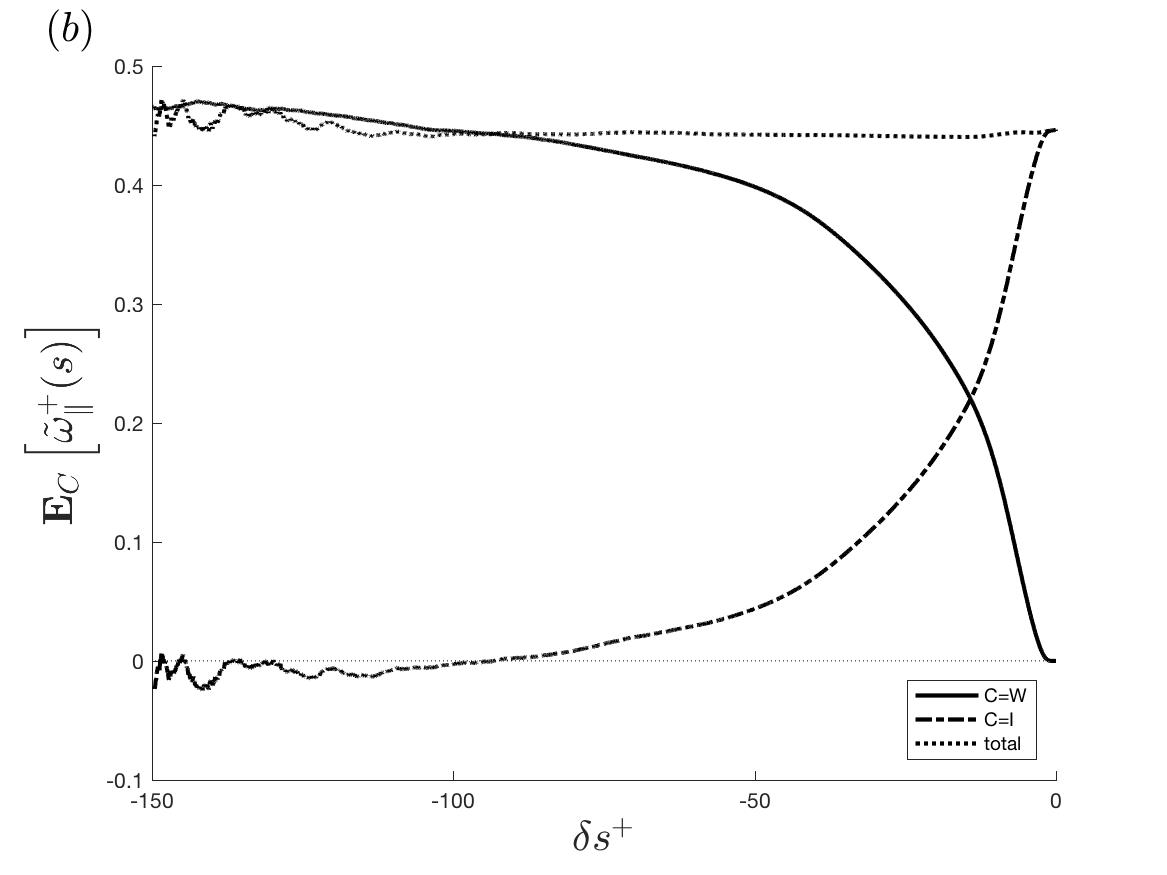}  
\end{subfigure}
\\
\begin{subfigure}[b]{\textwidth}
\centering
  \includegraphics[width=.65\linewidth]{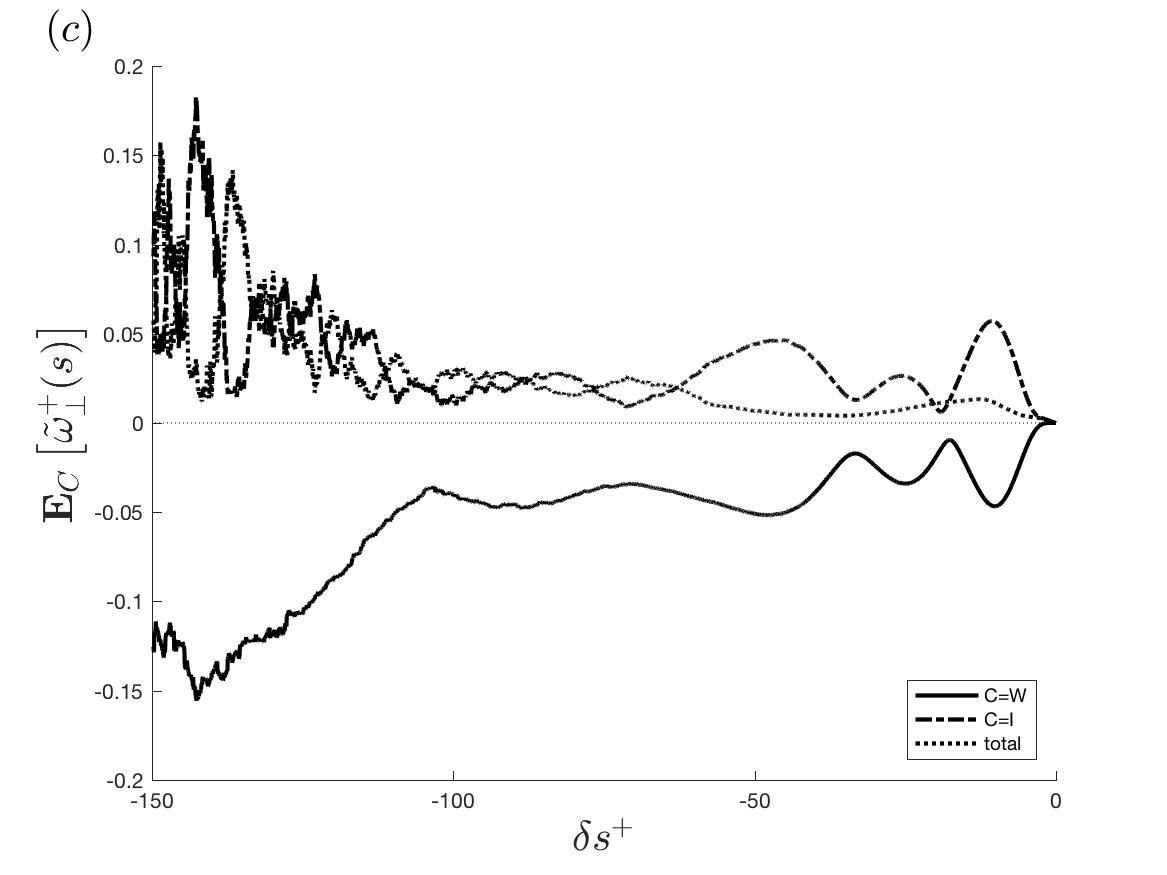}  
\end{subfigure}

\caption{(a) Fraction of particles at the wall, (b) partial means 
of the parallel component of the stochastic Cauchy invariant, and (c) partial means of the 
perpendicular component of the stochastic Cauchy invariant. All quantities in wall units 
plotted versus $\delta s^+.$}
\label{fig5-cauchycontribL}
\end{figure}

Numerical results on stochastic Lagrangian dynamics of vorticity from our Monte Carlo approach 
confirm the above theoretical expectations. The percentage of particles released at the selected point
at $y^+=5.35$ which have hit the wall going backward in time is plotted in Fig.~\ref{fig5-cauchycontribL}(a)  
versus $\delta s=s-t.$ The percentage grows faster in time for this fluid ejection event than it does for pure 
diffusion, because the lifting flow advects particles toward the wall backward in time. Nevertheless, 
hundreds of viscous times are required for nearly the entire ensemble of particles to hit the wall.
After an initial fairly abrupt rise to $\sim 75\%$ in the interval $-50<\delta s^+<0,$ less than 
$90\%$ of the particles have hit the wall even after 150 viscous times. An even more salient issue 
is the relative contributions to the stochastic Cauchy invariant which arise from particles that have arrived 
at the wall and from those still in the flow interior. We can define such partial contributions by 
$$ \bE_C\big[\widetilde{\omega}_{s\,i}(\bx,t)\big]:= \bE\big[{\bf 1}_C \widetilde{\omega}_{s\,i}(\bx,t)\big]$$   
where $C$ is a subset of stochastic trajectories satisfying conditions such as 
$$ \mbox{particle at wall at time $s$:} \quad W=\{ |\tilde{b}^s_t(\bx)|=h\} $$
$$ \mbox{particle in interior at time $s$:} \quad I=\{ |\tilde{b}^s_t(\bx)|<h\} $$
and ${\bf 1}_C$ is the indicator function of this subset, $=1$ on the set and $=0$ on its complement. 
The stochastic Cauchy invariants are vector quantities, and we consider here the intrinsic parallel 
and perpendicular components as defined in paper I, Eq.(3.14), or:
\be \widetilde{\omega}_{s\|}(\bx,t):=\widetilde{\bom}_{s}(\bx,t)\cdot \bn_\omega(\bx,t), \quad 
\widetilde{\bom}_{s\perp}(\bx,t):=\widetilde{\bom}_{s}(\bx,t)-\widetilde{\omega}_{s\|}(\bx,t) \bn_\omega(\bx,t),
\lb{eq5-1} \ee 
where $\bn_\omega(\bx,t)=\bom(\bx,t)/|\bom(\bx,t)|.$  
In Fig.~\ref{fig5-cauchycontribL}(b)-(c) we plot partial means for subsets $C=W,$ $I$ and for components 
$i=\|,$ $\perp$. We plot also in Fig.~\ref{fig5-cauchycontribL}(b)-(c)  the conserved total means 
of the stochastic Cauchy invariants, reproducing the results for $i=\|,$ $\perp$ in 
Fig. 2 of Paper I. 
As can be observed, the conserved means are the summed results of strongly time-dependent 
contributions separately from particles at the wall and in the interior and conservation for $i=\perp,$ in particular,
involves complete vector cancellations between the two contributions. (To make this cancellation 
visually obvious, we have plotted $+\left|\bE_I\big[\widetilde{\omega}_{s\perp}(\bx,t)\big]\right|$
and $-\left|\bE_W\big[\widetilde{\omega}_{s\perp}(\bx,t)\big]\right|$ in Fig.~\ref{fig5-cauchycontribL}(c).
The complete cancellation is more perfectly exhibited by the results on the coarse-grained Cauchy invariant 
presented in the SM, since the total perpendicular component remains closer to vanishing there.) 
For the parallel component plotted in Fig.~\ref{fig5-cauchycontribL}(b) there is a gradual crossover 
from the conserved mean arising from interior particles to instead arising from wall particles going
backward in time. The contribution from the wall particles is notably larger 
than the fraction of the particles located at the wall. Indeed, at $\delta s^+\simeq -100,$ almost 
100\% of the conserved parallel component arises from the wall contribution, even though only 
about 85\% of the particles have reached the wall.

\begin{figure} 
\centering
  \includegraphics[width=.8\linewidth]{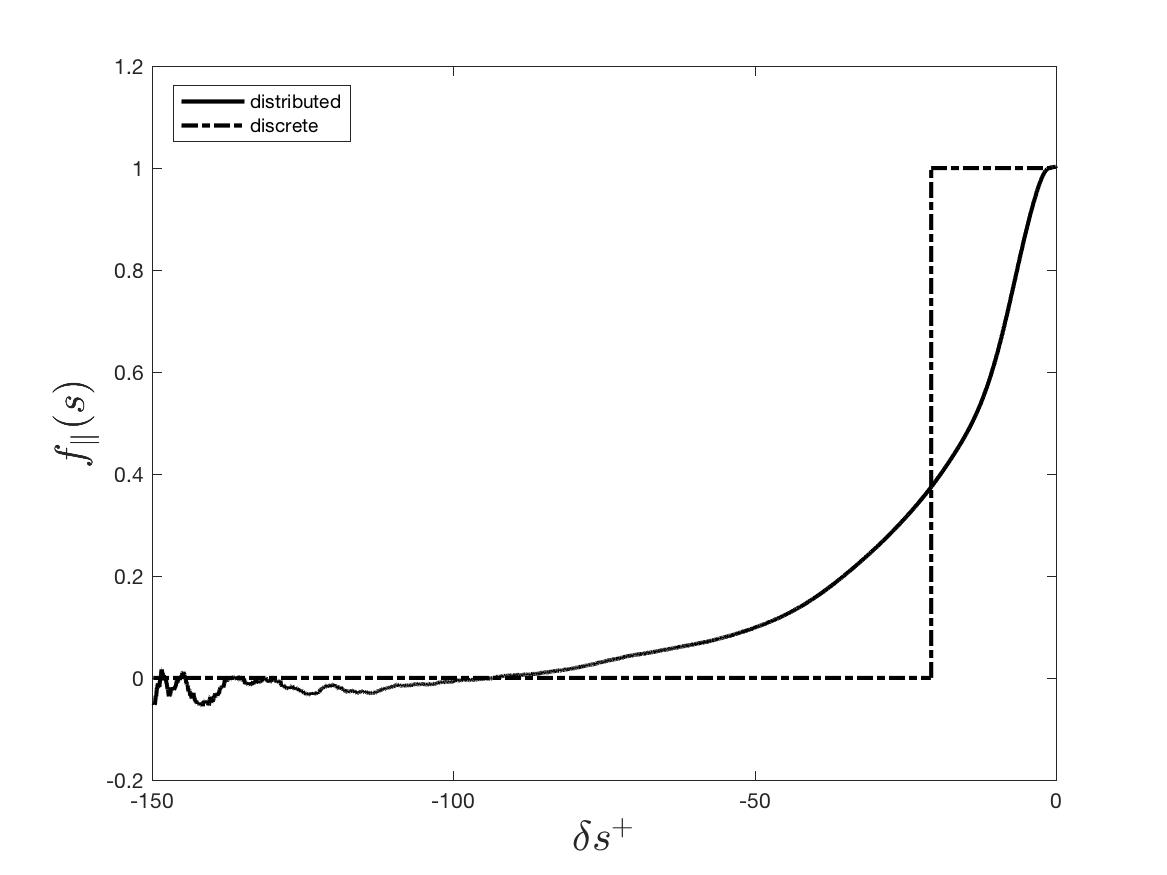}  
\caption{Fractional contribution of the parallel vorticity arising at times $s<t$ from interior particles,  
plotted versus $\delta s^+,$ and its discrete approximation as ``abrupt lifting''.}
\label{fig5-integraltimeL}
\end{figure}

Even so, however, the $\sim 100$ viscous times to get the entire parallel vorticity component 
originating from the wall is much longer than the few viscous times postulated by \cite{sheng2009buffer}. 
Lifting of vorticity from the wall is not an ``abrupt'' event but is instead a very prolonged process. 
To quantify the time required or the mean ``age'' of the vorticity vector $\bom(\bx,t)$, we can introduce an integral 
formation time for the parallel component 
\be   T_\|(\bx,t) := \int_{-\infty}^t ds \ f_\|(s;\bx,t), \quad 
f_\|(s;\bx,t):=\frac{\bE_I\big[\widetilde{\omega}_{s\|}(\bx,t)\big]}{|\bom(\bx,t)|}.
\lb{eq5-7} \ee
Here $f_\|(s;\bx,t)$ is the fractional contribution of the parallel vorticity arising at times $s<t$ from interior particles,
satisfying $f_\|(t;\bx,t)=1$ and $f_\|(-\infty;\bx,t)=0.$ Fig.~\ref{fig5-integraltimeL} plots as a function of $\delta s$ 
the fraction $f_\|(s;\bx,t)$ obtained from our numerical Monte Carlo method, together with the discrete approximation 
that corresponds to a Heaviside step function with jump at $\delta s=-T_\|(\bx,t)$. Numerical quadrature gives 
$T_\|^+(\bx,t)\simeq 20.7$ in wall units. The step function represents graphically the ``abrupt lifting'' 
proposed by \cite{sheng2009buffer}, pictured as a discrete event. This is not an unreasonable caricature 
of the actual vortex-lifting process, except that the integral time is about an order of magnitude larger 
than the heuristic time estimate by \cite{sheng2009buffer} and the discrete picture misses entirely the long,
slowly decaying tail. As a matter of fact, the true ``age'' of the vorticity $\bom(\bx,t)$ in our selected point 
is even much larger than estimated by $T_\|(\bx,t)$ because, as shown in Fig.~\ref{fig5-cauchycontribL}(c),
the interior particles contribute also a very slowly vanishing {\it perpendicular} component to the vorticity.
One could define an integral time for decay of this perpendicular component to zero, for example, by
\be  T_\perp(\bx,t) := \int_{-\infty}^t ds \ f_\perp(s;\bx,t), \quad 
f_\perp(s;\bx,t):=\frac{\left|\bE_I\big[\widetilde{\omega}_{s\perp}(\bx,t)\big]\right|}
{\max_s\left|\bE_I\big[\widetilde{\omega}_{s\perp}(\bx,t)\big]\right|}.
\lb{eq5-8} \ee   
We shall not attempt a quantitative evaluation of this quantity, because our numerical Monte Carlo scheme 
with $N=10^{7}$ particles does not yield converged results for the perpendicular Cauchy invariant at times 
$\delta s^+<-100.$ However, Fig.~\ref{fig5-cauchycontribL}(c) shows that the perpendicular component 
at $\delta s^+=-100$ still remains about $0.1|\bom|.$ It is thus clear from Fig.~\ref{fig5-cauchycontribL}(b)-(c) 
at least that $T_\perp(\bx,t) \gg T_\|(\bx,t).$

\begin{figure}
\hspace{-65pt} 
\includegraphics[width=1.4\linewidth]{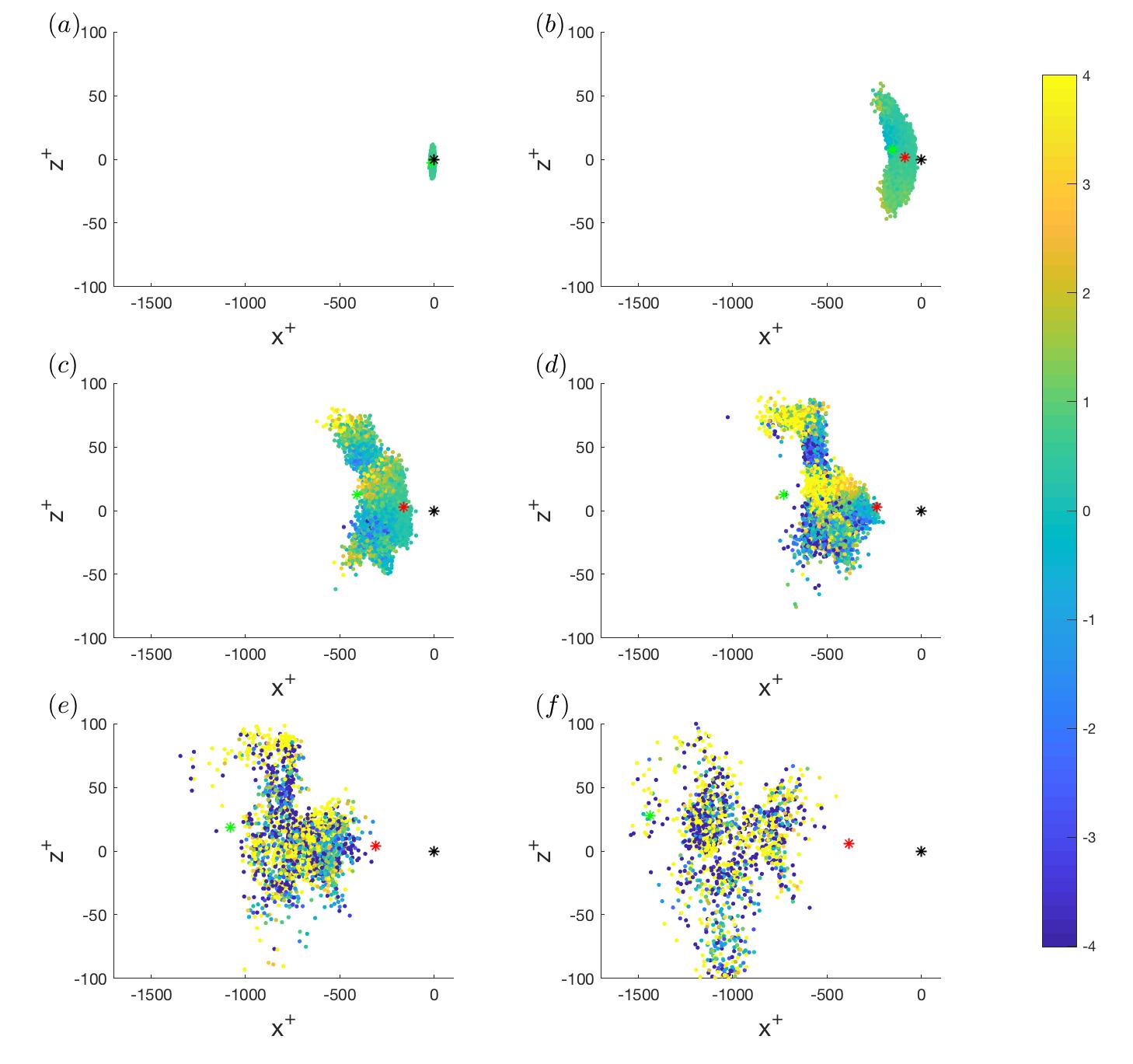}  
\caption{Scatterplot of $\tilde{\omega}_{s\|}^+(\bx,t)$ values for particles hitting the wall in 5-unit 
time intervals ending at: (a) $\delta s^+=-5,$ (b) $\delta s^+=-30,$ (c) $\delta s^+=-55,$  (d) $\delta s^+=-80,$  
(e) $\delta s^+=-105,$  (f) $\delta s^+=-130.$ Mean position of all particles ($\textcolor{red}{*}$) 
and only in interior ($\textcolor{green}{*}$). The resultant vorticity magnitude $|\bom^+(\bx,t)|\doteq 0.45.$\\ }
\label{fig5-movieL}
\end{figure}
                    
The broad space-time distribution and intricate Lagrangian dynamics of the vorticity generation process is 
further revealed by Fig.~\ref{fig5-movieL}. Plotted at the channel wall are realizations of the parallel component 
of the stochastic Cauchy invariant, $\tilde{\omega}_{s\|}(\bx,t),$ with values encoded by color/shade, at the position 
$(\tilde{a}_*(\bx,t),\tilde{c}_*(\bx,t))$ where the particle hits the wall and grouped in intervals of hitting time 
$ -5(5k+1)<\delta \tsigma_t^+(\bx)< -25k$ for $k=0,1,2,3,4,5,$ with $\delta \tsigma_t(\bx):=\tsigma_t(\bx)-t.$ 
This figure was created with a sub-ensemble of $10^6$ stochastic Lagrangian particles and each panel plots
Cauchy vorticity contributions for all of the particles in this sub-ensemble that hit the wall in the given 5-viscous-time 
interval.  In the SM, we provide a video with similar plots as frames, at a larger set of backward times $\delta s,$ 
with greater time resolution, and employing all $10^7$ available particles. Averaging over all of these wall contributions 
yields the resultant vorticity magnitude $|\bom(\bx,t)|$ at the selected space-time point $(\bx,t)$ in the ejection event. 
For spatial reference, Fig.~\ref{fig5-movieL} plots  also in each frame three points: (1) the wall-parallel position $(x,z)$ 
of the selected point, taken as the coordinate origin (0,0), plotted as a black asterisk ``$\textcolor{black}{*}$''; 
(2) the mean position of all particles in the entire ensemble at the given time $s,$ plotted as a red/dark-grey asterisk 
``$\textcolor{red}{*}$'';  and, (3) the mean position of all interior particles at the given time $s,$ plotted 
as a green/light-grey asterisk ``$\textcolor{green}{*}$''.  The green point drifts quickly upstream backward in time, since the   
``living'' particles are generally at higher elevations where the streamwise velocity is larger. The red point also 
drifts upstream, but more and more slowly as particles hit the wall and stop at their ``birth place.'' These points 
provide useful context in the figure. 

The wall-plots in Fig.~\ref{fig5-movieL} display clearly the space-time origin of the resultant vorticity. 
The particles newly arrived to the wall in each frame land roughly between the mean positions of all particles 
in the ensemble and of interior ``living'' particles. Over the range of times $-100<\delta s^+<0$ that contribute 
substantially to $\bom(\bx,t),$ the stochastic particles hit the wall in a region extending $\sim 1000$ wall units 
in the streamwise direction  and $\sim 200$ wall units in the spanwise direction. The ``cloud'' of points 
instantaneously hitting the wall in each frame also expands going backward in time, faster in the streamwise 
direction than in the spanwise. The faster growth of the particle cloud in the streamwise direction is explained 
by the dispersive effect of the mean-shear, which produces a super-ballistic $\sim (\delta s)^3$ growth of 
the mean-square streamwise extent of the cloud \citep{corrsin1953remarks}. Such super-ballistic growth 
was previously verified for stochastic Lagrangian particles in the buffer-layer of this same channel-flow
database \citep{drivas2017lagrangianII} and it is also observed here (see SM). The mean-square spanwise 
extent of the cloud also grows backward in time, but at a slower diffusive rate $\sim \delta s$ (see SM). 
We conclude that the vorticity in the vertical arch at  $y^+=5.35$ does not arise from a location $\sim 10$ wall units 
upstream, as conjectured by \cite{sheng2009buffer}, but instead from a region at the wall at least 
1-2 orders of magnitude larger in extent. 

The magnitude of the final vorticity in the arch is $|\bom^+|\doteq 0.45,$ but the individual contributions 
of the stochastic Cauchy invariant are much larger. The maximum values observed 
will, of course, increase with the number of samples $N$ employed in our calculation. For the $N=10^6$ 
ensemble used to prepare Fig.~\ref{fig5-movieL}, the largest values of $\tilde{\omega}_{s\|}^+(\bx,t)$ were seen 
to grow with increasing $|\delta s|$
at a slightly less than exponential rate over the interval $-150<\delta s^+<0,$ from values near $\pm 1$ at
small $\delta s^+$ to around $\pm 500$ at $\delta s^+=-150.$ There are substantial fluctuations from smooth 
(sub)exponential growth, however, and the largest values of the stochastic invariant $\tilde{\omega}_{s\|}^+$ 
encountered over the interval $-150<\delta s^+<0$ with $N=10^6$ were $\pm 10^4.$ It should be emphasized that
these large values do not correspond to the vorticity magnitudes sampled by the particles when they 
hit the wall. The wall-vorticity is pointed mainly in the spanwise direction, as illustrated in Fig.~\ref{fig4-pressL}(b), 
and has magnitude $\doteq 0.45$ in wall units. The large values are instead the consequence of exponential 
growth of the wall-vorticity as it is transported forward in time along the stochastic Lagrangian trajectories, 
consistent with the growth of variances observed in 
Paper I, Fig. 2(b) 
and with the strong Lagrangian 
chaos reported in the buffer-layer \citep{johnson2017analysis}. The mean value of the realizations $\tilde{\bom}^+_{s\|}$
arising from the wall can yield the order-unity value $|\bom^+|\doteq 0.45$ only if there is nearly complete cancellation
between contributions from opposite sign. This is clearly exhibited in Fig.~\ref{fig5-movieL}, where yellow/light  
indicates large positive values and blue/dark large negative values. The negative values arise from vorticity 
elements that start at the wall aligned in the negative spanwise direction with the mean vorticity, but whose 
parallel component is rotated $180^\circ$ as the vector is transported from its ``birth place'' at the wall to 
the final point $(\bx,t)$ on the arch. The extensive cancellation between oppositely signed contributions 
is the representation in our stochastic Lagrangian framework of strong viscous destruction of vorticity, 
which counterbalances the exponential growth of vorticity by strong Lagrangian chaos.         

The plots in Fig.~\ref{fig5-movieL} exhibit interesting and nontrivial structure of the parallel Cauchy invariant
plotted against wall-position, especially for intermediate values of $\delta s^+.$ The particles that hit the wall 
in the earliest time interval $-5<\delta s^+<0$ (panel (a)) contribute only positive values of order unity. 
The wall-vorticity in this early time is transported essentially by pure diffusion and without stretching or 
rotation. However, for more negative values of $\delta s^+,$ large opposite signs of $\tilde{\bom}^+_\|$
develop at the wall, with clear spatial organization. This order presumably reflects in part the well-known     
Eulerian vortex structures in the flow, such as the counter-rotating pair of streamwise vortices pictured in 
Fig.~\ref{fig4-lam2L}(a). However, these patterns involve also the Lagrangian evolution over time and 
become progressively more complex and fine-grained as $\delta s^+$ grows more negative. The 
complex, intertwined positive and negative values enhance the amount of cancellation. Eventually, for 
$\delta s^+<-100,$ the scatter of positive and negative values of $\tilde{\bom}^+_\|$ becomes 
essentially random and the cancellation is nearly complete (panel (f)). Particles continue to hit the wall 
backward in time for $\delta s^+\ll-100$ and the r.m.s. values of $\tilde{\bom}^+_\|$ grow larger,
but these very early contributions become less and less probable and cancel almost entirely, giving 
no net contribution to the resultant magnitude $|\bom(\bx,t)|$ at the top of the arch. 
                
\subsubsection{Relation to the Eulerian Vorticity Source}

\begin{figure}
\hspace{-65pt} 
\includegraphics[width=1.4\linewidth]{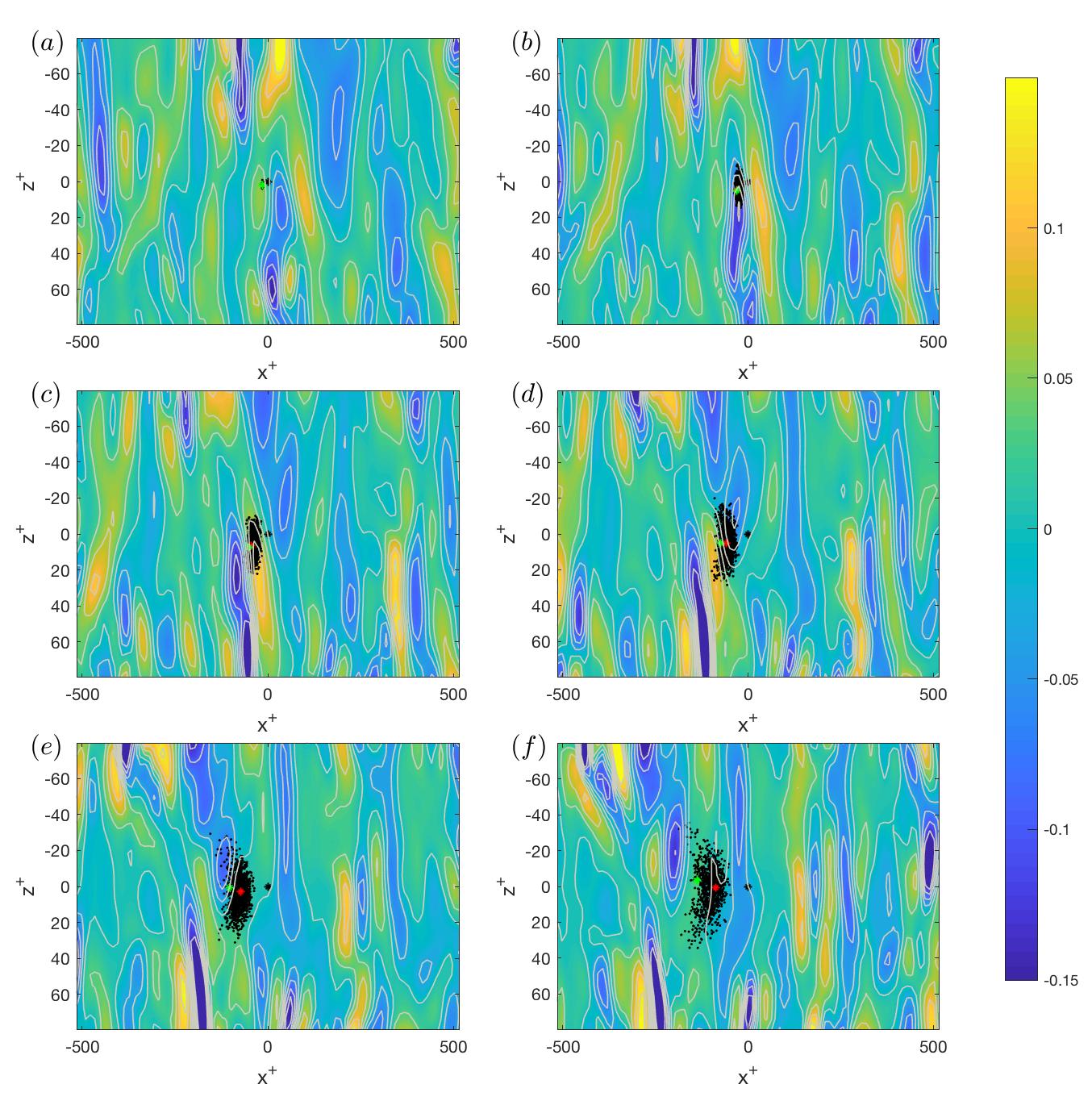}  
\caption{Color plots of negative spanwise vorticity source together with scatterplots of particles hitting the wall
in unit time intervals ending at: (a) $\delta s^+=-1,$ (b) $\delta s^+=-6,$ (c) $\delta s^+=-11,$  
(d) $\delta s^+=-16,$ (e) $\delta s^+=-21,$  (f) $\delta s^+=-26.$ The light grey lines are contour
levels $-\sigma_z^+=k/30$ for integers $|k|\leq 4.$  Also shown are the mean position of all particles 
($\textcolor{red}{*}$) and of only those in the interior ($\textcolor{green}{*}$).} 
\label{fig5-pressuremovieL}
\end{figure}          
          
To make connection with the Eulerian theory of \cite{lighthill1963boundary} and \cite{morton1984generation},
we present in Fig.~\ref{fig5-pressuremovieL} a plot of the negative spanwise vorticity source $-\sigma_z(\bx,s)$
at time $\delta s^+=-5k,$ together with a scatterplot of particles landing at the wall in the interval of hitting time 
$ -(5k+1)<\delta \tsigma_t^+(\bx)< -5k,$ for values $k=0,1,2,3,4,5$ in successive panels. To render more clear 
the pattern of the source underneath the particle markers we have added isolines $-\sigma_z^+=k/30$ for $|k|\leq 4.$
We go back in time by only about $T_\|^+\doteq 20.7,$ since Fig.~\ref{fig5-integraltimeL} shows that more than 50\% 
of the final parallel vorticity is generated from the wall in that interval. We provide in the SM a movie whose frames are plots of 
the same format, but with more frames and going back 50 viscous times. The first panel, Fig.~\ref{fig5-pressuremovieL}(a), 
is essentially the same as Fig.~\ref{fig4-pressL}(b) but showing a larger domain. The two regions with $\sigma_z<0$ 
just upstream of the stress-minimum and with $\sigma_z>0$ just downstream in Fig.~\ref{fig4-pressL}(b)
now appear as parts of larger ``band" structures. The spanwise vorticity source plotted in Fig.~\ref{fig5-pressuremovieL} 
exhibits an alternating pattern of such bands, each with streamwise thickness $L_x^+\sim 25-50$ and spanwise length 
$L_z^+\sim 50-100.$ The reverse sign in successive bands is the manifestation of the ``bipolarity'' 
of the vorticity source. Since it has been proposed in the literature that ejections should be associated 
with values $\sigma_z<0$ \citep{andreopoulos1996wall}, we would like to investigate whether the 
region with $\sigma_z<0,$ just upstream of the point marked with an asterisk ``$\ast$'', 
can be the source of the vertical vortex arch at height $y^+=5.35.$

A striking feature in Fig.~\ref{fig5-pressuremovieL} which is even more apparent in the associated 
movie (see SM) is that the band patterns seem to travel in the negative streamwise direction,
backward in time.  In fact, it is well-known that velocity and pressure structures at the wall
consist of travelling waves with streamwise velocities $\sim 10-15$ measured in units of the 
friction velocity $u_*$ \citep{kim1993propagation}, agreeing well with the velocity inferred  
by eye from Fig.~\ref{fig5-pressuremovieL}.  This means in particular that propagation speeds 
of the waves exceed the mean flow velocity $\overline{u}(y)$ for $y^+<15$ and increasingly so 
as $y^+$ decreases. A consequence is that the band with $\sigma_z<0$ just upstream of the ``$\ast$''
moves further upstream (backward in time) at a considerably faster speed than does the cloud of particles 
released at $y^+=5.35,$ and thus particles in that ensemble hit this moving band with very low probability.   
It therefore seems ruled out that the upstream band with $\sigma_z<0$ is the ``source'' of the vorticity 
in the arch at $y^+=5.35.$ Only particles released at very small heights $y^+\lesssim 1,$ well below the 
arch, hit that band with substantial probability and at those heights the vortex lines are flat, with almost 
no vertical component.

Instead, the particles released from the vortex arch at $y^+=5.35$ and going backward in time hit structures 
that were originally {\it downstream} of ``$\ast$'' in Fig.~\ref{fig5-pressuremovieL}(a) but which rapidly moved 
upstream of ``$\ast$'' in subsequent panels. In particular, there is a band with large negative values 
$\sigma_z^+\simeq -0.1$ about 100 wall units downstream of ``$\ast$'' in Fig.~\ref{fig5-pressuremovieL}(a) 
which particles hit with very high probability at times $-15<\delta s^+<-10$ (Fig.~\ref{fig5-pressuremovieL}(c-d)).
At this same interval in Fig.~\ref{fig5-integraltimeL} one sees a very sharp increase in the contribution from the wall 
to the final vorticity magnitude. The intense band 100 wall units downstream of ``$\ast$'' in Fig.~\ref{fig5-pressuremovieL}(a)  
is thus a more likely source of the enhanced negative spanwise vorticity in the vertical arch (Fig.~\ref{fig4-domzL}), 
or at least the part associated to ``abrupt lifting''. Of course, the negative spanwise vorticity
injected by this source is not all delivered to this one arch but is instead distributed more generally throughout the flow.
Fig.~\ref{fig5-pressuremovieL}(e-f) shows that the particles for $-26<\delta s^+<-21$ sample a broad region 
with smaller negative values of source $\sigma_z^+,$ corresponding to the turnover to a 
slowly decaying tail in Fig.~\ref{fig5-integraltimeL}.

\subsection{Lagrangian Description of Sweep Event}\lb{sec:lag_sweep}

We now discuss the results for the vorticity at the bottom of the trough in the ``sweep'' event, 
as pictured in Fig.~\ref{fig5-chosenline}(b). In particular, we investigate how the vorticity in the trough 
originates at the wall. We are especially interested to compare with the previously discussed results 
for the low arch in the ``ejection'' event  pictured in Fig.~\ref{fig5-chosenline}(a).

\subsubsection{Origin of Vorticity at the Wall}

\begin{figure}
\begin{subfigure}[b]{\textwidth}
  \centering
  \includegraphics[width=.65\linewidth]{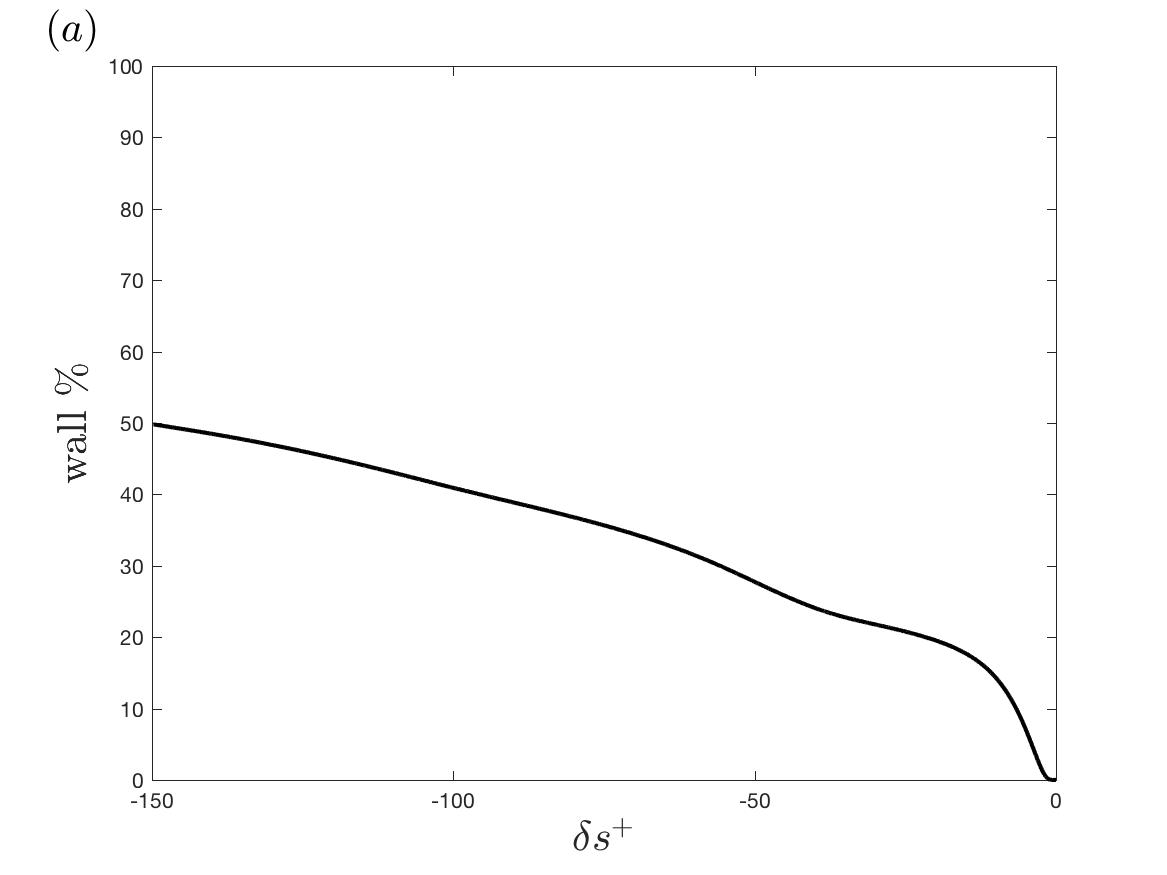}  
\end{subfigure}
\\
\begin{subfigure}[b]{\textwidth}
\centering
  \includegraphics[width=.65\linewidth]{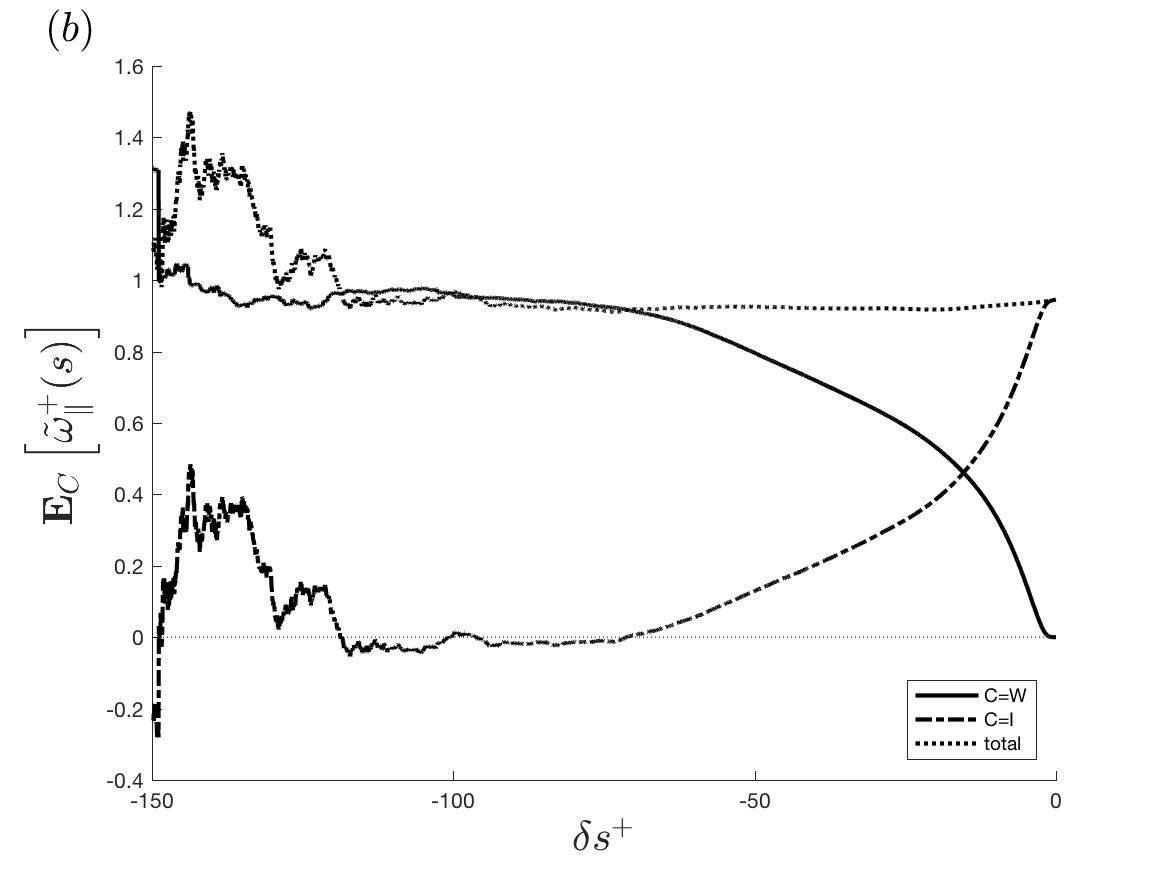}  
\end{subfigure}
\\
\begin{subfigure}[b]{\textwidth}
\centering
  \includegraphics[width=.65\linewidth]{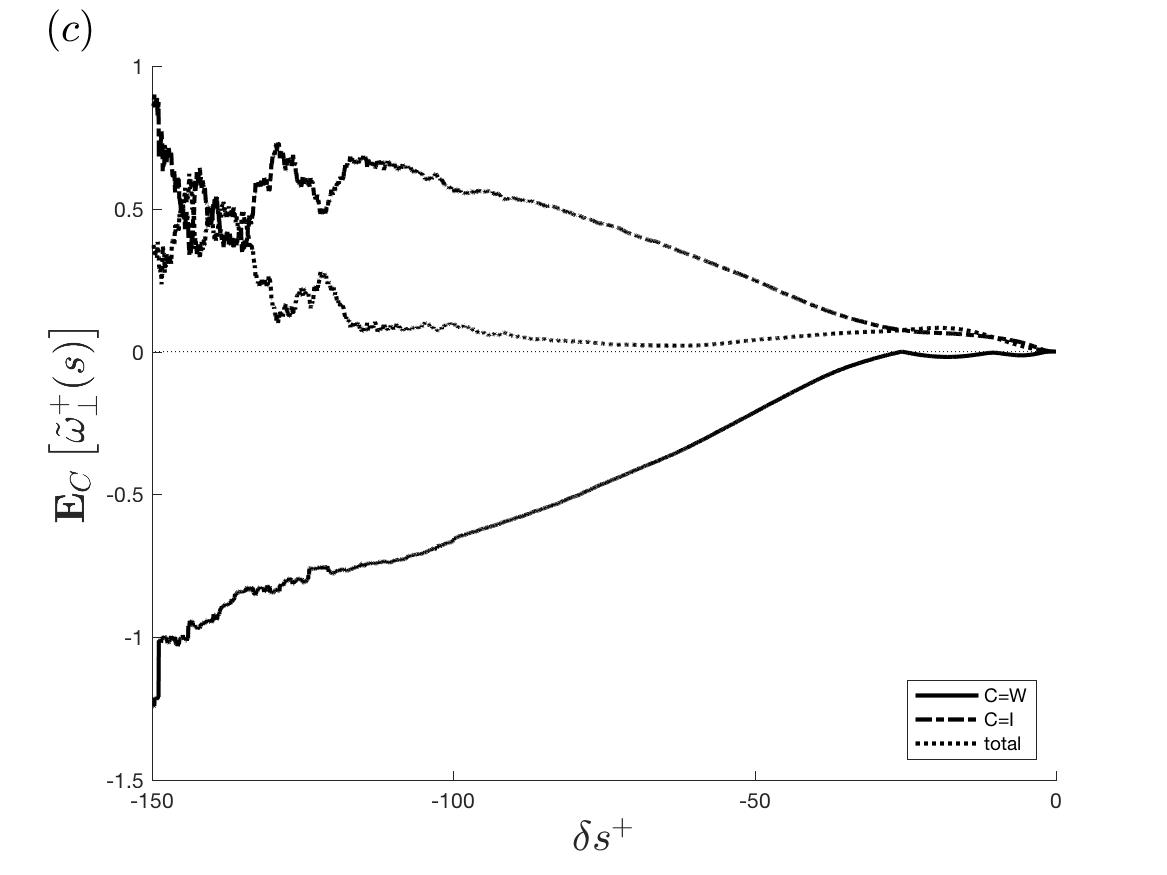}  
\end{subfigure}

\caption{(a) Fraction of particles at the wall, (b) partial means 
of the parallel component of the stochastic Cauchy invariant, and (c) partial means of the 
perpendicular component of the stochastic Cauchy invariant. All quantities in wall units 
plotted versus $\delta s^+.$}
\label{fig5-cauchycontribH}
\end{figure}

\begin{figure} 
\centering
  \includegraphics[width=.8\linewidth]{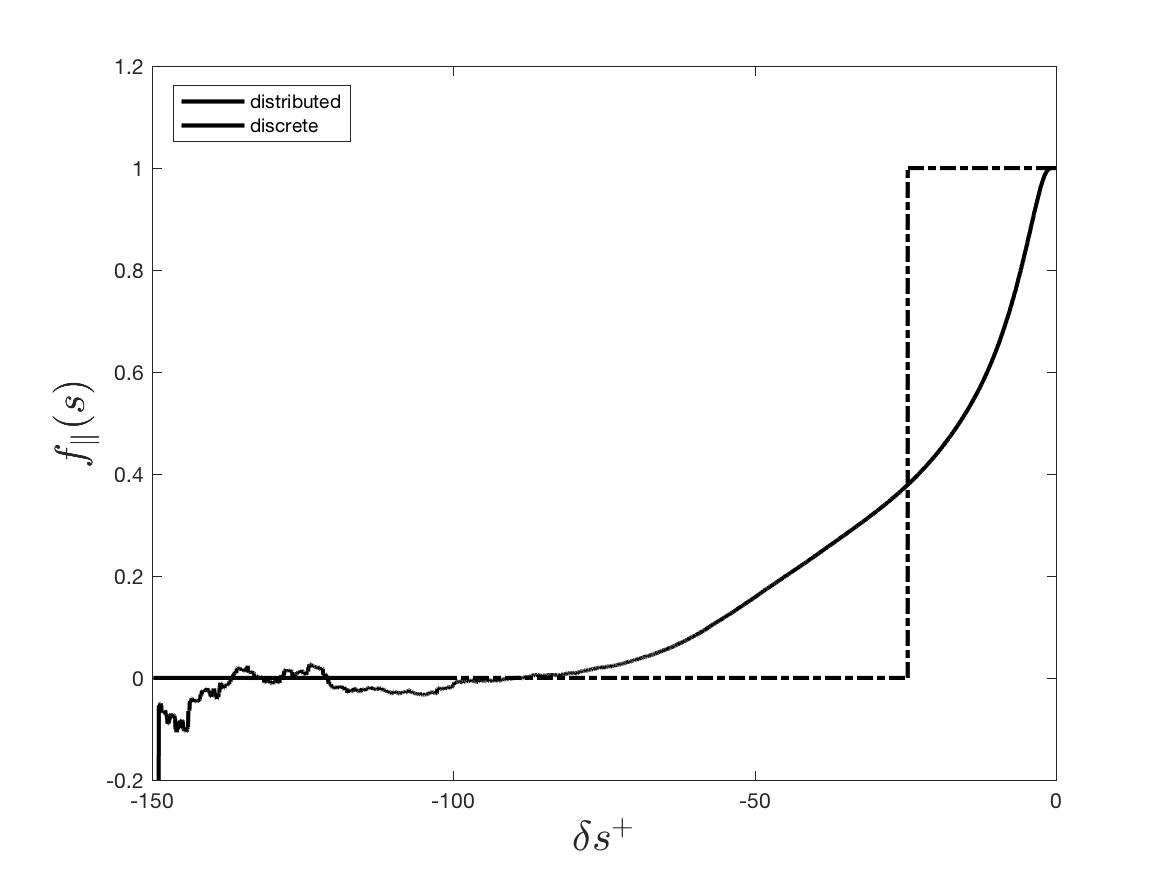}  
\caption{Fractional contribution of the parallel vorticity arising at times $s<t$ from interior particles,  
plotted versus $\delta s^+,$ and its discrete approximation as ``abrupt lifting''.}
\label{fig5-integraltimeH}
\end{figure}

Plotted in Fig.~\ref{fig5-cauchycontribH}(a) is the percentage of particles 
residing at the wall as a function of $\delta s=s-t,$ in wall units.
The percentage is more slowly rising  (backward in time) than for the ejection case 
pictured in  Fig.~\ref{fig5-cauchycontribL}(a) and, indeed, at $\delta s^+=-150$ has 
risen to only 50\%. This is consistent with the expectation that the vorticity is ``older'' in 
the the sweep than it is in the ejection, at the same height above the wall. Since the 
wall-normal velocity is downward in the sweep, the ensemble of particles moves 
generally upward going backward in time and fewer particles hit the wall over the same 
time interval than for the ejection. 

Plotted next in Fig.~\ref{fig5-cauchycontribH} are the partial contributions to the mean
Cauchy invariants arising from the wall $(C=W)$ and from interior $(C=I)$ as 
functions of $\delta s^+,$ for parallel component $(i=\|)$ in panel (b) and perpendicular 
component $(i=\perp)$ in panel (c). Just as for the ejection case, we see that the conservation 
of the mean invariant is non-trivial and involves detailed cancellations between contributions 
from the wall and from the interior. Indeed, for the perpendicular component 
of the stochastic Cauchy invariant in Fig.~\ref{fig5-cauchycontribH}(c) the two separate contributions 
are both {\it increasing} backward in time, roughly linearly in $\delta s^+$ over the range $-150<\delta s^+<0.$  
The entire perpendicular contribution  $\bom_{\perp,int-near}^+$ from the interior at time $\delta s^+=-150$ 
has a magnitude near $|\bom^+|$, the value of the ultimate parallel component. It is not surprising 
to see such a larger perpendicular contribution for the sweep.  The cloud of interior particles rises 
steadily backward in time from $y^+\simeq 5$ at $\delta s^+=0$ to $y^+\simeq 42.5$ at 
$\delta s^+=-150$ (see plot in SM),while simultaneously the number of interior particles as a percentage of the total 
drops from 100\% to 50\%. This subcloud of particles represents the vorticity brought down 
from the interior of the flow by the splatting motion, forward in time. Since the vorticity high 
in the buffer layer is more variable, it is natural that this interior contribution $\bom_{\perp,int-near}^+$ 
to the resultant vorticity is not mainly spanwise but points instead in an orthogonal direction. 
This orthogonal component is exactly cancelled by an equal and opposite contribution 
$\bom_{\perp,wall-near}^+=-\bom_{\perp,int-near}^+$ from the other 50\% of the particles that 
hit the wall in the near-time interval $-150<\delta s^+<0.$

A first conclusion of the results plotted in Fig.~\ref{fig5-cauchycontribH}(c) is that the vorticity 
in the sweep is indeed very ``old'' and arose from the wall in the distant past, as expected.
The quantity $\left|\bE_I\big[\widetilde{\omega}_{s\perp}(\bx,t)\big]\right|$
that appears in the definition \eqref{eq5-8} of the perpendicular integral time 
$T_\perp(\bx,t)$ is apparently increasing past $\delta s^+=-150$ and to values
$>|\bom|,$ before finally decaying to zero. We do not have results well enough converged, 
even with $N=10^7,$ in order to evaulate this time accurately, but it is clear at least that 
$T^+_\perp(\bx,t)\gg 100$ for the sweep.
A more surprising and puzzling conclusion follows, however, from the exact anti-correlation  
$\bom_{\perp,wall-near}^+=-\bom_{\perp,int-near}^+.$  
The interior contribution $\bom_{\perp,int-near}^+$ at time $\delta s^+=-150$ was itself the result of vorticity shed from 
the wall in the far distant past. Going far backward to distant times where nearly 100\% of the particles 
have hit the wall must reproduce it. Thus, $\bom_{\perp,int-near}^+=\bom_{\perp,wall-dist}^+,$
where the latter is the contribution from particles that hit the wall in the distant past
$\delta s^+<-150.$ The immediate implication is that there is also a perfect anti-correlation
 $\bom_{\perp,wall-dist}^+= -\bom_{\perp,wall-near}^+$. 
 In other words, the vorticity shed from the wall in the far distant 
past $\delta s^+<-150$ is making a very sizable contribution to streamwise and wall-normal components of vorticity 
in the trough, with magnitude $|\bom_{\perp,wall-dist}^+|>1,$ but this is {\it exactly} cancelled by a large, exactly 
anti-parallel contribution $\bom_{\perp,wall-near}^+$
from the wall in the near past $-150<\delta s^+<0$!
This very long-range temporal correlation is required by mean conservation
of the value 0 of the perpendicular Cauchy invariant, but it seems a bit counterintuitive, fluid dynamically. 
A possible explanation is that $\bom_{\perp,wall-near}^+$ arises from secondary vorticity induced 
by the strong interaction with the wall of the primary interior vorticity $\bom_{\perp,int-near}^+$ as it is advected 
downward. This picture may help make plausible the exact anti-correlation 
$\bom_{\perp,wall-near}^+=-\bom_{\perp,int-near}^+.$   In any case, our findings emphasize not only 
the great ``age'' of the vorticity vector in the ``trough''  but also its very complex origin at the wall.

\begin{figure}
\hspace{-65pt} 
\includegraphics[width=1.4\linewidth]{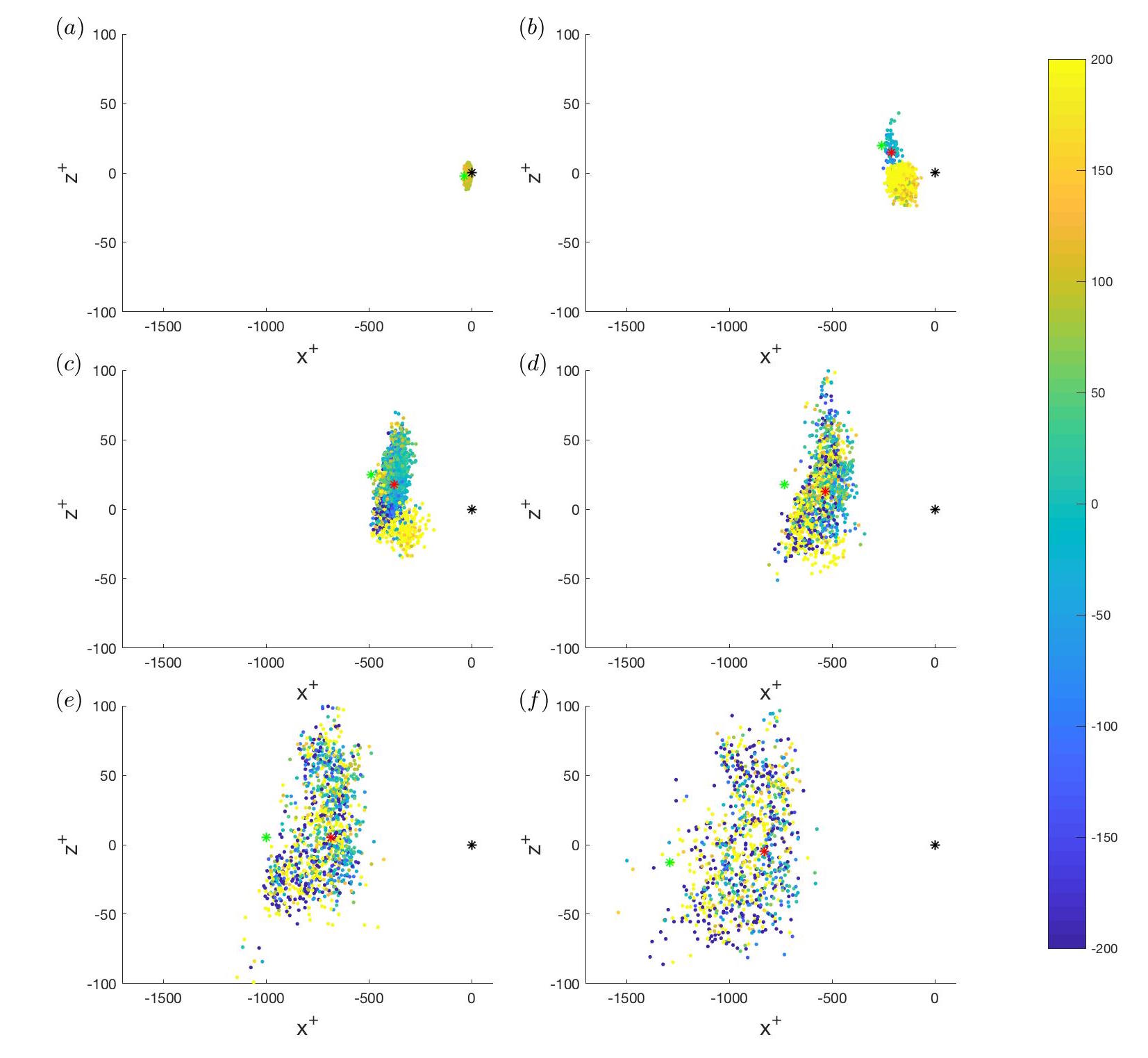}  
\caption{Scatterplot of $\tilde{\omega}_{s\|}^+(\bx,t)$ values for particles hitting the wall in 5-unit 
time intervals ending at: (a) $\delta s^+=-5,$ (b) $\delta s^+=-30,$ (c) $\delta s^+=-55,$  (d) $\delta s^+=-80,$  
(e) $\delta s^+=-105,$  (f) $\delta s^+=-130.$ Mean position of all particles ($\textcolor{red}{*}$) 
and only in interior ($\textcolor{green}{*}$). The resultant vorticity magnitude $|\bom^+(\bx,t)|=0.95.$
}
\label{fig5-movieH}
\end{figure}

A surprise in the opposite direction is that the final parallel component of the vorticity vector 
in the ``trough'', which is pointed almost spanwise, is just about as ``old'' as the 
vorticity vector at the same height on the ``arch'' in the ejection case. Indeed,
the history of formation of the parallel component out of vorticity shed from the wall 
is remarkably similar for the ``ejection'' and the ``sweep'', as can be seen by 
comparing the results in Fig.~\ref{fig5-cauchycontribL}(b) and Fig.~\ref{fig5-cauchycontribH}(b).    
Except for the different magnitudes of the parallel components in the two cases, the 
plots otherwise agree quite closely. This fact is underlined by Fig.~\ref{fig5-integraltimeH},
which plots for the vorticity vector in the ``trough'' the fractional contribution to the parallel 
component arising from the interior. This plot is almost identical to that in  Fig.~\ref{fig5-integraltimeL}
for the ejection case. The integral time of formation of the parallel component calculated 
from  Eq.\eqref{eq5-7} must therefore be similar also for the two cases. Indeed the value 
obtained by numerical quadrature for the sweep, $T_\|^+(\bx,t)\doteq 24.6,$ is just slightly larger 
than the value  $T_\|^+(\bx,t)\doteq 20.7$ for the ejection.

To further explore this issue we have made for the sweep case a figure of the same type as Fig.~\ref{fig5-movieL} 
for the ejection, again plotting realizations of the parallel component of the stochastic Cauchy invariant. 
See Fig.~\ref{fig5-movieH}, which uses the same time intervals and the same size subensemble  of $10^6$
particles as in the earlier plot. In the SM, we provide a video with greater time resolution and 
employing all $10^7$ available particles. This plot only deepens the mystery, however. 
The plots in Fig.~\ref{fig5-movieH} are broadly similar to those in Fig.~\ref{fig5-movieL} for the ejection, 
but also show significant differences. The clouds of particles are clearly more compact for the 
sweep case and disperse less quickly going backward in time. Furthermore, the spatial pattern
at the wall of the parallel Cauchy invariant values is strikingly ``bipartite'' for the sweep at times
$-60<\delta s^+<0,$ with large positive values in the lower half of the particle cluster and 
large negative values in the upper half. This pattern is presumably due to the rotation of vorticity 
vectors by the pair of low-lying quasi-streamwise vortices pictured in Fig.~\ref{fig4-lam2H}(a),
and it is much more ordered than the pattern for the ejection case in Fig.~\ref{fig5-movieL}. 
Despite these clear differences in the spatial patterns in the two plots, the summed results 
yield the time variations plotted in Fig.~\ref{fig5-integraltimeL} and Fig.~\ref{fig5-integraltimeH},
which are remarkably similar.

\subsubsection{Relation to the Eulerian Vorticity Source}

\begin{figure}
\hspace{-65pt} 
\includegraphics[width=1.4\linewidth]{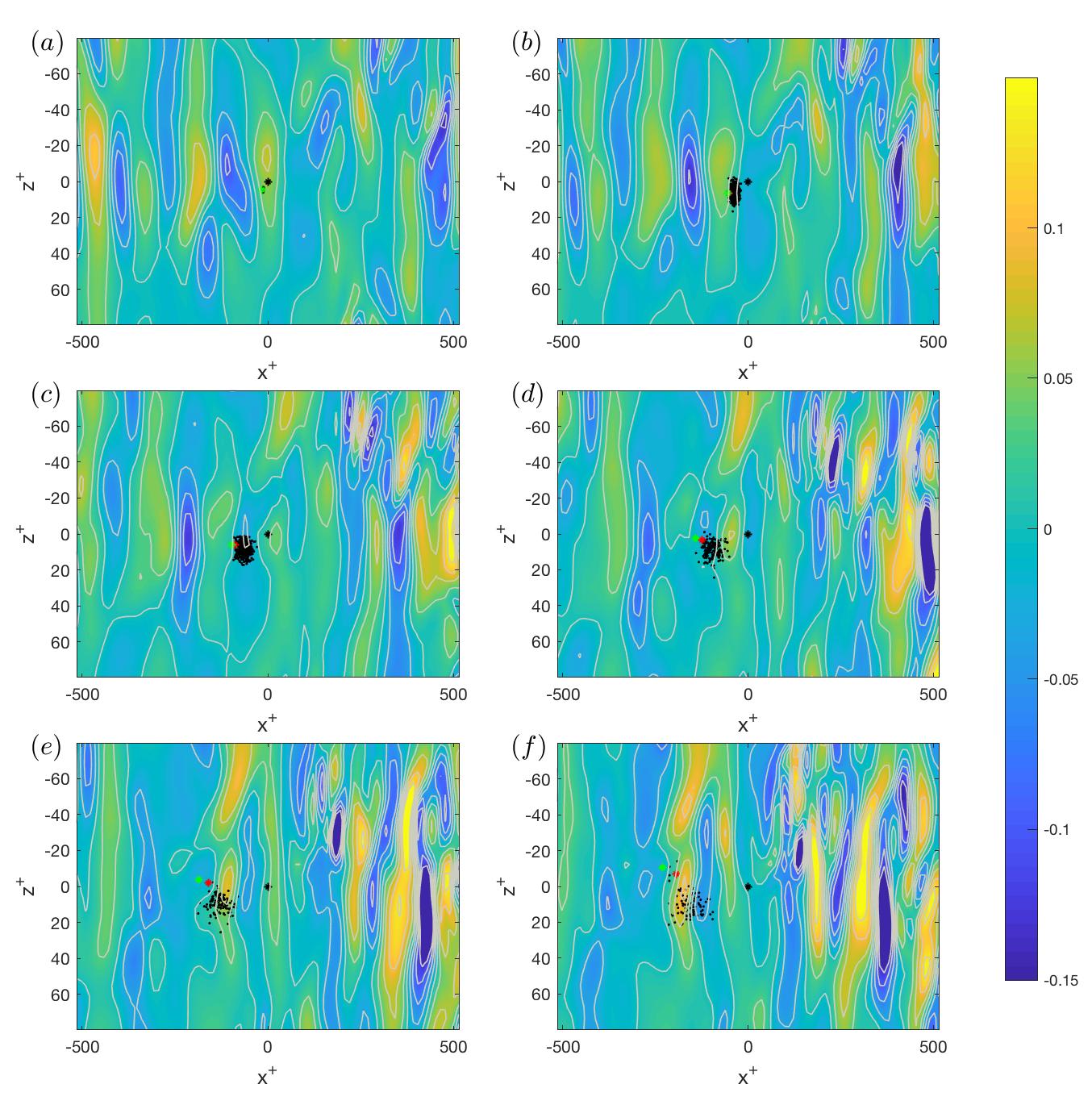}  
\caption{Color plots of negative spanwise vorticity source together with scatterplots of particles hitting the wall
in unit time intervals ending at: (a) $\delta s^+=-1,$ (b) $\delta s^+=-6,$ (c) $\delta s^+=-11,$  
(d) $\delta s^+=-16,$ (e) $\delta s^+=-21,$  (f) $\delta s^+=-26.$ The light grey lines are contour
levels $-\sigma_z^+=k/30$ for integers $|k|\leq 4.$  Also shown are the mean position of all particles 
($\textcolor{red}{*}$) and of only those in the interior ($\textcolor{green}{*}$). }
\label{fig5-pressuremovieH}
\end{figure}

Some further insight may be gained by considering the Eulerian picture. In Fig.~\ref{fig5-pressuremovieH} 
we plot the negative spanwise vorticity source $-\sigma_z(\bx,s)$ together with a scatterplot of particles 
landing at the wall. We consider the same set of times $\delta s^+=-(5k+1),$ $k=0,1,2,3,4,5$ as in 
Fig.~\ref{fig5-pressuremovieL} for the ejection, going backward by 26 viscous times. This is close in 
value to the integral shedding time $T_\|^+\doteq 24.6$ for the sweep. A movie with greater 
time-resolution and going back further in time is provided in the SM. As before, the first panel, 
Fig.~\ref{fig5-pressuremovieH}(a), is essentially the same as Fig.~\ref{fig4-pressH}(b) but over a larger 
spatial domain. It was conjectured by \cite{andreopoulos1996wall} that positive wall sources $\sigma_z>0$ 
should be associated with sweeps. Indeed, just upstream of the point marked with an asterisk ``$\ast$''
in Fig.~\ref{fig5-pressuremovieH}(a) there is a band with large values $\sigma_z\doteq 0.1.$  However, 
as shown by the subsequent panels (or by the movie, in more detail) the particles hit that region of the wall 
with negligible probability backward in time, since this band with $\sigma_z>0$ retreats upstream 
with high velocity $\sim 10u_*.$  Instead, the particles hit mainly regions with $\sigma_z<0$ over 
the time interval shown in Fig.~\ref{fig5-pressuremovieH}. In particular, there is a space band with 
moderately negative values originally about 200 wall units downstream of the point marked ``$\ast$'' 
which travels rapidly upstream backward in time and which intensifies to values $\sigma_z\doteq -0.1.$
The particles released from the depressed vortex line at height $y^+=4.90$ are sampling 
mainly points in this band of strong negative source $\sigma_z\doteq -0.1$ over the time interval 
$-25<\delta s^+<-16$. In fact, the pattern of vorticity source sampled by the particles in this sweep 
case is rather similar to that sampled in the ejection case, pictured in Fig.~\ref{fig5-pressuremovieL}.  
This seems to contradict the proposal of \cite{andreopoulos1996wall}. Here it should be noted 
that nonlinear advection contributes also in the sweep case a {\it negative} flux of spanwise 
vorticity away from the wall over much of the region $y^+<15.$  As illustrated in Fig.\ref{fig4-domzH},
positive fluctuations of spanwise vorticity $\omega_z'>0$ are here often associated pointwise
with downward fluctuations $v'<0$ of wall-normal velocity.     

These results suggest a possible explanation for the surprising agreement of Fig.~\ref{fig5-integraltimeL} 
for the ejection and Fig.~\ref{fig5-integraltimeH} for the sweep. This agreement could be 
expected if viscous diffusion dominates the transport of spanwise vorticity in the viscous sublayer, 
not only on average \citep{klewicki2007physical,eyink2008turbulent}, but also in instantaneous realizations 
of the flow. The structure of in-wall vortex lines is very similar for the ejection and sweep, with 
magnitudes $|\bom^+|\doteq 1$ and pointed mainly in the spanwise direction (or negative spanwise
direction for rotated visualizations). Of course, these nearly uni-directional wall vorticity vectors are strongly 
stretched and rotated by the stochastic Lagrangian flow and yield both large positive and large negative
values of the parallel stochastic Cauchy invariant, as illustrated in Figs.~\ref{fig5-movieL}, \ref{fig5-movieH}
for the ejection and sweep, respectively. It is possible however that these large values almost completely cancel
and leave only the contributions of viscous diffusion and advection by the mean velocity. Here we note that 
the viscous fluxes/wall sources of spanwise vorticity sampled by the stochastic particles are also quite large. 
The values $\sigma_z^+\doteq -0.1$ sampled in both the ejection and sweep events for this $Re_\tau=1000$ 
simulation are $\sim 100$ times larger than the mean value $\langle \sigma_z^+\rangle \doteq -1\times 10^{-3}.$
The very similar temporal profiles in Fig.~\ref{fig5-integraltimeL} and Fig.~\ref{fig5-integraltimeH} may just 
reflect a common origin of the spanwise vorticity at height $y^+\doteq 5$ in the parallel-transport of initially 
spanwise wall-vorticity by viscous diffusion and by advection due to the mean shear-velocity.

In summary, our results for the sweep rather dramatically contradict the naive idea that vortex lines are 
approximately ``frozen-in'' and advected by the flow. If that idea were correct, the spanwise vorticity at the bottom 
of the depressed vortex lines over the high-speed streak would have been swept down from the interior of the flow. 
We have found that there is indeed a large non-spanwise vorticity contribution swept down from the interior 
in the near past, but this contribution is exactly cancelled by an equal and opposite vorticity originating
from the wall in that same period. This cancelling contribution from the wall can be plausibly explained 
to arise from opposite-signed vorticity induced by interactions of the solid wall and the downswept interior vorticity. 
In contrast to this ``old'' interior vorticity, the resultant (mainly spanwise) vorticity at the bottom of the depressed 
vortex line at $y^+=5$ is much ``younger'', arising from vorticity shed by the wall in the near past 
and perhaps transported primarily by viscous diffusion. 
 




\section{Conclusions and Prospects}\lb{sec:conclusion} 

We have presented a first concrete application of our Monte Carlo numerical Lagrangian 
method to channel-flow turbulence, using an online database of a high-$Re$ channel-flow
simulation \citep{graham2016web}. We have analyzed the Lagrangian vorticity dynamics 
for two specific events in the near-wall buffer layer, a pair selected as generic examples of 
an ``ejection'' and a ``sweep''. We find that the growth of vorticity magnitude due to nonlinear 
Lagrangian chaos is compensated by viscous cancellation of oppositely-signed vector components, 
or viscous reconnection in a generalized sense. We may refer to this as ``virtual reconnection'',
because spanwise anti-parallel components of the Eulerian vorticity field almost never appear in the 
viscous sublayer but arise by nontrivial Lagrangian dynamics in an average over virtual, stochastic 
processes. We find also that vortex-lifting from the wall is a highly distributed development 
in space-time, not an abrupt, discrete event. Consistent with expectations, we find that, 
at the same distance from the wall,  the vorticity in the ``sweep'' event is older 
than in the ``ejection'' event, being birthed at the wall in the more distant past.
Surprisingly, however, the greater age is evidenced only by a persistent orthogonal contribution 
from the wall, which requires many hundreds of viscous times to decay, whereas the vorticity
in the ultimate parallel direction is assembled over $\sim$100 viscous times in a similar manner for both events. 

Future numerical studies should examine many more such events, to verify whether such features 
are typical and to reach statistically relevant conclusions. Such studies should illuminate the detailed 
Lagrangian mechanisms of the organized transport of spanwise vorticity away from the wall, which is 
required for turbulent drag and dissipation. More empirical studies are required even of the Eulerian
turbulent vorticity flux, which has been examined much less than momentum transport and Reynolds 
stress, in order to understand which flow structures sustain the required flow of spanwise 
vorticity. As emphasized by \cite{brown2012turbulent}, ``The subject stands at the beginning of a new 
era in which both LES and DNS calculations can provide details of the vorticity field and the fluxes 
of vorticity (vortex force).'' The stochastic Lagrangian methods developed in our work can provide 
even deeper insight into the dynamics, especially when tightly integrated with the Eulerian picture.
We take some further steps in this direction in the following paper of \cite{eyink2020stochastic},
which develops a stochastic Lagrangian representation in which the Eulerian vorticity 
source of \cite{lighthill1963boundary}-\cite{morton1984generation} is incorporated as Neumann 
boundary conditions for the Helmholtz equation via stochastic particle trajectories that are reflected
from the wall. 
 
A remarkable aspect of the stochastic Lagrangian theory is the many unifying features that 
emerge naturally between classical and quantum fluids.   
Even based upon our preliminary results, we can make some relevant comparisons with 
quantum turbulence in superfluids \citep{barenghi2014introduction}. The regime with the 
closest correspondence to the classical case is two-fluid turbulence in ${\,\!}^4$He.
Forced flows of superfluid ${\,\!}^4$He in the two-fluid regime through smooth-wall tubes 
and square channels at high Reynolds numbers suffer a pressure-drop in reasonable
agreement with classical friction laws \citep{swanson2000turbulent,fuzier2001steady} and velocity profiles determined 
from particle-imaging velocimetry exhibit a near-wall turbulent boundary layer \citep{xu2007particle}.  
A factor in favor of such classical correspondences is the locking or coupling of the 
two fluid components \citep{vinen2000classical,kivotides2007relaxation}, but complete 
understanding of the similar behaviors is still lacking. We believe that the theory developed
in this paper and paper I may assist in developing such explanations, because of the several connections it 
exposes between classical and quantum fluids. 

To emphasize this point, we briefly summarize here some of the common  features. 
The \cite{kuzmin1983ideal}-\cite{oseledets1989new} 
formulation of Navier-Stokes dynamics in terms of a continuous distribution of infinitesimal 
vortex-rings is very similar to the intuitive picture of a quantized vortex tangle proposed by  
\cite{campbell1972critical}, as a superposition of small vortex rings, which was invoked 
by \cite{schwarz1982generation} to explain intuitively the phase-slip process in superfluid 
turbulence. See also \cite{kuzmin1999vortex}. \cite{huggins1994vortex} has already emphasized 
that constant flux of a conserved vorticity current is necessary for dissipative drag in both classical 
and superfluid turbulent channel-flows (see also \cite{eyink2008turbulent}). The mechanism 
in quantum turbulence proposed by \cite{schwarz1988three,schwarz1990phase}, based 
on his vortex-filament simulations, was the cross-stream ``ballooning'' of ring vortices, which 
are ultimately driven to annihilate at the wall. As observed by \cite{huggins1994vortex}, an 
equivalent flux in the classical case results from vorticity creation at the wall and subsequent 
transport to the channel center, where opposite orientations cancel.
Last but not least, \cite{eyink2010stochastic} showed that mean conservation of 
the stochastic Cauchy invariants and stochastic circulations (Kelvin Theorem) for 
incompressible Navier-Stokes solutions arise from particle relabelling symmetry 
in a stochastic least-action principle. These conservation laws hold, in close analogy to those 
for ideal Euler equations, although viscosity vitiates the standard ``frozen-in'' property 
of vorticity and permits vortex reconnection. Similarly, it has been shown for superfluids, 
both in the zero-temperature Gross-Pitaevskii model \citep{nilsen2006velocity} and in strongly 
rotating, chiral flows \citep{wiegmann2019quantization}, that the Kelvin theorem holds 
while simultaneously quantized vortices are {\it not} frozen into the flow. For the chiral flows 
the Kelvin theorem is derived as a consequence of particle relabelling symmetry \citep{wiegmann2019quantization}, 
while this issue seems open for Gross-Pitaevskii.  The contrary finding of \cite{kedia2018helicity}
with the Thomas-Fermi approximation neglects the quantum pressure, which is crucial to determine  
the correct motion of quantized vortices \citep{nilsen2006velocity} and to permit vortex 
reconnection \citep{koplik1993vortex}.

This underlying unity has importance because fewer differences
between classical and quantum fluids appear starker than the differences in reconnection physics. 
Vortex reconnection is believed to be a crucial element of superfluid turbulence, as discussed 
in works of \cite{schwarz1982generation,schwarz1988three,schwarz1990phase}, and we have 
argued that reconnection in some generalized sense is an essential feature also of classical channel-flow 
turbulence. In a classical fluid, vortex lines are continuously distributed in space and cannot be 
unambiguously tracked in time. Vorticity may be attributed a stochastic law of motion 
and viscous reconnection then results from cancellations in averaging random contributions, just 
as for the similar case of resistive magnetic reconnection in plasmas \citep{eyink2013flux}. The vorticity of 
a superfluid is quantized, on the other hand, and individual segments of vortex-lines are topological defects 
that may be followed objectively and deterministically.  We believe, however, that stochastic laws 
of motion similar to those for classical viscous fluids will hold also in superfluid turbulence for coherent 
``bundles'' of quantized vortex lines \citep{lvov2007bottleneck,baggaley2012coherent}. There 
seems no reason to doubt that individual vortex lines in a superfluid regime with a Kolmogorov 
energy spectrum \citep{nore1997decaying,barenghi2014introduction} will exhibit ``spontaneous stochasticity'' due to 
turbulent Richardson dispersion \citep{bernard1998slow,drivas2017lagrangianI}.  Because of such 
explosive dispersion effects and ubiquitous microscopic reconnection, the motion and collective reconnection 
of vortex bundles in superfluids \citep{alamri2008reconnection} should appear stochastic just 
as in classical viscous fluids. Stochastic Lagrangian invariants would naturally arise in a dissipative 
effective action for coarse-grained fields preserving relabelling symmetry.    

\section*{Acknowledgements}

We are grateful to Joseph Katz, Joseph Klewicki, Demosthenes Kivotides, Charles Meneveau, and Pablo Mininni 
for useful discussions.  We thank Paul Weigmann for informing us 
of his unpublished work. We acknowledge the NSF grant BigData:OCE-1633124 for support and G.E. 
also acknowledges the Simons Foundation through Targeted Grant in MPS-663054 for partial support. 
This research project was conducted using simulation data from the Johns Hopkins Turbulence 
Database (JHTDB) and scientific computing services at the Maryland Advanced Research Computing 
Center (MARCC).

\bibliographystyle{jfm}

\bibliography{bibliography}

\end{document}